\documentclass[11pt,twoside,a4paper,cmspaper,final,collab]{cms-tdr}
\def\svnVersion{98927bf-D}\def\svnDate{2026/05/13}\def\cmsCernNoTag{CERN-EP-2026-041}\def\cmsCernDate{\today}\def\cmsMessage{Submitted to the Journal of High Energy Physics}
\begin{document}\cmsNoteHeader{SMP-24-017}

\newcommand{\WWW}{\ensuremath{\PW\PW\PW}\xspace}
\newcommand{\WWZ}{\ensuremath{\PW\PW\PZ}\xspace}
\newcommand{\WZZ}{\ensuremath{\PW\PZ\PZ}\xspace}
\newcommand{\ZZZ}{\ensuremath{\PZ\PZ\PZ}\xspace}
\newcommand{\VVV}{\ensuremath{\PV\PV\PV}\xspace}
\newcommand{\VVZ}{\ensuremath{\PV\PV\PZ}\xspace}
\newcommand{\WVV}{\ensuremath{\PW\PV\PV}\xspace}
\newcommand{\ttV} {\ensuremath{\PQt\PQt\PV}\xspace}
\newcommand{\WZ}{\ensuremath{\PW\PZ}\xspace}
\newcommand{\fatjet}{\PV-tagged jet\xspace}
\newcommand{\fatjets}{\PV-tagged jets\xspace}
\providecommand{\ParticleNet}{{\textsc{ParticleNet}}\xspace}
\providecommand{\combine}{{\textsc{Combine}}\xspace}
\providecommand{\DeepTau}{{\textsc{DeepTau}}\xspace}
\providecommand{\DeepJet}{{\textsc{DeepJet}}\xspace}
\newcommand{\ST} {\ensuremath{S_\text{T}}\xspace}
\newcommand{\STnoMET} {\ensuremath{S_\text{T, no~\ptmiss}}\xspace}
\newcommand{\ptfatjet} {\ensuremath{p_{\text{T, \PV~jet}}}\xspace}
\newcommand{\ptjet} {\ensuremath{p_{\text{T, jet}}}\xspace}
\newcommand{\ptlep} {\ensuremath{p_{\text{T,\Pell}}}\xspace}
\newcommand{\pttau} {\ensuremath{p_{\text{T,\PQt}}}\xspace}
\newcommand{\mVVV} {\ensuremath{m_{\VVV}}\xspace}
\newcommand{\MJJlnu}{\ensuremath{m_{\mathrm{JJ}\Pell\PGn}}\xspace}
\newcommand{\PTlepW}{\ensuremath{p_\text{T,lepW}}\xspace}
\newcommand{\Mll}{\ensuremath{m_{\Pell\Pell}}\xspace}
\newcommand{\MZ}{\ensuremath{m_{\PZ}}\xspace}
\newcommand{\ISOL}{\ensuremath{I_{\Pell}}\xspace}
\newcommand{\mSD} {\ensuremath{m_{\text{SD}}}\xspace}
\newcommand{\mSDone} {\ensuremath{m_{\text{SD,1}}\xspace}}
\newcommand{\mSDtwo} {\ensuremath{m_{\text{SD,2}}\xspace}}
\newcommand{\mSDthree} {\ensuremath{m_{\text{SD,3}}\xspace}}
\newcommand{\noLEPtwoFJ}        {SR-0\Pell-2VTJ\xspace}
\newcommand{\noLEPthreeFJ}      {SR-0\Pell-3VTJ\xspace}
\newcommand{\oneLtwoFJ}         {SR-1\Pell-2VTJ\xspace}
\newcommand{\oneLEPtwoFJ}       {SR-1\Pell-2VTJ\xspace}
\newcommand{\twoLEPOSoneFJ}     {SR-2\Pell-OS-1VTJ\xspace}
\newcommand{\twoLEPOSOFFZoneFJ} {SR-2\Pell-OSoffZ-1VTJ\xspace}
\newcommand{\twoLEPOSONZoneFJ}  {SR-2\Pell-OSonZ-1VTJ\xspace}
\newcommand{\twoLEPOStwoFJ}     {SR-2\Pell-OS-2VTJ\xspace}
\newcommand{\twoLEPOSOFoneFJ}   {SR-2\Pell-OSDF-1VTJ\xspace}
\newcommand{\twoLEPSSoneFJ}     {SR-2\Pell-SS-1VTJ\xspace}
\newcommand{\oneLEPoneTAUoneFJ} {SR-1\Pell-1\tauh-1VTJ\xspace}
\newcommand{\twoLEPoneTAUnoFJ}  {SR-2\Pell-1\tauh-0VTJ\xspace}
\newcommand{\nbinstotal}  {\ensuremath{37}\xspace}
\newcommand{\cW}        {\ensuremath{c_\PW}\xspace}
\newcommand{\cWvalue}   {0.123}
\newcommand{\cHqthree}  {\ensuremath{c_{\text{Hq3}}}\xspace}
\newcommand{\cHqone}    {\ensuremath{c_{\text{Hq1}}}\xspace}
\newcommand{\cHu}       {\ensuremath{c_{\text{Hu}}}\xspace}
\newcommand{\cHd}       {\ensuremath{c_{\text{Hd}}}\xspace}
\newcommand{\cHW}       {\ensuremath{c_{\text{HW}}}\xspace}
\newcommand{\cHWB}      {\ensuremath{c_{\text{HWB}}}\xspace}
\newcommand{\cHlthree}  {\ensuremath{c_{\text{H}\Pell3}}\xspace}
\newcommand{\cHB}       {\ensuremath{c_{\text{HB}}}\xspace}
\newcommand{\cllone}    {\ensuremath{c_{\Pell\Pell1}}\xspace}
\newcommand{\cHsquare}  {\ensuremath{c_{\text{H}\square}}\xspace}
\newcommand{\cHDD}      {\ensuremath{c_{\text{H}DD}}\xspace}
\newcommand{\fT}[1]     {\ensuremath{f_{\text{T},#1}}\xspace}
\newcommand{\fM}[1]     {\ensuremath{f_{\text{M},#1}}\xspace}
\newcommand{\fS}[1]     {\ensuremath{f_{\text{S},#1}}\xspace}
\newcommand{\LSM}       {\ensuremath{L_{\text{SM}}}\xspace}
\newcommand{\muSM}      {\ensuremath{\mu_{\text{SM}}}\xspace}
\newcommand{\twoDNLL}   {{\ensuremath{2\Delta}{\text{NLL}}}\xspace}
\newcommand{\PPHI}      {\Phi}
\newcommand{\PPHIdagger}{\PPHI^{\dagger}}
\newcommand{\Dlra}      {\overleftrightarrow{D}}
\newlength\cmsTabSkip\setlength{\cmsTabSkip}{1ex}

\cmsNoteHeader{SMP-24-017}
\title{Search for new physics in triple boson production in proton-proton collisions at \texorpdfstring{$\sqrt{s} = 13\TeV$}{sqrt(s) = 13 TeV} using the effective field theory approach}

\date{\today}

\abstract{
A search for new physics in the production of three massive gauge bosons (\VVV, where \PV is a \PW or \PZ boson) is presented. The event selection is most effective in the Lorentz-boosted regime in which all three bosons have a transverse momentum (\pt) above 200\GeV. Standard model (SM) processes contribute few events in this regime. When a boosted \PW or \PZ boson decays hadronically, the decay products tend to form a large-radius jet with substructure that reflects the presence of two quarks from the decay; such jets are called \PV-tagged jets. Special techniques to reconstruct and select \PV-tagged jets are applied. Events are categorized according to the number and kinematic features of charged leptons and \PV-tagged jets. Event yields are obtained in bins of a suitable kinematic variable such as the scalar \pt sum of the reconstructed objects in the event. No excess over SM expectations is observed. Bounds are placed on Wilson coefficients for a set of mass dimension-6 and -8 operators in the framework of SM effective field theory. The two most stringent bounds placed by this analysis are $-0.13 < \cW/\Lambda^2 < 0.12\TeV^{-2}$ and $-0.24 < \cHqthree/\Lambda^2 < 0.21\TeV^{-2}$ at 95\% \CL, where \cW and \cHqthree are dimension-6 Wilson coefficients in the Warsaw basis and $\Lambda$ is the mass scale of new physics.}

\hypersetup{
pdfauthor={CMS Collaboration},
pdftitle={Search for new physics in triple boson production in proton-proton collisions at sqrt(s) = 13 TeV using the effective field theory approach},
pdfsubject={CMS},
pdfkeywords={CMS, EFT, multiboson production}}

\maketitle

\section{Introduction}
\label{sec:Intro}

The standard model (SM) predicts that the production of three massive gauge bosons (collectively referred to
as \VVV, where \PV is a \PW or \PZ boson) occurs in proton-proton ($\Pp\Pp$) collisions at the CERN LHC.
The production of these rare events is sensitive to both triple and quartic gauge couplings, making a
measurement of this production process an interesting probe of the electroweak sector of the SM.
New particles and interactions at mass scales well above 1\TeV could modify these couplings resulting
in changes to certain kinematic distributions. This sensitivity motivates the search for new physics (NP)
beyond the SM presented in this paper. We formulate this search in the framework of SM effective field
theory (SMEFT)~\cite{Grzadkowski:2010es,Brivio:2020onw}. The most significant evidence for NP might be
obtained in a kinematic regime in which all three vector bosons are Lorentz-boosted, \ie, typically they
have high transverse momentum $\pt > 200\GeV$.

In this paper, \VVV production includes \WWW, \WWZ, \WZZ, and \ZZZ production.
Combined \VVV production has been observed by the CMS Collaboration including individual evidence for the production of
\WWW and \WWZ in $\sqrt{s} = 13\TeV$ $\Pp\Pp$ collisions~\cite{CMS:2020hjs}
and separate evidence for \WWZ at 13.6\TeV~\cite{CMS:2025hlu}.
The ATLAS Collaboration has reported observations of \WWW production~\cite{ATLAS:2022xnu},
\VVZ production~\cite{ATLAS:2025dir}, and evidence for \WVV production~\cite{ATLAS:2025dir}.
The sensitivity to the individual contributions to \VVV production differs from channel to channel,
ranging from an observed significance of more than five standard deviations for \WWW, to an upper limit of 5.4 times the
SM prediction for \ZZZ production.

This analysis is based on a sample of $\Pp\Pp$ collisions
produced by the LHC at a center-of-mass energy of 13\TeV and recorded with the CMS detector in 2016--2018.
This data set corresponds to an integrated luminosity of 138\fbinv.
Events are categorized based on the numbers of charged leptons and \PV-tagged jets in the event.
A \PV-tagged jet is a large-radius jet with $\pt > 200\GeV$ and jet substructure characteristic of the
hadronic decay of a boosted vector boson.

From a theoretical standpoint, the observable of choice would be the \VVV invariant mass, {\mVVV}.
The experimental reconstruction of \mVVV is often not possible, however, due to the presence of one or more neutrinos
in many of the final states selected for analysis. We therefore approximate \mVVV using \ST, the scalar \pt sum
of the leptons and jets in an event. The definition of \ST varies slightly depending on the final state.
The \ST distributions are binned and range from a
low-\ST regime dominated by SM processes to a high-\ST regime where a signal could emerge above a very small SM background.
The observed yield in each bin, the expected SM contributions, and the predicted signal as a function of
one or more Wilson coefficients characterizing NP are used as inputs to a maximum-likelihood fit
based on the CMS statistical analysis tool
\combine and the CMS anomalous coupling interface~\cite{Combine:2024csbi,Boldrini-PhD}.
This fit enables us to place bounds on Wilson coefficients at 95\% confidence level (\CL).

This paper is structured as follows.
Section~\ref{sec:EFTs} presents the effective field theory (EFT) framework used in this analysis;
this is followed by a brief description of the CMS detector in Section~\ref{sec:Detector}.
The Monte Carlo simulations are described in Section~\ref{sec:Samples}.
The offline reconstruction of events and the requisite tools are discussed in Section~\ref{sec:Reconstruction}.
Section~\ref{sec:Strategy} presents the analysis strategy and is followed by the details of the
individual signal channels in Section~\ref{sec:Channels}.
The sources and assessments of systematic uncertainty are discussed in Section~\ref{sec:Systematics}.
Our results are presented in Section~\ref{sec:Results}, and we provide a summary in Section~\ref{sec:Conclusion}.
Tabulated results are provided in the HEPData record for this analysis~\cite{hepdata}.

\section{EFTs and triboson production}
\label{sec:EFTs}

Effective field theories offer a flexible and general theoretical framework for analyses seeking indirect evidence for new physics.
The starting point is the SM Lagrangian, \LSM, which has mass dimension four. This Lagrangian can be systematically extended by adding terms of higher mass dimension to \LSM. In this analysis, we consider Wilson operators of dimensions six and eight (referred to as dim-6 and dim-8):
\begin{equation}
 L = \LSM +
 \sum_{\text{dim-}6} \frac{c_i}{\Lambda^2} {\cal{O}}_i +
 \sum_{\text{dim-}8} \frac{f_j}{\Lambda^4} {\cal{O}}_j .
\label{eq:Lagrangian}
\end{equation}
Here, $\Lambda$ is the mass scale of new physics, ${\cal{O}}$ is a particular Wilson operator,
and $c_i$ and $f_j$ are dim-6 and dim-8 Wilson coefficients whose absolute values should not be
very large and are zero in the SM.
The goal of this analysis is to determine the Wilson coefficients. A value for one or more of them that is significantly different from zero would be a sign of new physics. The Wilson operators ${{\cal{O}}_i}$ are not uniquely defined.
We follow LHC conventions and use the Warsaw~\cite{Grzadkowski:2010es} and Eboli~\cite{Corbett:2024yoy} bases
for dim-6 and -8 operators.

The Lagrangian in Eq.~(\ref{eq:Lagrangian}) leads, for any specific process, to a matrix element that can be written
\begin{equation}
 M = M_{\text{SM}} + c\, M_{\text{NP}}
\label{eq:MatrixElement}
\end{equation}
where $M_{\text{SM}}$ is the SM matrix element, $M_{\text{NP}}$ is the matrix element for the terms with mass dimension
six or eight, and $c$ is a Wilson coefficient.
It follows that any differential cross section has the following dependence on $c$,
\begin{equation}
 \sigma(c) = \sigma_{\text{SM}} + c \sigma_{\text{lin}} + c^2 \sigma_{\text{quad}} ,
\label{eq:CrossSection}
\end{equation}
which leads to a quadratic dependence of the signal yields on $c$.
A term independent of all Wilson coefficients represents the SM expectation and is treated as a background.
A term quadratic in the coefficients represents a purely NP contribution, and a linear term represents an
interference between the SM and NP matrix elements.
In general, the linear term may dominate the quadratic term, or vice versa.
For the analysis presented in this paper, the sensitivity is dominated by the quadratic term.
As \VVV production is an SM process, \VVV events are not a signal per se.
A signal for new physics is associated with $\abs{c} > 0$.

The twelve dim-6 and twenty dim-8 operators studied in this analysis are listed in
Tables~\ref{tab:dim6} and~\ref{tab:dim8}.
In these tables, $\PW_{\mu\nu}^{I}$ is the SU(2)$_{L}$ field strength,
$B_{\mu\nu}$ is the U(1)$_{Y}$ field strength,
and $\Phi$ is the Higgs doublet.
Additionally, $q$ and $\Pell$ represent the left-handed quark and lepton doublet fields, respectively.
This list is obtained by varying all SMEFT operators one at a time while keeping others fixed to zero,
and examining the predicted $\mVVV$ spectrum for significant deviations,
in both its rate and shape, relative to the SM prediction.

\begin{table}[htb]
\centering
\topcaption{\label{tab:dim6} The set of 12 dim-6 operators studied in this analysis}
\begin{tabular}{lccl}
\hline
Operator ${\cal{O}}_i$ & Wilson coefficient\\
\hline
\multicolumn{2}{c}{Gauge boson self-interaction} \\
$\epsilon^{IJK}\PW_{\mu}^{I\nu}\PW_{\nu}^{J\rho}\PW_{\rho}^{K\mu}$ & $\cW$ \\
$\PPHIdagger\PPHI\, \PW_{I\mu\nu}\PW^{I\mu\nu}$ & $\cHW$\\
$\PPHIdagger\PPHI\, \PB_{\mu\nu} \PB^{\mu\nu}$ & $\cHB$\\
$\PPHIdagger\tau^{I}\PPHI\,\PW_{\mu\nu}^{I} \PB^{\mu\nu}$ & $\cHWB$ \\
$(\PPHIdagger\PPHI) \Box (\PPHIdagger\PPHI)$ & $\cHsquare$\\
$(\PPHIdagger D_{\mu}\PPHI)^{\star}\,(\PPHIdagger D^{\mu}\PPHI)$ & $\cHDD$\\[\cmsTabSkip]
\multicolumn{2}{c}{Gauge boson and fermion interaction} \\
$({\PPHIdagger}i \Dlra_{\mu}^{I}\PPHI)(\bar{q}_{p}\tau^{I}\gamma^{\mu}q_{r})$ & $\cHqthree$\\
$({\PPHIdagger}i \Dlra_{\mu}\PPHI)(\bar{q}_{p}\gamma^{\mu}q_{r})$ & $\cHqone$\\
$({\PPHIdagger}i \Dlra_{\mu}\PPHI)(\bar{u}_{p}\gamma^{\mu}u_{r})$ & $\cHu$ \\
$({\PPHIdagger}i \Dlra_{\mu}\PPHI)(\bar{d}_{p}\gamma^{\mu}d_{r})$ & $\cHd$ \\
$({\PPHIdagger}i \Dlra_{\mu}^{I}\PPHI)(\bar{\ell}_{p}\tau^{I}\gamma^{\mu}\ell_{r})$ & $\cHlthree$\\[\cmsTabSkip]
\multicolumn{2}{c}{Four-fermion interaction} \\
$(\bar{\ell}_{p}\gamma_{\mu}\ell_{r})(\bar{\ell}_{s}\gamma^{\mu}\ell_{t})$ & $\cllone$\\
\hline
\end{tabular}
\end{table}

\begin{table}[htb]
\centering
\topcaption{\label{tab:dim8} The set of 20 dim-8 operators studied in this analysis. H.c. stands for Hermitian conjugate.}
\begin{tabular}{lccl}
\hline
Operator ${\mathcal{O}}_j$ & Wilson coefficient\\
\hline
\multicolumn{2}{c}{Longitudinal operators} \\
$[(D_\mu\Phi)^{\dagger}D_\nu\Phi] \, [(D^\mu\Phi)^{\dagger}D^\nu\Phi]$ & $\fS{0}$ \\
$[(D_\mu\Phi)^{\dagger}D^\mu\Phi] \, [(D_\nu\Phi)^{\dagger}D^\nu\Phi]$ & $\fS{1}$ \\
$[(D_\mu\Phi)^{\dagger}D_\nu\Phi] \, [(D^\nu\Phi)^{\dagger}D^\mu\Phi]$ & $\fS{2}$\\[\cmsTabSkip]
\multicolumn{2}{c}{Transverse operators} \\
$\Tr[\hat{\PW}_{\mu\nu}\hat{\PW}^{\mu\nu}] \, \Tr[\hat{\PW}_{\alpha\beta}\hat{\PW}^{\alpha\beta}]$ & $\fT{0}$\\
$\Tr[\hat{\PW}_{\alpha\nu}\hat{\PW}^{\mu\beta}] \, \Tr[\hat{\PW}_{\mu\beta}\hat{\PW}^{\alpha\nu}]$ & $\fT{1}$\\
$\Tr[\hat{\PW}_{\alpha\mu}\hat{\PW}^{\mu\beta}] \, \Tr[\hat{\PW}_{\beta\nu}\hat{\PW}^{\nu\alpha}]$ & $\fT{2}$\\
$\Tr[\hat{\PW}_{\mu\nu}\hat{\PW}_{\alpha\beta}] \, \Tr[\hat{\PW}^{\alpha\nu}\hat{\PW}^{\mu\beta}]$ & $\fT{3}$ \\
$\Tr[\hat{\PW}_{\mu\nu}\hat{\PW}_{\alpha\beta}] \, \PB^{\alpha\nu}\PB^{\mu\beta}$ & $\fT{4}$ \\
$\Tr[\hat{\PW}_{\mu\nu}\hat{\PW}^{\mu\nu}] \, \PB_{\alpha\beta}\PB^{\alpha\beta}$ & $\fT{5}$ \\
$\Tr[\hat{\PW}_{\alpha\nu}\hat{\PW}^{\mu\beta}] \, \PB_{\mu\beta}\PB^{\alpha\nu}$ & $\fT{6}$\\
$\Tr[\hat{\PW}_{\alpha\mu}\hat{\PW}^{\mu\beta}] \, \PB_{\beta\nu}\PB^{\nu\alpha}$ & $\fT{7}$ \\
$\PB_{\mu\nu}\PB^{\mu\nu} \, \PB_{\alpha\beta}\PB^{\alpha\beta}$ & $\fT{8}$\\
$\PB_{\alpha\mu}\PB^{\mu\beta} \, \PB_{\beta\nu}\PB^{\nu\alpha}$ & $\fT{9}$\\[\cmsTabSkip]
\multicolumn{2}{c}{Mixed operators} \\
$\Tr[\hat{\PW}_{\mu\nu}\hat{\PW}^{\mu\nu}] \, [(D_\beta\Phi)^{\dagger}D^\beta\Phi]$ & $\fM{0}$\\
$\Tr[\hat{\PW}_{\mu\nu}\hat{\PW}^{\nu\beta}] \, [(D_\beta\Phi)^{\dagger}D^\mu\Phi]$ & $\fM{1}$ \\
$[\PB_{\mu\nu}\PB^{\mu\nu}] \, [(D_\beta\Phi)^{\dagger}D^\beta\Phi]$ & $\fM{2}$\\
$[\PB_{\mu\nu}\PB^{\nu\beta}] \, [(D_\beta\Phi)^{\dagger}D^\mu\Phi]$ & $\fM{3}$\\
$[(D_\mu\Phi)^{\dagger} \hat{\PW}_{\beta\nu}D^\mu\Phi] \, \PB^{\beta\nu}$ & $\fM{4}$\\
$[(D_\mu\Phi)^{\dagger} \hat{\PW}_{\beta\nu}D^\nu\Phi] \, \PB^{\beta\mu} + \text{H.c.}$ & $\fM{5}$ \\
$[(D_\mu\Phi)^{\dagger} \hat{\PW}_{\beta\nu} \hat{\PW}^{\beta\mu} D^\nu\Phi]$ & $\fM{7}$\\
\hline
\end{tabular}
\end{table}

\section{The CMS detector}
\label{sec:Detector}

The CMS detector's central feature is a superconducting solenoid with an internal diameter of 6 meters, generating a magnetic field of 3.8\unit{T}. Enclosed within the solenoid are a silicon pixel and strip tracker, a lead tungstate crystal electromagnetic calorimeter (ECAL), and a brass and scintillator hadron calorimeter (HCAL), all comprising a barrel and two endcap sections. Forward calorimeters extend the pseudorapidity ($\eta$) coverage of the barrel and endcap detectors. Muons are identified using gas-ionization detectors embedded in the steel flux-return yoke outside the solenoid.

Events of interest are selected using a two-tier trigger system~\cite{CMS:2016ngn}.
The first level, consisting of custom hardware processors, uses data from the
calorimeters and muon detectors to identify the most significant events within a fixed interval of less than $4\mus$.
The second level, known as the high-level trigger, consists of a farm of processors running a version of the
full event reconstruction software optimized for fast processing, and reduces the event rate to a few kHz before
data storage~\cite{CMS:2016ngn,CMS:2024aqx}.
A more comprehensive description of the CMS detector, including the coordinate system and relevant kinematic variables, is available in Ref.~\cite{CMS-Experiment,CMSRun3}.

\section{Monte Carlo simulations}
\label{sec:Samples}

Monte Carlo (MC) simulations are employed to model signal processes, optimize the event selection, and
estimate some of the backgrounds. For all processes, the detector response is simulated using
the \GEANTfour package~\cite{Agostinelli:2002hh} and a detailed description of the CMS detector.
The \MGvATNLO~2.6.5 generator~\cite{MadGraph} is used to generate SM \VVV events in the next-to-leading order (NLO) mode.
The same generator is used at leading order (LO) with MLM jet matching~\cite{Alwall:2007fs} to generate QCD multijet,
$\PW+$jets, $\PZ+$jets, and diboson events. The \ttbar, $\ttbar+X$ ($X = \PW$, \PZ, \PH), and the
single top quark processes are generated at NLO with \POWHEG~2.0~\cite{Nason:2004rx,Frixione:2007vw,Alioli:2010xd}.
The NNPDF 3.1~\cite{NNPDF31} set of parton distribution functions (PDFs) is used.
The simulated event samples are normalized using the most precise cross section calculations available;
they usually correspond to NLO or next-to-next-to-leading order (NNLO) accuracy. For parton
showering and hadronization, all generated samples are interfaced with \PYTHIA~8.230~\cite{PYTHIA} and the parameters
controlling the simulation of the underlying event are set according to the CP5 tune~\cite{CMS:2019csb}.
Particles produced in additional $\Pp\Pp$ collisions (pileup) are taken into account by adding minimum-bias events
simulated with \PYTHIA8 to the hard scattering process. The distribution of the number of vertices
in the simulation matches that of the data.

The strategy for generating signal events accounts for several important considerations typical of EFT analyses.
The operators listed in Tables~\ref{tab:dim6} and~\ref{tab:dim8} are incorporated into the signal sample using
the reweighting feature of the \MGvATNLO event generator.
The reweighting procedure is based on two key elements: selecting an appropriate reference point and defining
a grid of Wilson coefficient values over which event weights are calculated.
Two analysis strategies steer the generation of two EFT signal samples: (1)~allowing one Wilson coefficient at a time to vary with
all others fixed at zero to derive bounds on single Wilson coefficients, and (2)~allowing pairs of Wilson coefficients to vary
simultaneously in a multi-parameter fit and obtain bounds on pairs of Wilson coefficients.
Closure tests ensure consistency between the two samples; comparisons of generator-level distributions show excellent
agreement between them, thereby validating the reweighting approach.

For this analysis, a LO SM model fails to adequately capture the interference effects and
shows large discrepancies with respect to NLO predictions, particularly at high \mVVV.
To address this inadequacy, the signal generation is modified to include an additional parton,
which approximates NLO behavior~\cite{Goldouzian:2020ekx} well.
Two generator-level parameters, {\tt xqcut} and {\tt qcut}, are scanned to find the values
producing the smoothest differential jet rate.
Here, {\tt xqcut} determines which partons should be generated at matrix element level and
which should be generated as part of a shower; {\tt qcut} is the matching scale used in \PYTHIA.
The MLM matching scheme~\cite{Alwall:2007fs} is used.
Validation studies show that these LO plus one-parton samples agree well with full NLO predictions
across multiple variables and final states.

Higgs boson interactions with gauge bosons appear in a subset of Feynman diagrams for \VVV production.
These subprocesses have a very small cross section and are taken to be background.

\section{Event reconstruction}
\label{sec:Reconstruction}

Event reconstruction is based on the particle-flow (PF) algorithm~\cite{CMS-PRF-14-001} which
combines information from the tracker, calorimeters, and muon systems to identify charged and neutral hadrons,
photons, electrons, and muons, known as PF candidates. Each selected event must contain at least
one $\Pp\Pp$ interaction vertex. The reconstructed vertex with the largest value of summed charged
$\pt^{2}$ is taken to be the primary $\Pp\Pp$ interaction vertex.

Electron identification is performed using a multi-variate analysis (MVA) algorithm~\cite{CMS:eID2021} that utilizes shower shape,
track-cluster matching, and track quality variables.
A selection algorithm~\cite{CMS:eID2015} distinguishes prompt electrons originating from hard-scattering processes from misidentified charged hadrons and secondary electrons resulting from photon conversions.
Muons are reconstructed~\cite{CMS:muID2012,CMS:muID2018} using information from the tracker and muon chambers.
The energy deposited in the HCAL is required to be small. At the initial stage of event selection,
electron (muon) candidates must satisfy $\pt > 10\GeV$ and $\abs{\eta} < 2.5$ $(2.4)$. Electrons with
$1.444 < \abs{\eta} < 1.566$, which corresponds to the transition region between the ECAL barrel and endcaps, are excluded because
the reconstruction in this region is sub-optimal.

A prompt lepton is one that comes from the decay of a \PW or \PZ boson; these are the leptons we target for this analysis.
Electron and muon candidates must satisfy isolation criteria that
suppress nonprompt leptons, such as leptons arising from decays of heavy-flavor hadrons and jets misidentified as leptons.
The isolation variable \ISOL is defined as the ratio of the scalar \pt sum of charged and neutral PF candidates to
the lepton's \pt. Only those PF candidates falling within a cone of radius of 0.3 (0.4) in $\eta$-$\phi$ space
are used to define \ISOL for electrons (muons). The sum excludes the lepton candidate itself and any charged particles
originating from pileup occurring in the same or nearby bunch crossings.
Loose and tight collections of muons and electrons are defined using the isolation criteria.
The isolation criterion for the loose (tight) muon collection is $\ISOL < 0.25$ ($0.15$), while for
electrons the isolation is included in the MVA. A requirement for the loose (tight) electron collection
is applied to the output of the MVA to achieve an efficiency of 90\% (80\%) for prompt electrons.

Hadronic \PGt lepton decays \tauh are reconstructed using the hadrons-plus-strips algorithm~\cite{TAU-16-003,TAU-20-001}, designed to identify one- and three-prong hadronic \PGt decays, including up to two neutral pions.
The artificial neural network algorithm called \DeepTau~\cite{CMS:2025kgf}
is used to select \tauh candidates from those reconstructed from the hadrons-plus-strips algorithm; \DeepTau employs classifiers to distinguish between \tauh candidates and jets, electrons, and muons.

PF candidates are clustered to form jets using the anti-\kt jet clustering algorithm
as implemented in the \FASTJET package~\cite{Cacciari:2011ma,Cacciari:2005hq}.
When the distance parameter is $R = 0.4$, a collection of ``narrow'' jets is obtained,
whereas for $R = 0.8$ a collection of ``large-radius'' jets is formed.
Both narrow and large-radius jets must satisfy loose selection criteria based on the
relative contributions of electromagnetic and hadronic energy.
Only jets with $\pt > 30\GeV$ and $\abs{\eta} < 3$ are retained for analysis. Jet energy corrections are applied
to account for nonuniform detector response and to ensure that the measured energy of jets matches on average that
of particle-level jets.
The \DeepJet algorithm~\cite{Bols:2020bkb} is used to identify jets originating from \PQb quarks. The medium working point of the \PQb tagging algorithm is used, which, depending on the data-taking year, corresponds to a tagging efficiency of 70--80\% for \PQb jets and 1\% for jets initiated by light quarks or gluons.

The effects of pileup are mitigated at the reconstructed-particle level using the pileup-per-particle identification algorithm (PUPPI)~\cite{Bertolini:2014bba,CMS:2020ebo}. This algorithm uses local energy distributions, event pileup properties, and tracking information to identify and remove pileup contributions.
Charged particles not originating from the primary interaction vertex are discarded.
The momenta of neutral particles are rescaled according to the probability that they originate from the primary
interaction vertex as determined by the local shape variable, eliminating the need for jet-based pileup corrections.

Given that signal jets are merged products of \PV boson decays, we require the jet mass to be consistent with the \PV boson mass to reduce the significant background from QCD multijet events, which exhibit a steeply falling jet mass distribution.
The separation between signal and background is enhanced by applying the ``soft-drop'' jet grooming algorithm~\cite{Larkoski:2014wba}. This algorithm recursively removes soft, wide-angle radiation from anti-\kt jets. The soft-drop mass \mSD is calculated from the sum of the four-momenta of the remaining jet constituents.

The signal can be enhanced relative to the background by applying the \ParticleNet algorithm to the
large-radius jets~\cite{CMS:2020poo}. This algorithm is based on a neural network
and provides a score that indicates whether a jet originated from a boosted \PV boson decay. The mass-decorrelated
version of this algorithm is used to avoid distorting the shape of the \mSD distribution.
Requirements are placed on \mSD and the output of the \ParticleNet algorithm to select jets for analysis.
If the distance from a jet to the closest lepton is less than 0.8 in $\eta$-$\phi$ space, then the jet is excluded.
Jets passing these criteria are referred to as \fatjets.

The missing transverse momentum vector \ptvecmiss is defined as the negative vector sum of the momenta of all PF candidates. The magnitude of \ptvecmiss is denoted \ptmiss. Corrections to jet energies arising from the nonuniformity of the detector response are propagated to \ptmiss~\cite{CMS:2019ctu}.

\section{Analysis strategy}
\label{sec:Strategy}

This analysis focuses on fully hadronic and partially leptonic final states which have relatively high branching fractions,
in contrast to the approach of Ref.~\cite{CMS:2020hjs}.
Several distinct channels are defined, broadly classified according to charged-lepton multiplicity.
There is no overlap of events selected in each channel thanks to a common set of loose lepton identification
and reconstruction criteria, stricter requirements are subsequently applied to reduce backgrounds as
needed for the signal region (SR) in each channel.

The zero-lepton channels accept events with no reconstructed electrons or muons and targets final states where all three vector bosons decay hadronically. A distinction is made for events with two \fatjets (\noLEPtwoFJ) and three or more \fatjets (\noLEPthreeFJ) to improve the signal to background ratio. The main background contribution originates from QCD multijet production, with subdominant contributions stemming from $\PV+$jets, single top quark, diboson, and \ttV processes.

The one-lepton channel (\oneLEPtwoFJ) accepts events with one electron or muon and two \fatjets, primarily targeting \WWW and \WWZ processes. In this channel, one vector boson decays leptonically while the other two undergo hadronic decays. The background comes primarily from $\PW+$jets and \ttbar production.

Events with two electrons or muons are classified based on the charge of the leptons: the same-sign (SS) channel (\twoLEPSSoneFJ) targets final states with leptonic decays of two SS \PW bosons, requiring exactly one \fatjet for the hadronic decay of the third vector boson. The largest background contribution stems from \ttbar events with one nonprompt lepton.
Events with opposite-sign (OS) dileptons (\twoLEPOSoneFJ) are divided into four subchannels based on
whether the two leptons have the same or different flavors and whether they are consistent
with the decay of a \PZ~boson. A second dilepton channel (\twoLEPOStwoFJ) requires at least two \fatjets.

Finally, we designate two channels focused on hadronically decaying \PGt leptons (\tauh). The first channel (\oneLEPoneTAUoneFJ)
includes events with one \tauh decay, one lepton, and one \fatjet. The other channel (\twoLEPoneTAUnoFJ) requires
two leptons, one \tauh candidate, and no \fatjets, and is designed for final states with all three gauge bosons
decaying leptonically.

A discriminating variable, generically called \ST, is defined for each channel.
This variable, which is correlated with \mVVV,
varies from channel to channel according to the physics objects available,
the composition of the backgrounds, and the requirement that the variable is accurately modeled in the simulation.
The \ST definitions are summarized in Table~\ref{tab:discrvar}.
The \ST distributions are binned to maximize the sensitivity of the analysis to new physics.
The greatest sensitivity for a given channel typically arises from the highest \ST bin.

\begin{table}[hbtp!]
\centering
\topcaption{ \label{tab:discrvar}
Definitions of the discriminating kinematic variables.
Here, \twoLEPOSoneFJ stands for \twoLEPOSOFFZoneFJ, \twoLEPOSONZoneFJ, and \twoLEPOSOFoneFJ (OSDF stands for
opposite-sign, different-flavor). In the definition of $\MJJlnu$, $p$ stands for a four-vector.
}
\begin{tabular}{cc}
\hline
Analysis channels & Discriminating variable \\
\hline
  \noLEPtwoFJ and \noLEPthreeFJ &
  $\ST = \sum \ptfatjet + \sum \ptjet$ \\
  \oneLEPtwoFJ & $\MJJlnu = \sqrt{\smash[b]{(p_{\Pell} + p_{\nu} + p_{\text{\PV~jet,1}}+p_{\text{\PV~jet,2}})^{2}}}$ \\
  \twoLEPOSoneFJ &
  $\ST = \sum \ptfatjet + \sum \ptjet + \sum \ptlep$ \\
   \twoLEPOStwoFJ and \twoLEPSSoneFJ &
  $\ST = \sum \ptfatjet + \sum \ptjet + \sum \ptlep + \ptmiss$ \\
  \oneLEPoneTAUoneFJ and \twoLEPoneTAUnoFJ & $\ST=\sum \ptlep + \sum \ptjet + \pttau $ and BDT scores\\
\hline
\end{tabular}
\end{table}

Background yield estimates are derived using MC simulations and techniques based on control samples in data.
For each analysis channel, we define one or more control regions (CRs) that are enriched in the leading background processes
using selection criteria that mimic those of the SR. In these regions, MC simulations are compared
directly to data, and if necessary, correction factors are determined and applied to the MC background yield
in the SR.

\begin{figure}[hbpt]
    \centering
    \includegraphics[width=0.48\textwidth]{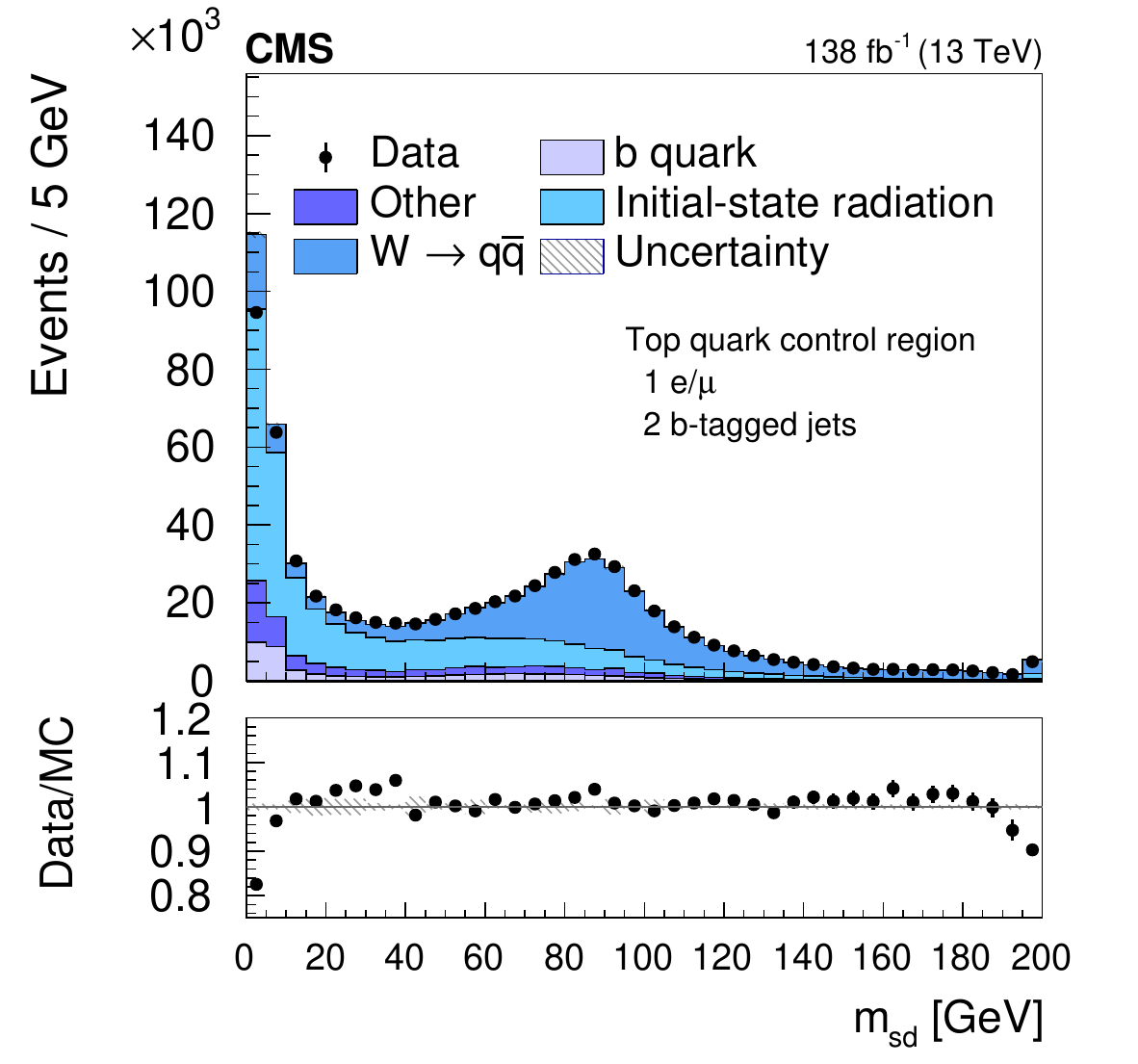}
    \caption{Comparison of the \mSD distribution in data and simulation for events in a control region
      dominated by \ttbar production. $\PW\to\qqbar$ represents \fatjets that match the
      hadronic decay of a \PW boson; a prominent peak at the \PW boson mass is seen.
      The contribution marked Initial-state radiation corresponds to \fatjets matched to gluons emitted in the initial state.
      Jets containing single \PQb quarks will sometimes be selected as \fatjets.
      Both ISR and \PQb jets peak at small \mSD but not near the \PW boson mass.
      A small contamination from non-\ttbar events is marked as Other in the plot.
      The data are represented by black dots with error bars.
      The shaded band in the data/MC ratio plot shows the MC statistical uncertainty.
      This plot shows the MC prediction after fitting, \ie, these are post-fit distributions.
    }
    \label{fig:1l2b}
\end{figure}

Several analysis channels use \mSD sidebands to derive corrections to MC background estimates.
Consequently, it is essential to validate the MC modeling of the \mSD distribution for key background processes
such as \ttbar and $\PV+$jets production.
Figure~\ref{fig:1l2b} shows a comparison of \mSD distributions from data and simulation
in a control region dominated by \ttbar production and defined by the presence of one charged lepton and two \PQb-tagged jets.
A clear \PW boson peak is observed, which contrasts sharply with falling distributions from gluons associated with
initial-state radiation (ISR) and jets that contain single \PQb quarks.
The data and simulation agree within 10\% across the entire \mSD range, and the agreement is better than 10\%
in the region most relevant to this analysis: $40 < \mSD < 120\GeV$.

The Wilson coefficients are extracted through a simultaneous fit to the observed yields in all SR bins,
using a parameterized model describing the signal and individual background contributions.
The variation of background estimates with Wilson coefficients is found to be negligible, and is not included in the fit.
The SR distributions shown in the following sections correspond to post-fit results, with event yields normalized
to the fit outcomes, whereas the CR distributions are shown pre-fit.

\section{Analysis channels}
\label{sec:Channels}

\subsection{Zero-lepton (all hadronic) channel}
Events are selected online using a logical "OR" of hadronic trigger criteria. These criteria require the scalar \pt sum of the jets (\HT)
to exceed a threshold that ranges from 800 to 1050\GeV, depending on the data-taking period. To increase the efficiency, an additional set of triggers is employed that require the presence of a high-energy large-radius jet in the event. The minimum \pt of the jet in the online selection varies from 360 to 500\GeV, depending on the data-taking year and whether additional conditions on the jet mass are imposed.
The trigger is fully efficient for events that satisfy the offline selection criteria.

The presence of at least two large-radius jets with $\pt > 200\GeV$ and the absence of electrons and muons is required. To classify as \fatjets,
the jets must have $40 < \mSD < 150\GeV$ and pass the medium working point
of the \ParticleNet tagger. Events are further divided into two nonoverlapping SRs based on the multiplicity of \fatjets.

The \noLEPtwoFJ channel is defined by events containing exactly two \fatjets.
The \fatjet or leptons corresponding to the third boson are not identified, or the third boson is a \PZ boson
and decays into neutrinos. The leading \fatjet is required to have $\pt > 600\GeV$ to ensure that the selected events
are at the plateau of the trigger efficiency.

Reflecting characteristics of the signal, the scalar sum of the \fatjet energies and \ptmiss
is required to exceed 1.1\TeV. Additionally, the following variable is introduced:
\begin{equation}
D_{2}=\sqrt{({\mSDone} - 85\GeV)^2 + ({\mSDtwo} - 85\GeV)^2},
\end{equation}
where $\mSDone$ and $\mSDtwo$ represent the soft-drop masses of the leading and subleading \fatjets.
Following an optimization procedure, the condition $D_{2} < 17.5\GeV$ is imposed to ensure that the
two \fatjets originate from either \PW or \PZ bosons, corresponding to a mass distribution that peaks near the average of
their masses.

\begin{figure*}[htpb]
    \centering
    \includegraphics[width=0.48\textwidth]{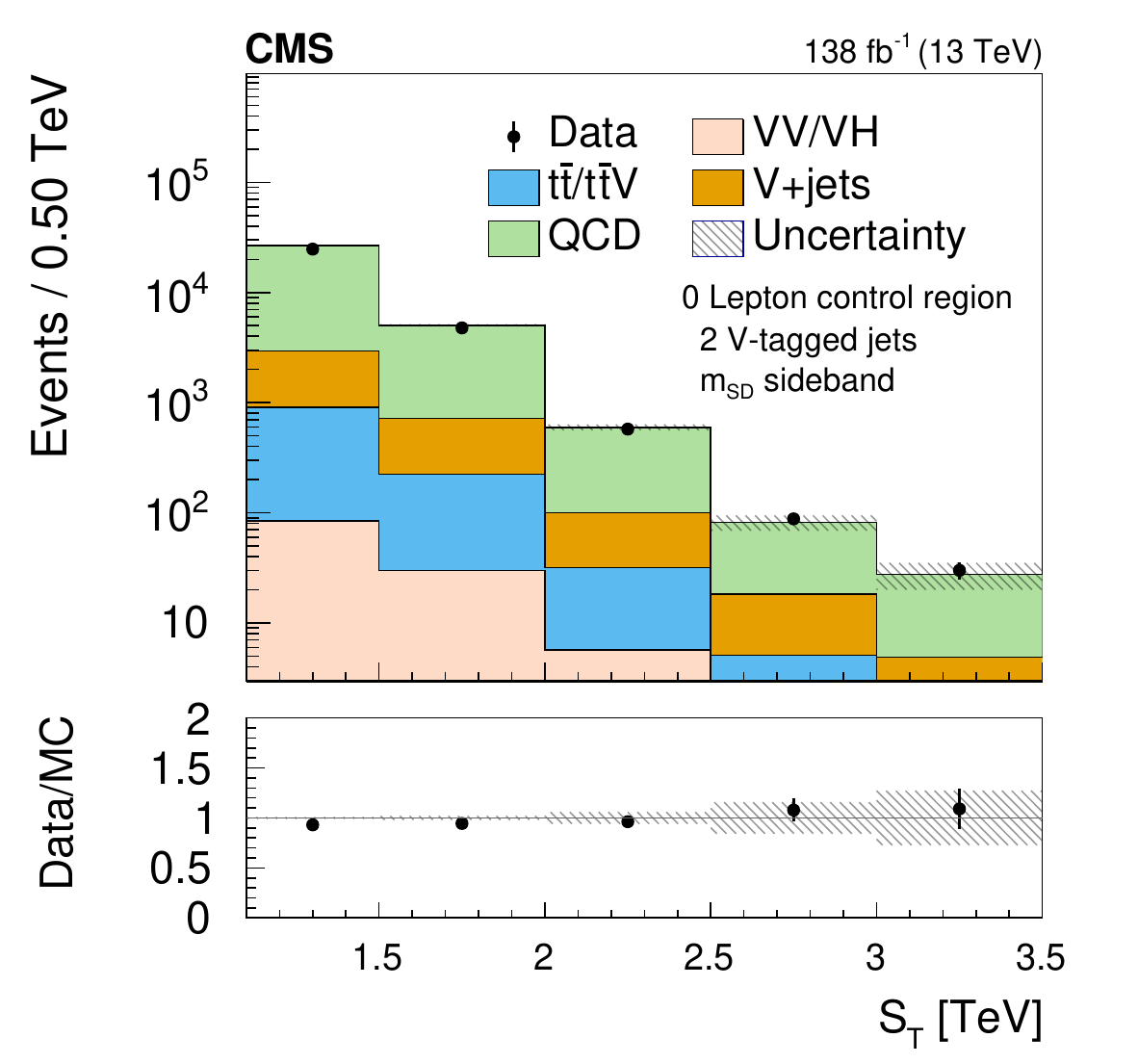}
    \includegraphics[width=0.48\textwidth]{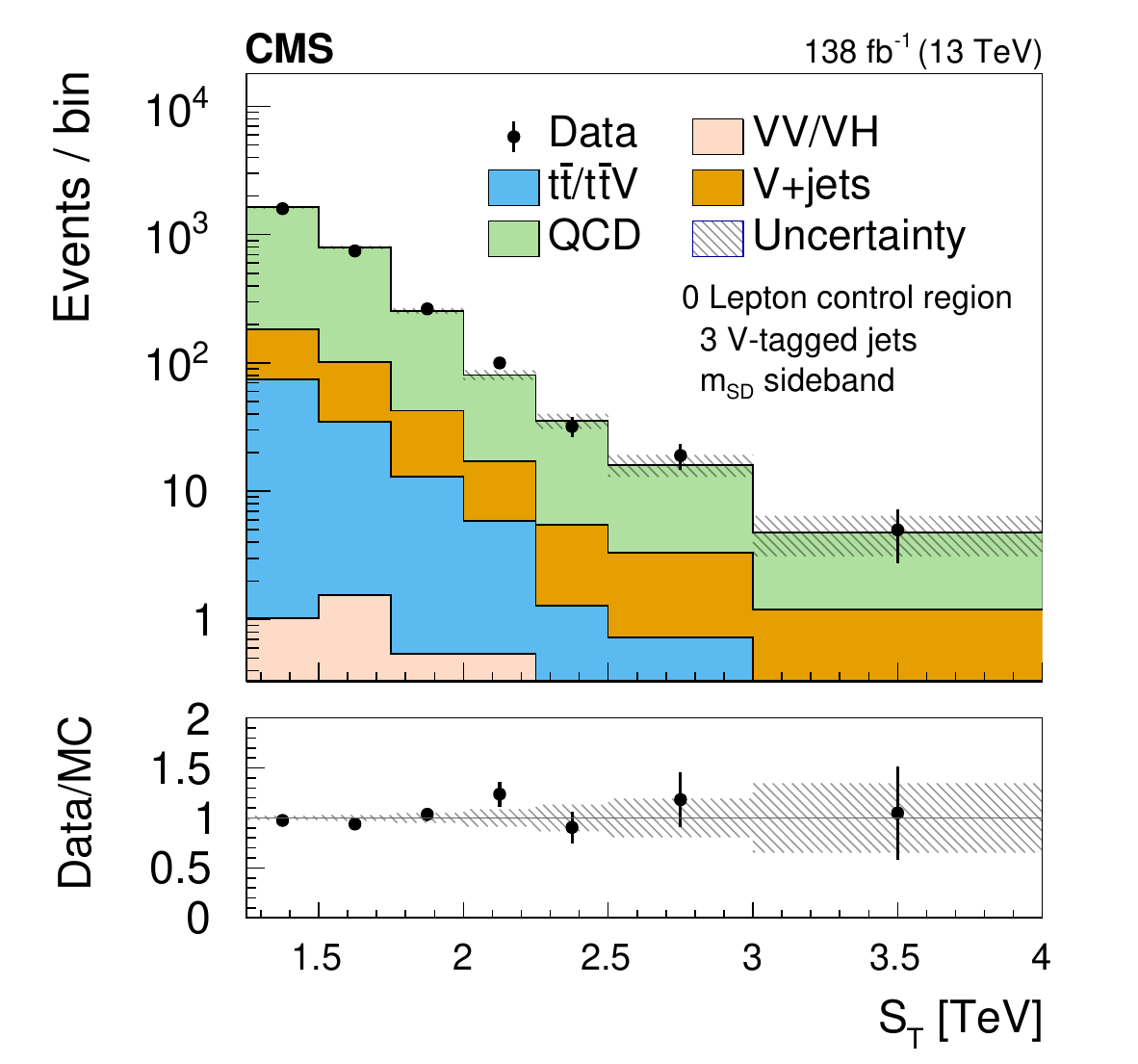}
    \caption{Tests of the ABCD method in the \noLEPtwoFJ (left) and \noLEPthreeFJ (right) channels.
      The validation regions are dominated by QCD multijet backgrounds.
      The ABCD method is used to predict the QCD multijet background and the total SM background
      is compared to the data, showing good agreement.
      The shaded band in the ratio plot shows the MC statistical uncertainty.
      The black dots with error bars represent the data with statistical uncertainties.
      These are pre-fit distributions.
    }
    \label{fig:0l2fjSRCDEFValidation}
\end{figure*}

The dominant background in \noLEPtwoFJ is QCD multijet events with subdominant contributions
from $\PW+$jets, $\PZ+$jets, and top quark backgrounds.
Since the high-energy tails of the $\ST$ distribution cannot be modeled reliably by the QCD MC simulations, this background must be estimated using methods based on control samples in data. An ``ABCD method''~\cite{ABCD-CDF} is used, employing regions with $D_{2} > 50\GeV$ and sidebands of the \ParticleNet tagging score distribution. Contributions from non-QCD processes to the background are taken directly from simulation. An additional $D_{2}$ sideband region with $17.5 < D_{2} < 50\GeV$ is used to validate the ABCD method. Figure~\ref{fig:0l2fjSRCDEFValidation} (left) shows good agreement in the validation region for the data and the prediction of the QCD multijet background.

Events with three or more \fatjets belong in \noLEPthreeFJ, with criteria similar to those of the \noLEPtwoFJ channel.
The leading \fatjet must have $\pt > 600\GeV$ while all other \fatjets must have $\pt > 200\GeV$.
Furthermore, the sum of the \pt of the \fatjets must be greater than 1.25\TeV.

Each of the three \fatjets must have $40 < \mSD < 150\GeV$ and pass the loose working point of the \ParticleNet tagger.
The loose working point was chosen because the SM background yields in this SR remain low even before applying this requirement.
We extend the definition of the variable $D_2$ to handle events with three or more \fatjets:
\begin{equation}
D_{3} = \sqrt{({\mSDone} - 85\GeV)^2 + ({\mSDtwo} - 85\GeV)^2 + ({\mSDthree} - 85\GeV)^2},
\end{equation}
and require $D_3 < 35\GeV$ in accord with an optimization procedure.

The primary background in this channel is QCD multijet production, and the estimation strategy closely follows the approach developed for the \noLEPtwoFJ channel. An ABCD method is employed, utilizing regions with $D_{3} > 50\GeV$ and the sidebands of the \ParticleNet tagging score distribution.
Figure~\ref{fig:0l2fjSRCDEFValidation} (right) demonstrates a good agreement between the data and the QCD multijet prediction within the validation region.

\begin{figure*}[hbpt]
    \centering
    \includegraphics[width=0.48\textwidth]{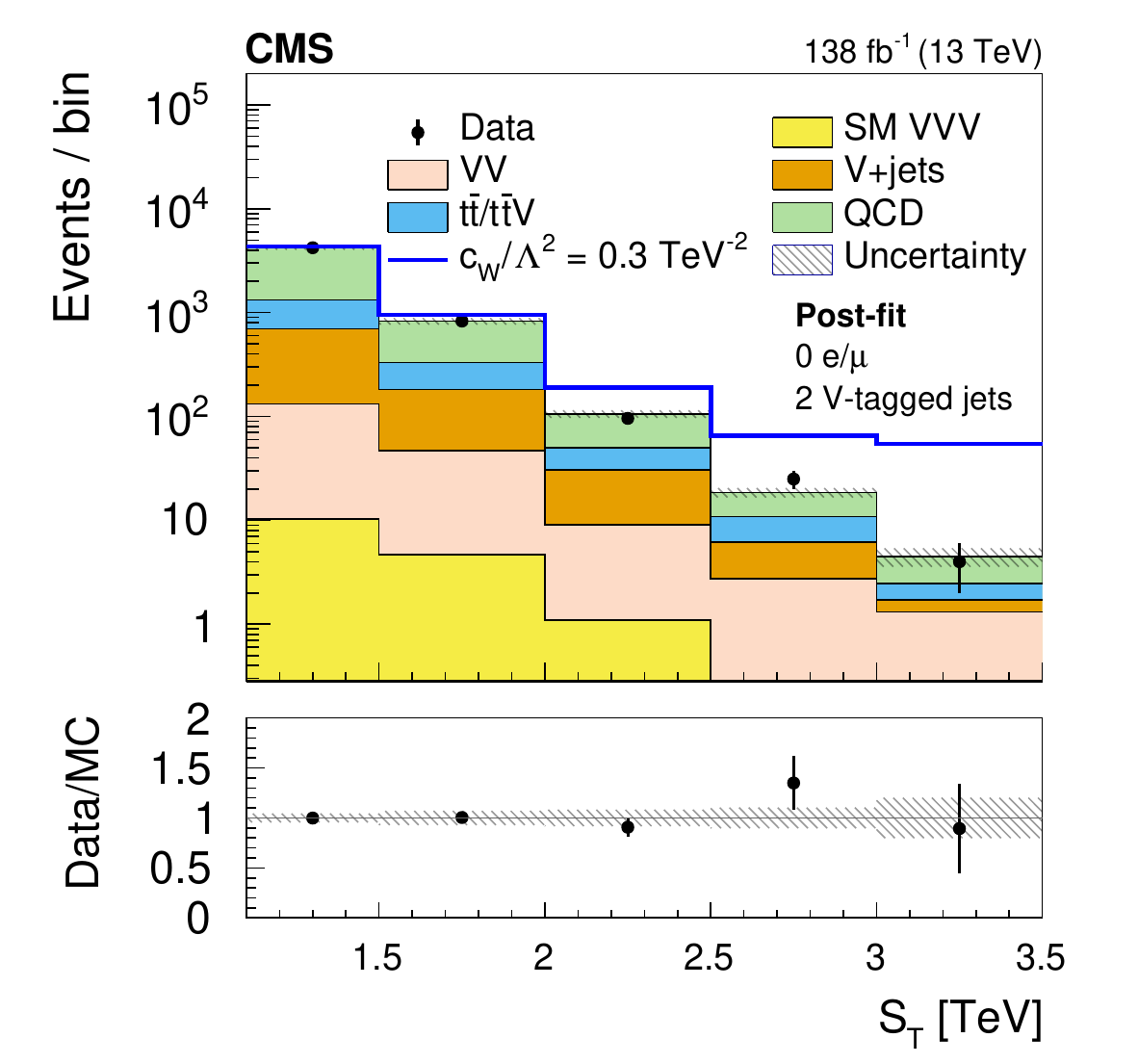}
    \includegraphics[width=0.48\textwidth]{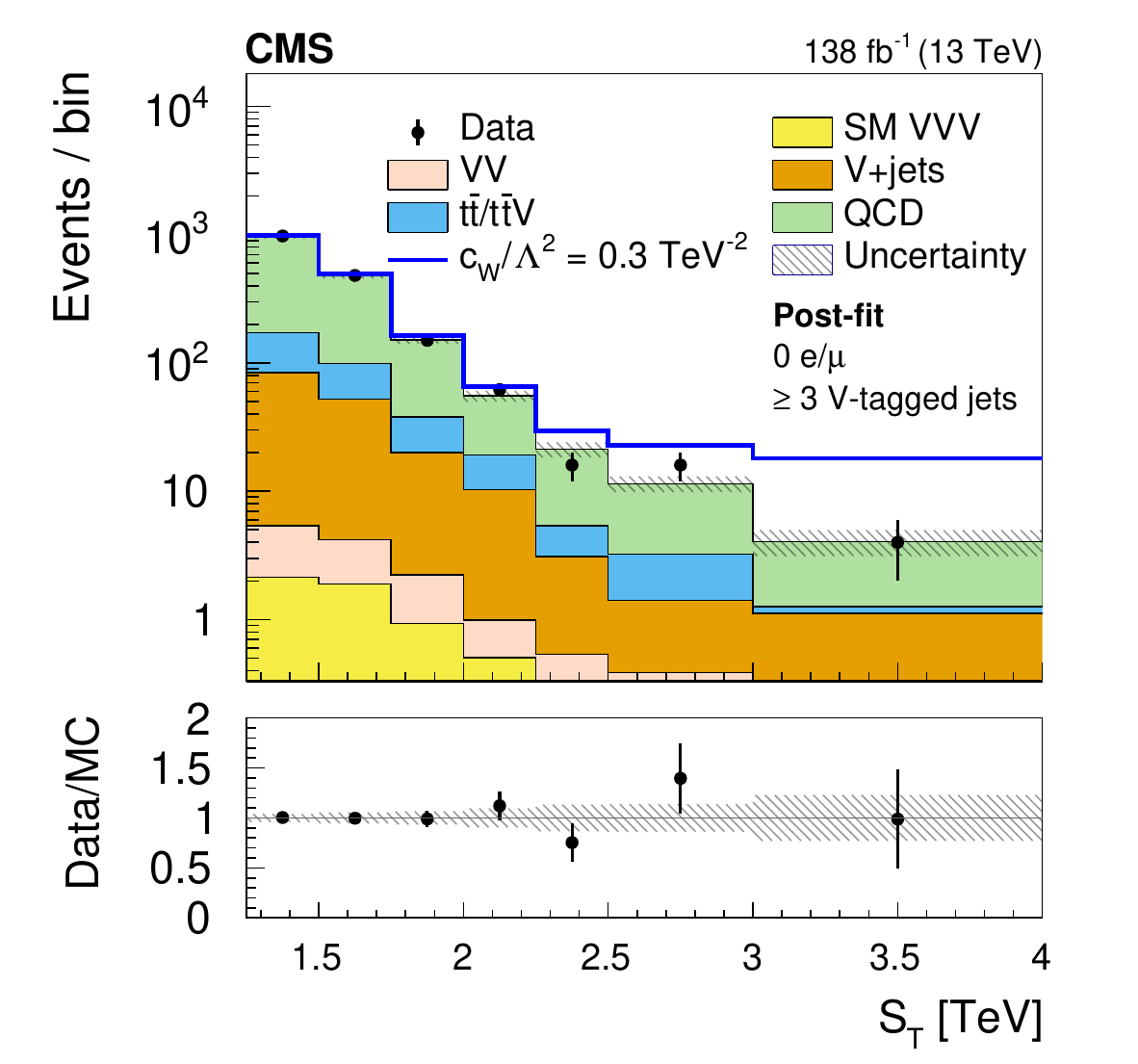}
    \caption{Comparison of the \ST distributions for events in the zero-lepton signal regions
      with two \fatjets (left) and three \fatjets (right).
      The shaded band in the ratio plot shows the total uncertainty.
      The black dots with error bars represent the data with statistical uncertainties.
      These distributions are made after the fit, \ie, they are post-fit distributions.
    }
    \label{fig:0lSR}
\end{figure*}

Figure~\ref{fig:0lSR} (left) and (right) display the \ST distributions for
the \noLEPtwoFJ and \noLEPthreeFJ signal regions including backgrounds and a hypothetical signal contribution.
The data are consistent with SM predictions.

\subsection{One-lepton channel}
Data for this channel are collected using single-lepton triggers that require at least one electron or muon.
Leptons passing the triggers must have high \pt and meet loose isolation criteria. The \pt threshold
for the electron (muon) trigger ranges from 27 to 35\GeV (24 to 27\GeV), depending on the data-taking year.

Events in \oneLtwoFJ must contain exactly one electron or muon with $\pt > 55\GeV$ and
at least two large-radius jets with $\pt > 200\GeV$.
Of these large-radius jets, at least two must pass the loose working point of the \ParticleNet tagger.
The transverse momentum of the leptonically decaying \PW boson, \PTlepW, is reconstructed using \ptmiss
as a proxy for the neutrino \pt.
The requirement $\PTlepW > 150\GeV$ is applied to reduce QCD multijet background. Top quark backgrounds
are suppressed by rejecting events with one or more \PQb-tagged narrow jets.

$\PW+$jets events enter the SR when one of the two \fatjets originates from ISR.
A $\PW+$jets CR is defined by applying the same selection
as the SR except that the \mSD requirement is modified to use the sidebands;
the ranges 50--65\GeV and 105--135\GeV define these CRs.
Approximately 80\% of these events are attributed to the $\PW+$jets process.

Top quark production events enter the SR when a \PQb-jet from a top quark decay is not correctly \PQb-tagged.
A \ttbar CR is defined by inverting the \PQb-jet selection criterion and requiring at least one \PQb-tagged
jet in the event. Approximately 85\% of the events in this CR come from \ttbar production.

\begin{figure*}[htpb]
    \centering
    \includegraphics[width=0.48\textwidth]{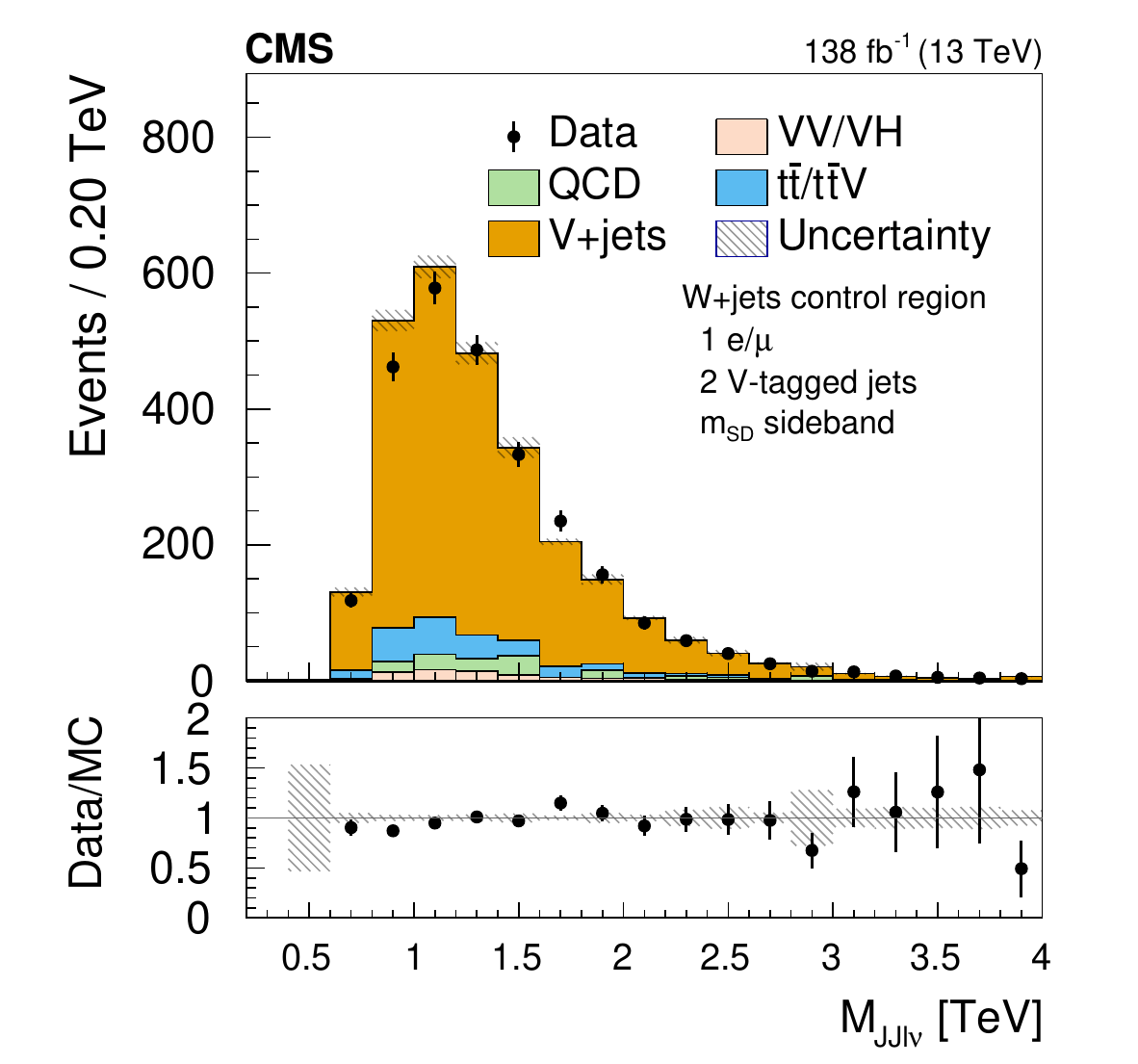}
    \includegraphics[width=0.48\textwidth]{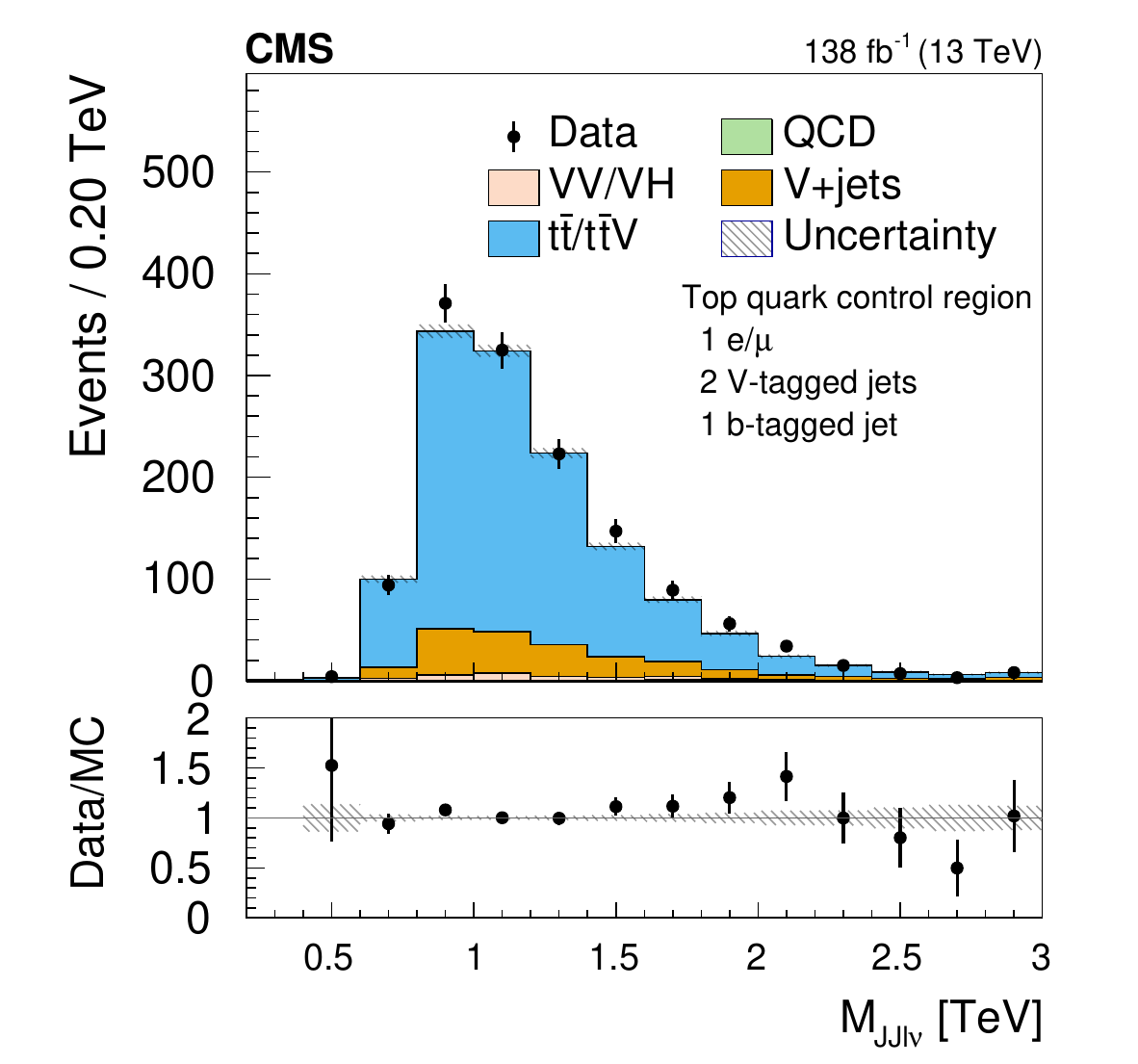}
    \caption{Comparison of the pre-fit $\MJJlnu$ distributions for the one-lepton control regions
      for $\PW+$jets (left) and \ttbar (right) backgrounds.
      The shaded band in the ratio plot represents the MC statistical uncertainty.
      The black dots with error bars represent the data with statistical uncertainties.
    }
    \label{fig:1l2fjCR}
\end{figure*}

Comparisons of data and simulation in the $\PW+$jets CR are shown in Fig.~\ref{fig:1l2fjCR} (left), and for the top CR in Fig.~\ref{fig:1l2fjCR} (right). Good agreement is observed between data and simulation. To account for any residual discrepancies, a bin-by-bin correction factor is derived for the $\PW+$jets and \ttbar backgrounds in their respective CRs and applied to the SR yields.
These correction factors range from $0.9$ to $2.0$ ($1.0$ to $1.4$) for $\PW+$jets (\ttbar) background.

\begin{figure}[htpb]
    \centering
    \includegraphics[width=0.48\textwidth]{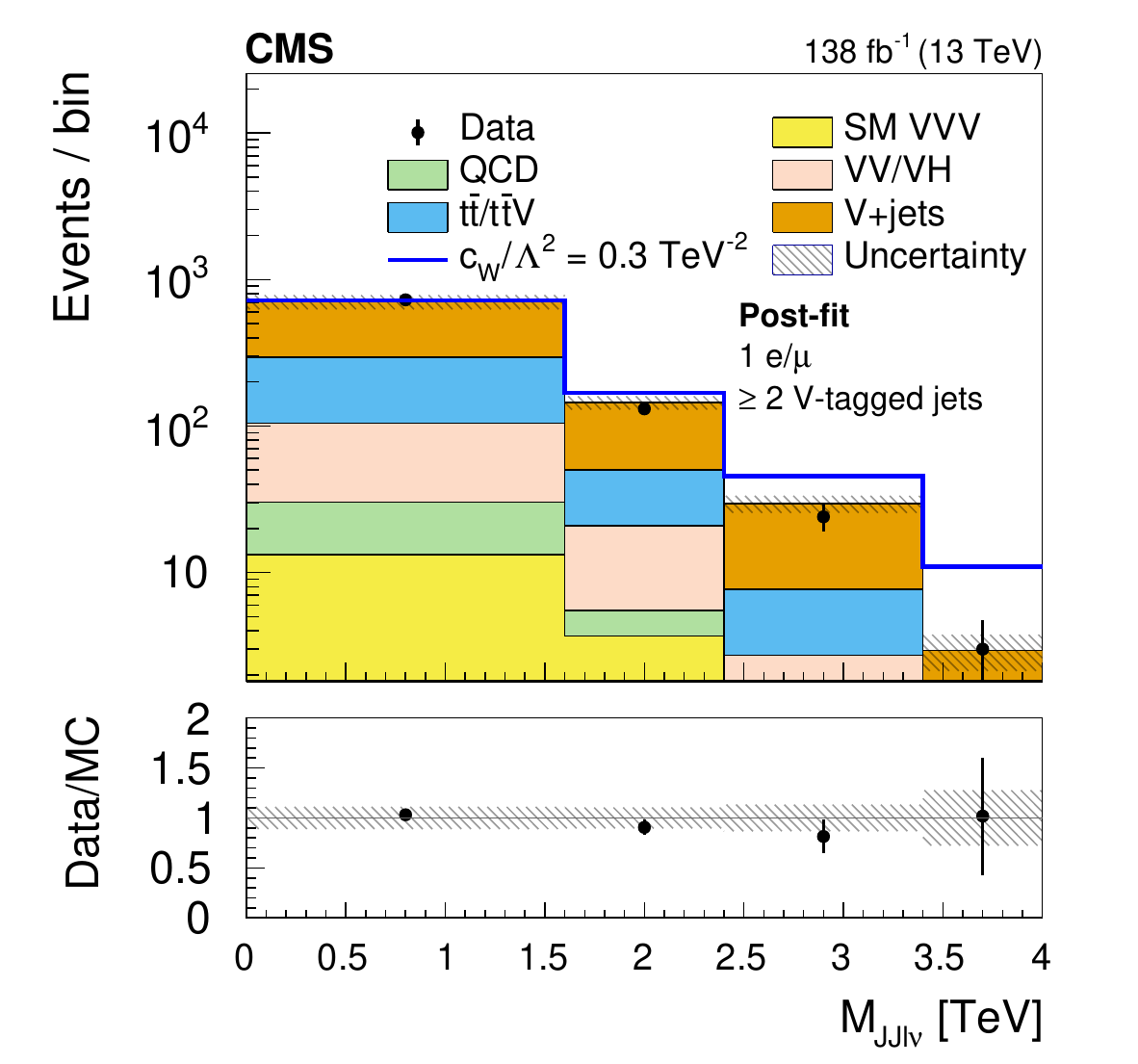}
    \caption{Comparison of the post-fit $\MJJlnu$ distributions for the one-lepton and two \fatjets
      (\oneLEPtwoFJ) signal region.
      The shaded band in the ratio plot represents the total uncertainty.
      The black dots with error bars represent the data with statistical uncertainties.
    }
    \label{fig:1lSR}
\end{figure}

Figure~\ref{fig:1lSR} shows the distributions of the discriminating variable $\MJJlnu$ in the data for
\oneLEPtwoFJ along with the predicted background and a hypothetical signal.
The data show good agreement with SM expectations.

\subsection{Opposite-sign dilepton channel}
The data in the opposite-sign dilepton channel are collected using dilepton triggers that require two electrons, two muons, or one electron and one muon. As in the case of single-lepton triggers, these triggers require the leptons to have sufficient \pt and to satisfy loose isolation criteria.
For the dielectron trigger, the leading electron (dimuon) must have $\pt > 23\,(17)\GeV$ and
the subleading electron (muon) must have $\pt > 12\,(8)\GeV$.
In the case of the electron+muon trigger, the leading lepton must have $\pt > 23\GeV$, while the subleading lepton must have
$\pt > 12\,(8)\GeV$ if it is an electron (muon).

Signal regions require either one or two \fatjets.
For events containing a single \fatjet, the signal selection requires the leading (subleading) lepton to have $\pt > 25\,(15)\GeV$,
and the \fatjet to have $\pt > 200\GeV$, while meeting the tight \ParticleNet working point criteria.
Events with one or more \PQb-tagged narrow jets are excluded to suppress the significant \ttbar background.

Selected events are further categorized into three nonoverlapping signal regions based on the flavor of the two leptons: different-flavor leptons (\twoLEPOSOFoneFJ), same-flavor leptons with an invariant mass consistent with the \PZ boson mass, $\abs{\Mll - \MZ} < 20\GeV$ (\twoLEPOSONZoneFJ),
and same-flavor leptons with an invariant mass inconsistent with the \PZ boson resonance (\twoLEPOSOFFZoneFJ).

In \twoLEPOSOFoneFJ, the dominant background comes from \ttbar events, whereas in \twoLEPOSONZoneFJ,
the primary background source is $\PZ+$jets production. The \twoLEPOSOFFZoneFJ region contains roughly equal
contributions from both backgrounds. Subdominant contributions include single- and double-\PV production and
associated Higgs boson production.
Methods based on control samples in data are employed to estimate contributions from the two leading background processes.
A CR enriched in \ttbar events is defined by inverting the \PQb tagging requirement.
For the $\PZ+$jets process, an ABCD method based on \mSD (the nominal SR and a lower sideband
$40 < \mSD < 65\GeV$) and the \ParticleNet tagging score (passing and failing the tight working point) is employed.
An upper sideband ($105 < \mSD < 130\GeV$) serves as a validation region to test the reliability of the method.
The closure of the ABCD method is validated using an MC simulation of $\PZ+$jets production.
Correction factors in bins in \ST are derived for \ttbar ($\PZ+$jets) production range from $0.98$ to $1.02$ ($0.4$ to $1.6$).

The analysis also includes an SR with opposite-sign dileptons and two or more \fatjets (\twoLEPOStwoFJ).
The lepton and \fatjet \pt requirements remain the same as in the single \fatjet case. However, due to lower
background levels and the desire for higher signal efficiency, the jets are required to meet the medium \ParticleNet
working point instead of the tight selection. Since this SR is specifically sensitive to \VVV processes
involving a leptonically decaying \PZ boson, only events with $\abs{\Mll - \MZ} < 20\GeV$ are considered. The primary
backgrounds in this region arise from $\PZ+$jets and \WZ production, with a smaller contribution from \ttbar events.
Validation regions with two opposite-sign dileptons and two \fatjets enriched in $\PZ+$jets and \ttbar
events are used to verify the accuracy of the background predictions.

\begin{figure*}[htpb]
    \centering
    \includegraphics[width=0.48\textwidth]{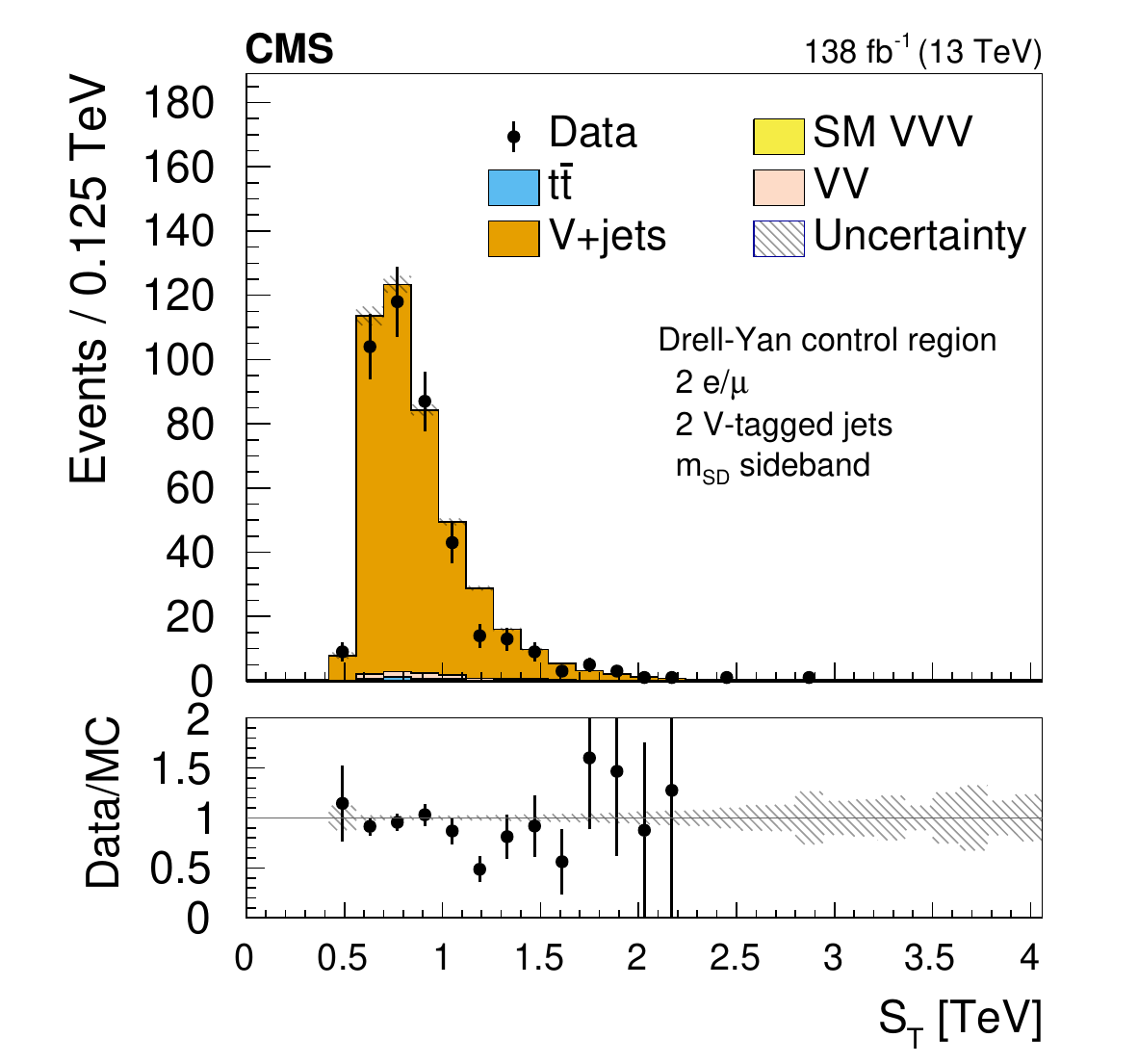}
    \includegraphics[width=0.48\textwidth]{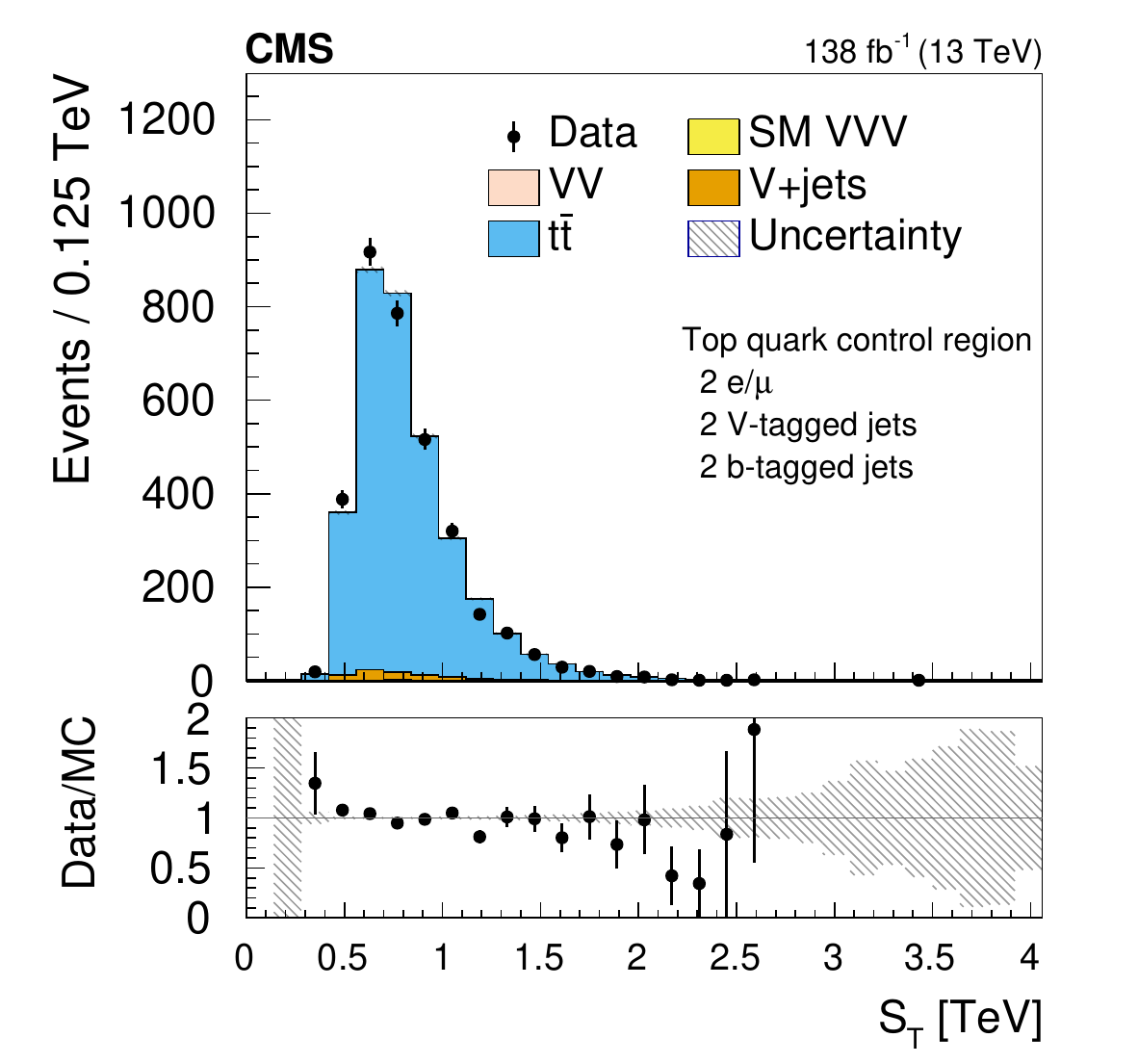}
    \caption{Comparison of the pre-fit \ST distributions for the opposite-sign dilepton plus two \fatjets (\twoLEPOStwoFJ)
      control regions for $\PZ+$jets (left) and \ttbar (right) backgrounds.
      The shaded band in the ratio plot represents the MC statistical uncertainty.
      The black dots with error bars represent the data with statistical uncertainties.
    }
    \label{fig:2l2fjCR}
\end{figure*}

For the $\PZ+$jets process, modeling is studied in several regions using the sidebands of the \mSD distribution. For the \ttbar process, events are selected by requiring at least two \PQb-tagged jets and inverting the dilepton mass selection to $\abs{\Mll - \MZ} > 20\GeV$. These selections provide nearly pure $\PZ+$jets and \ttbar samples. The observed agreement between data and simulation is good, as shown in Fig.~\ref{fig:2l2fjCR}.

\begin{figure*}[htpb]
    \centering
    \includegraphics[width=0.48\textwidth]{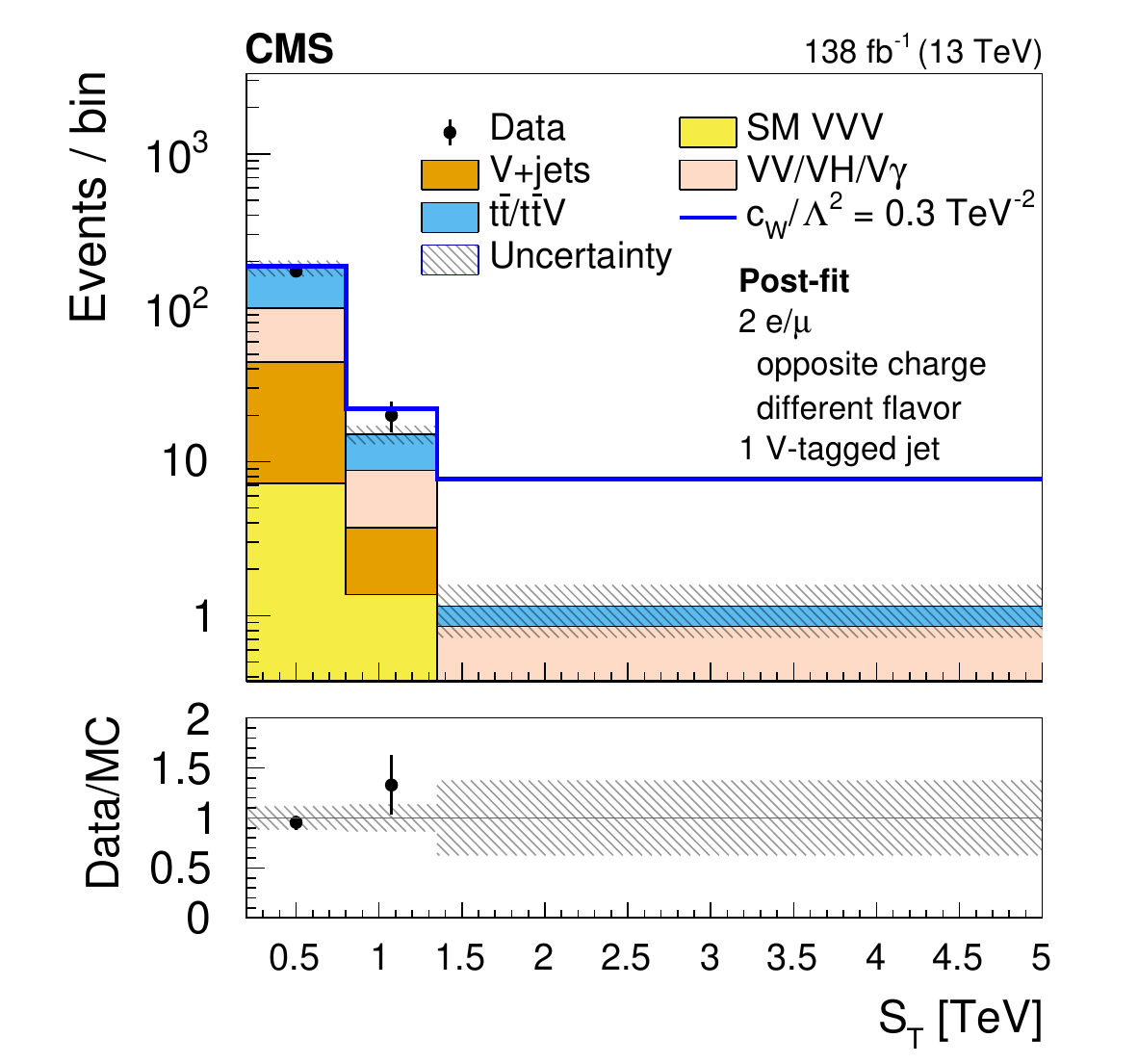}
    \includegraphics[width=0.48\textwidth]{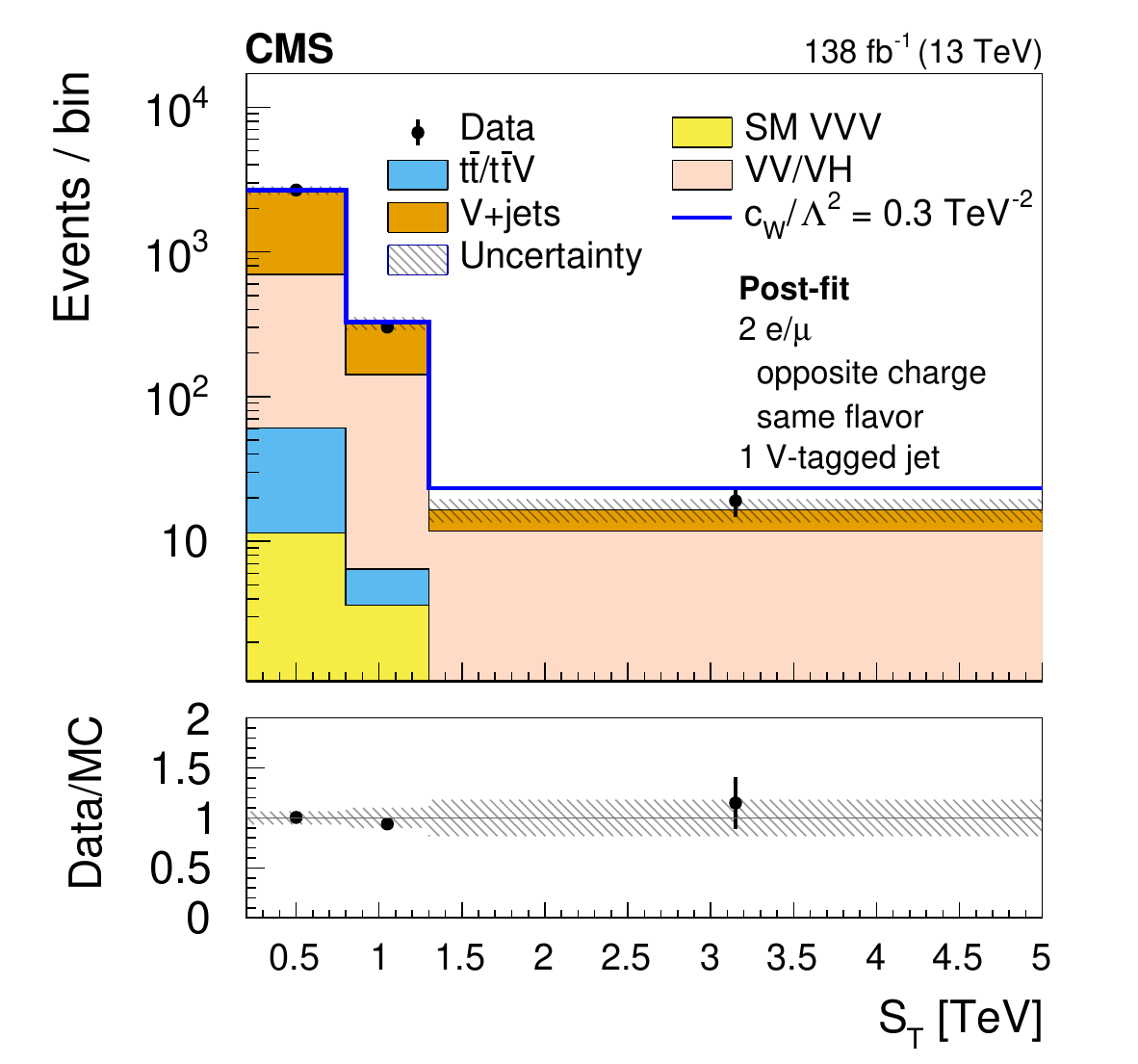}
    \includegraphics[width=0.48\textwidth]{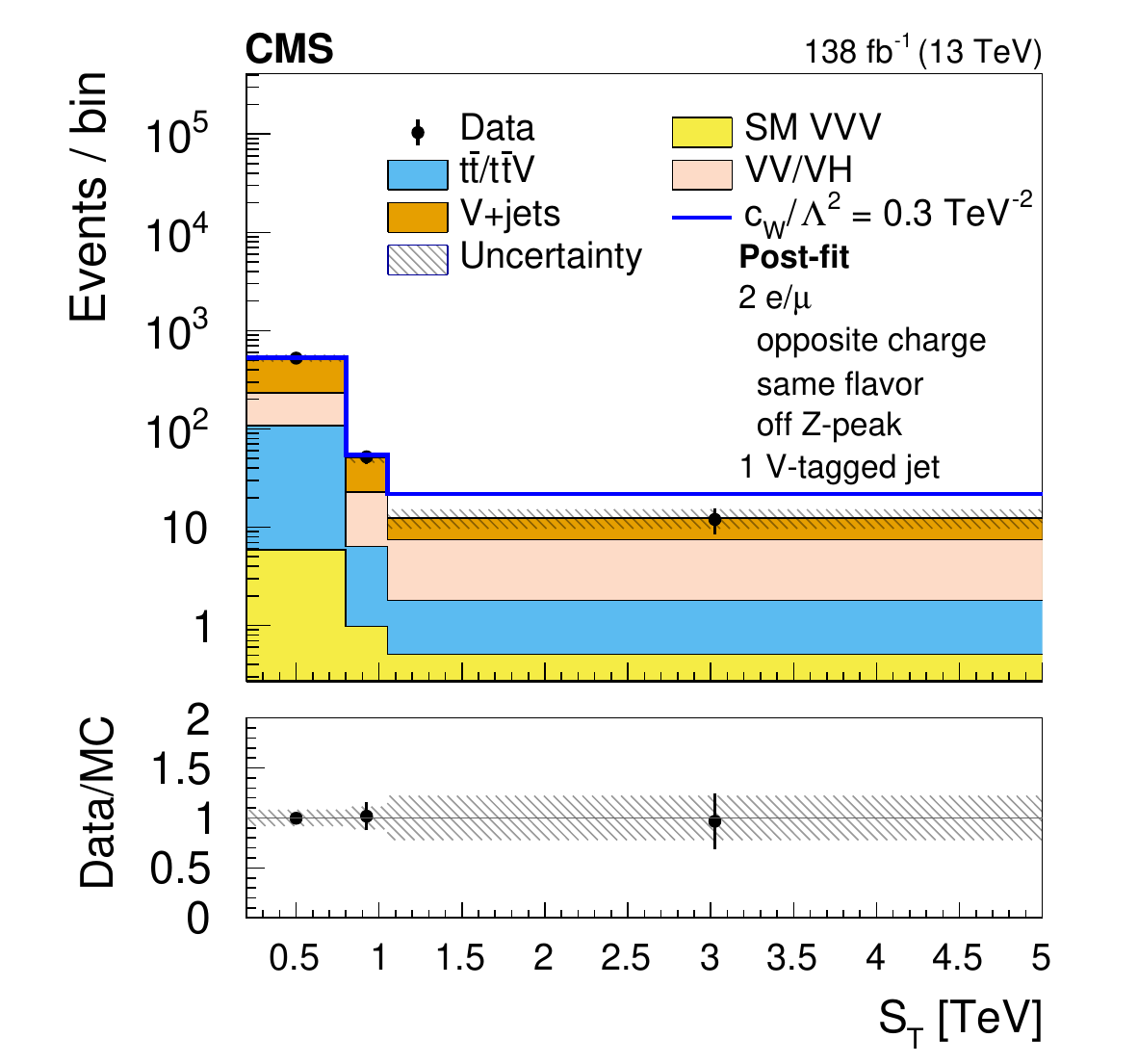}
    \includegraphics[width=0.48\textwidth]{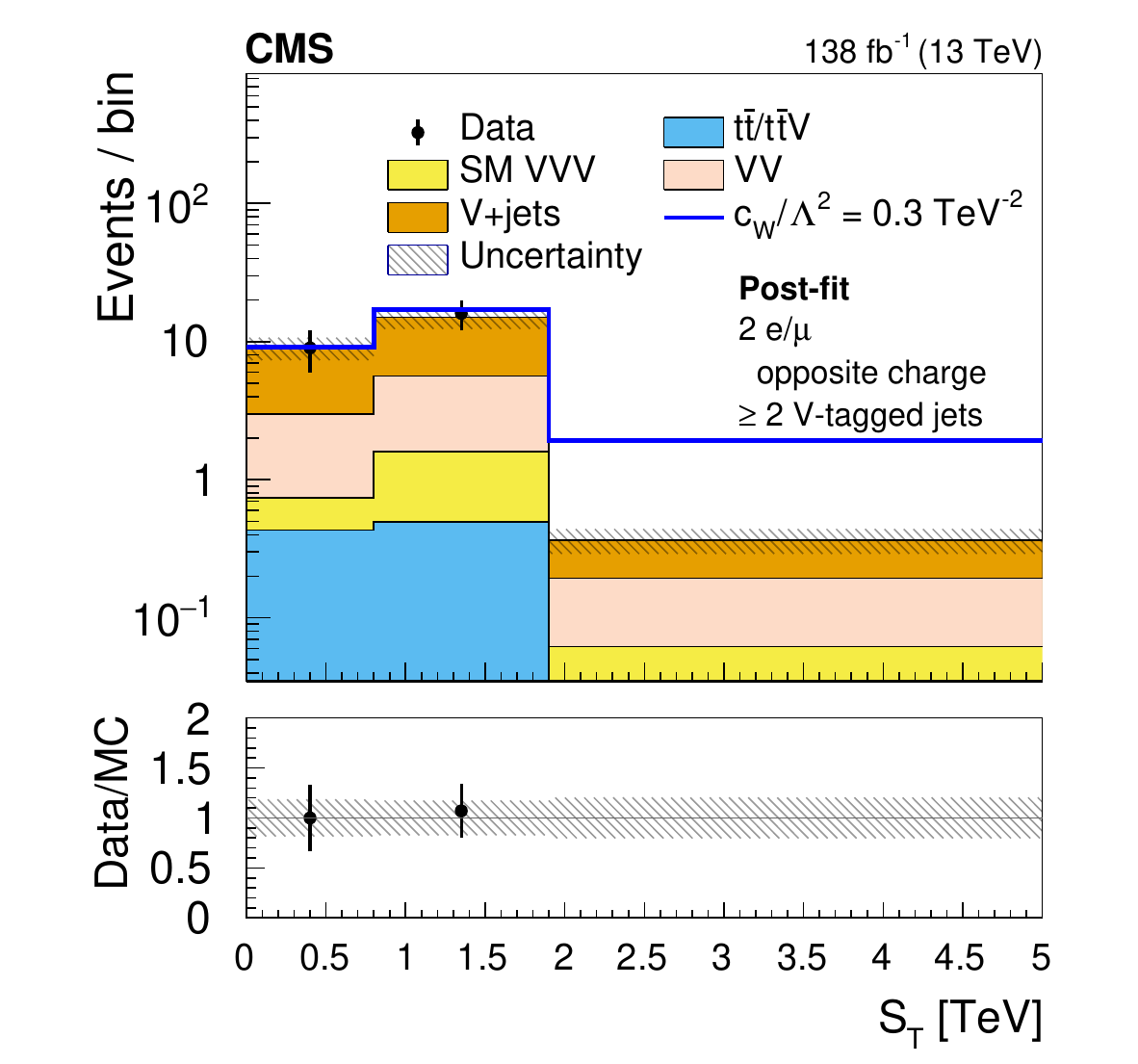}
    \caption{Comparison of the post-fit \ST distributions.
      The upper plots and the lower left plot correspond to the
      opposite-sign dilepton and one \fatjet (\twoLEPOSoneFJ) channel,
      while the lower right plot corresponds to the
      opposite-sign dilepton and two or more \fatjets (\twoLEPOStwoFJ) channel.
      The shaded bands in the ratio plots represent the total uncertainties.
      The black dots with error bars represent the data with statistical uncertainties.
    }
    \label{fig:2lOSSR}
\end{figure*}

Figure~\ref{fig:2lOSSR} shows the distributions of the corresponding discriminating variables in the data for the four signal SRs corresponding to the opposite-sign dilepton channel. The data in all SRs are in agreement with the predictions of the SM.

\subsection{Same-sign dilepton channel}
Events for the SS dilepton channel (\twoLEPSSoneFJ) are collected using the same dilepton triggers as in the OS channel.
The leading (subleading) lepton must have $\pt > 40\,(30)\GeV$ and there must be at least one \fatjet that meets the
\ParticleNet medium working point criterion.
The relatively high \pt requirement for leptons is intended to reduce backgrounds where a jet is misidentified as a lepton.
To further suppress the significant \ttbar background, events containing one or more \PQb-tagged narrow jets are excluded.
$\PZ+$jets events can enter the selection if a lepton is reconstructed with the wrong charge, which occurs more frequently
for electrons than for muons. To mitigate this background, the requirement $\abs{m_{\Pe\Pe} - \MZ} > 20\GeV$ is imposed.
To reject trident events, in which a lepton radiates a photon that converts in the detector,
the distance between leptons in $\eta$-$\phi$ space is required to exceed~1.2.
The primary backgrounds in the SS dilepton SR come from \ttbar and \WZ processes.
Control regions are defined to check the estimated contributions of these backgrounds.

The \ttbar CR is established by requiring at least one \PQb-tagged jet;
all other selection criteria are kept the same as those in the SR.
This region is approximately 90\% pure in \ttbar events.

The \WZ background enters the SR when at least one ISR jet passes
the \fatjet selection criteria, and both bosons decay leptonically but one of the leptons
from the \PZ boson decay is not identified.
To evaluate this background, a CR is defined
by selecting events with exactly three electrons or muons, two of which must be
consistent with a \PZ boson decay. The remaining lepton, assumed to originate from
the \PW boson, must have $\pt > 30\GeV$. Since there is no third boson in the \WZ process, the large-radius
jet is not required to meet the \PV-tagged jet substructure criteria. A range of soft-drop mass
$20 < \mSD < 60\GeV$ is used to avoid signal contamination in this CR. The event sample
is approximately 80\% pure in \WZ events, with the remaining events coming primarily from
\ttbar and $\PZ+$jets processes.

\begin{figure*}[htpb]
    \centering
    \includegraphics[width=0.48\textwidth]{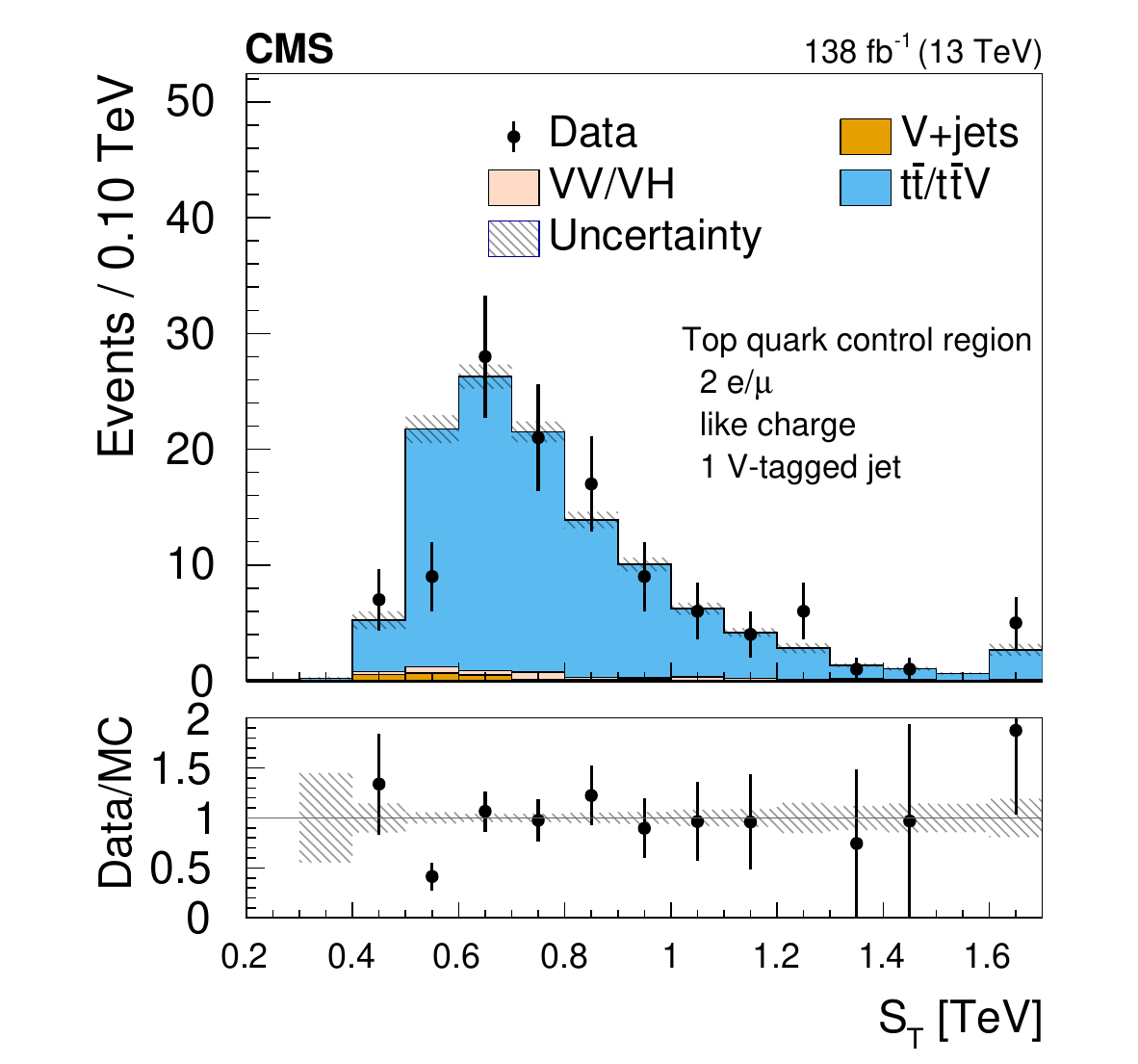}
    \includegraphics[width=0.48\textwidth]{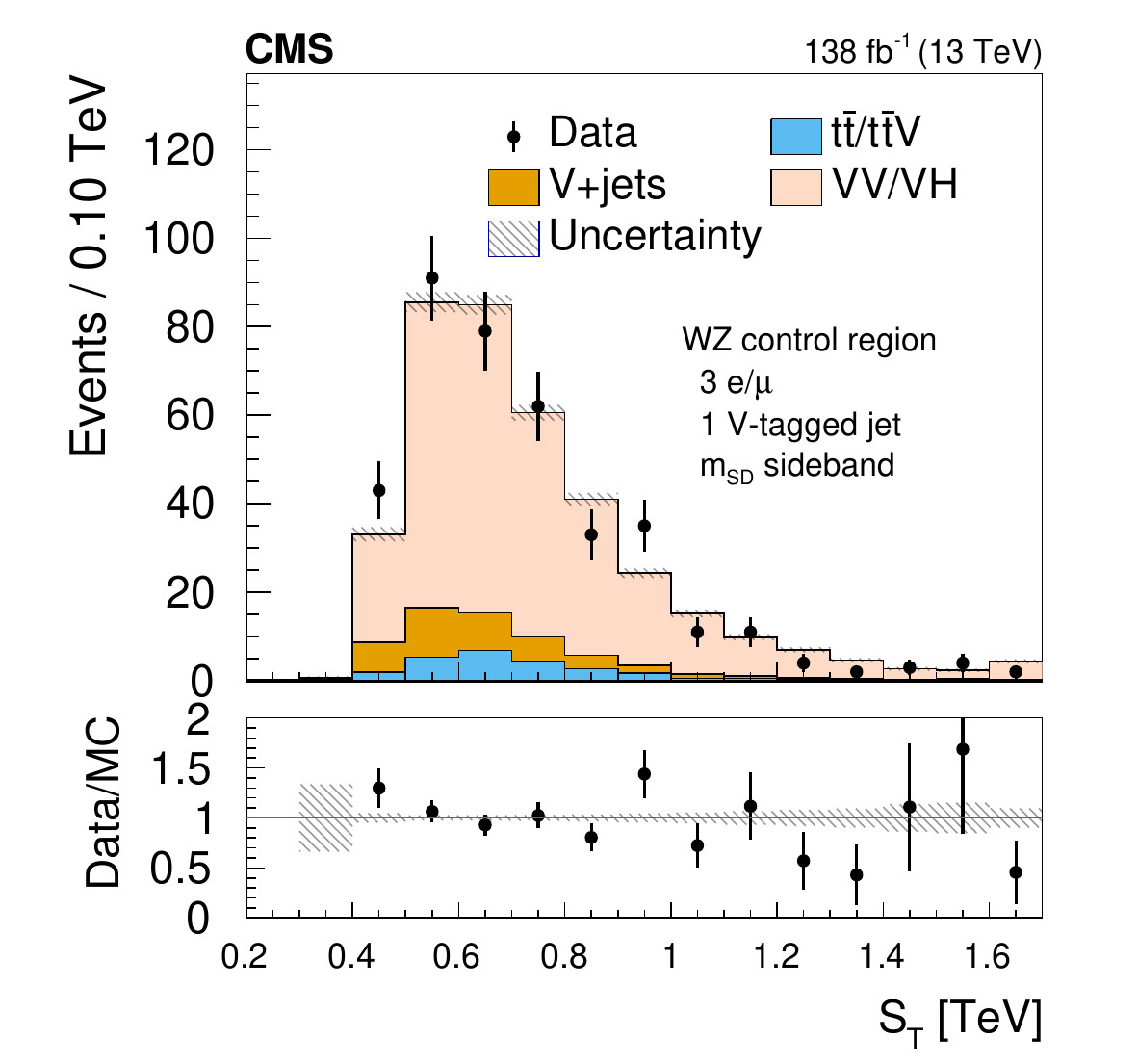}
    \caption{Comparison of pre-fit \ST distributions for the \ttbar (left) and \WZ (right) control regions
      in the \twoLEPSSoneFJ channel.
      The shaded band in the ratio plot represents the MC statistical uncertainty.
      The black dots with error bars represent the data with statistical uncertainties.}
    \label{fig:2l2fjSSCR}
\end{figure*}

Good agreement between data and simulation is observed in both the \ttbar and \WZ CRs. A background estimation is performed by computing the bin-by-bin ratio between data and simulation in these CRs, after subtracting subdominant backgrounds and applying this ratio to the MC predicted yields in the SR. Figure~\ref{fig:2l2fjSSCR} shows a comparison of data and simulation in the \ttbar and \WZ CRs.

\begin{figure}[htpb]
    \centering
    \includegraphics[width=0.48\textwidth]{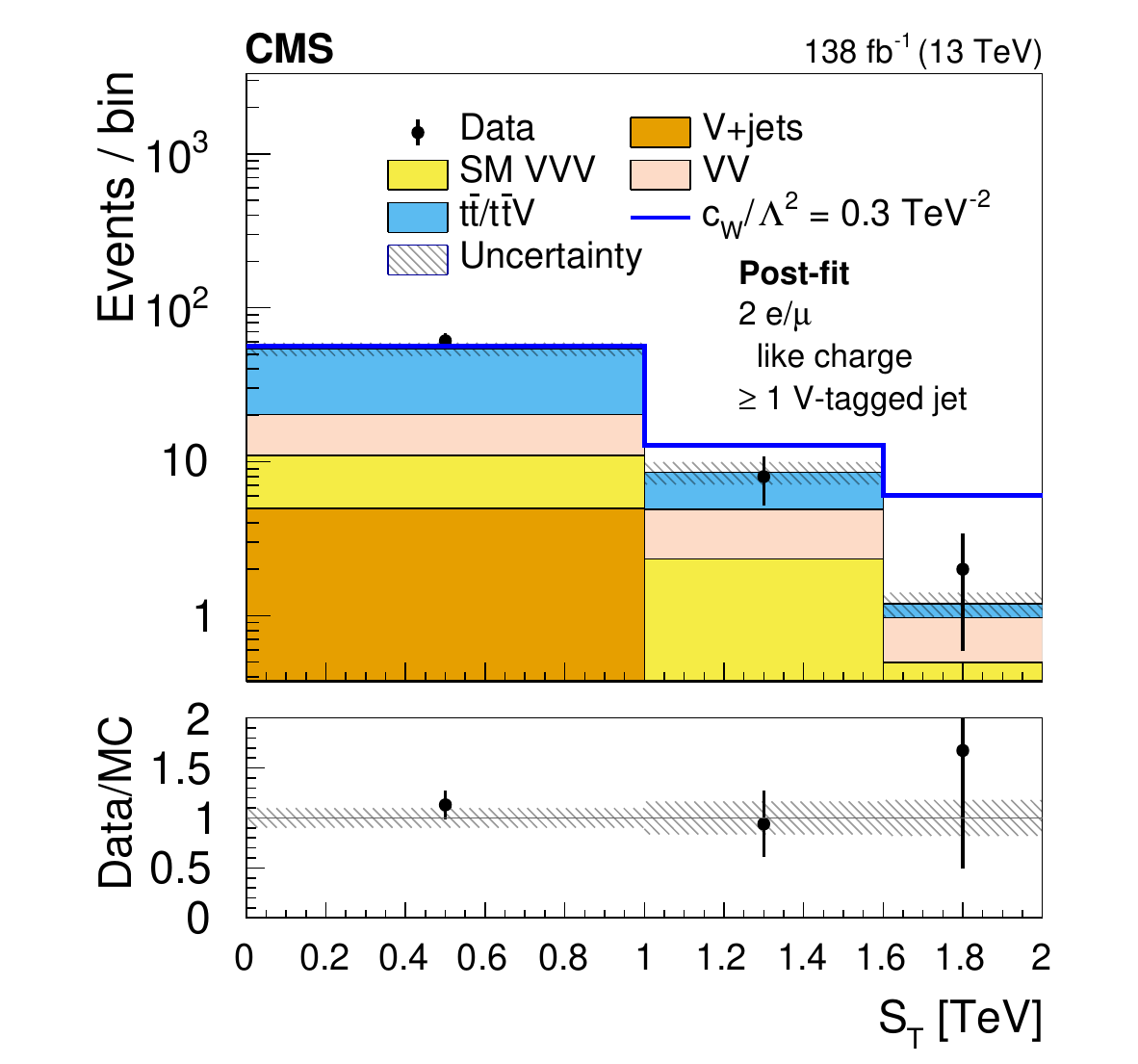}
    \caption{Comparison of the post-fit \ST distributions for the
      same-sign dilepton plus one \fatjets (\twoLEPSSoneFJ) signal region.
      The shaded band in the ratio plot represents the total uncertainty.
      The black dots with error bars represent the data with statistical uncertainties.
    }
    \label{fig:2lSSSR}
\end{figure}

The \ST distribution in \twoLEPSSoneFJ, shown in Fig.~\ref{fig:2lSSSR}, is compared to the predicted background
and a hypothetical signal. The data agree with the SM prediction.

\subsection{Channels with tau leptons that decay hadronically}
These channels are characterized by the presence of a single $\tauh$ candidate, originating from a \PW or \PZ boson decay. Since \PGt leptons decay rapidly, leptons from \PGt decays are difficult to distinguish from prompt leptons. Consequently, \PW and \PZ boson decays involving leptonic \PGt decays are accounted for in the leptonic channels described above. The SRs described in this section focus on the remaining 64.8\% of taus that decay hadronically~\cite{PDG24}.

All events in this channel must contain one \tauh candidate and either one or two electrons or muons.
Events in the $\tauh+$lepton category are required to include a \fatjet corresponding to the
decay of one of the three gauge bosons. Events in the $\tauh+$dilepton category must contain no \fatjets,
consistent with the assumption that all three gauge bosons decay leptonically.

Additional selection criteria are applied in the $\tauh+$dilepton channel: the leptons must be spatially separated
by a distance of at least 0.4 in the $\eta$-$\phi$ plane, and the sum of the charges of both leptons and the \tauh candidate must equal $\pm1$.
To suppress $\PZ+$jets background, events with OS SF leptons and $\abs{\Mll - \MZ} < 20\GeV$ are rejected.

The primary background in these channels arises from events with nonprompt leptons and a~\tauh candidate.
Since simulating the rate of nonprompt leptons is difficult, this background contribution is estimated using the tight-to-loose (TL)
method~\cite{CMS:2010uxk}, which accounts for the kinematic properties and the flavor of the parent parton of the nonprompt lepton.
This method utilizes two CRs: the measurement region, where the TL ratio ($f$) is determined, and the application region, where $f$ is applied to estimate the nonprompt-lepton background contribution to the signal region. The ratio $f$ is defined as the fraction of events in the measurement region where the loose lepton also satisfies the tight lepton selection criteria. It is calculated as a function of \pt and $\eta$. The selection criteria for the $f$ measurement region depend on the lepton type. To calculate the electron (muon) fake ratio, events with two OS muons (electrons) are selected.
To determine the \tauh fake rate, events containing two OS SF leptons are used. The application regions are defined similarly to the SRs, with the key difference that one of the leptons satisfies the loose selection but fails the tight selection. These regions are primarily composed of nonprompt leptons, while smaller contributions from prompt-lepton events are estimated using simulation and subtracted. The background contribution is then determined by weighting each event by $f$, and summing over different fake lepton combinations.
Other background sources include Drell--Yan, $\PW+$jets, \ttbar, and diboson production and are estimated using MC simulations.

Signal discrimination is enhanced by using boosted decision trees (BDTs) which take the kinematic quantities
of reconstructed particles as input. Separate BDTs are trained for the single-lepton and dilepton categories and for
each of the data-taking years. Hyperparameters are optimized to maximize analysis sensitivity while avoiding overfitting.
Each BDT is trained on a benchmark signal and subsequently validated to ensure its effectiveness across other signal points. Variables with the most discriminating power include the lepton \pt and \ptvecmiss.

\begin{figure*}[htpb]
    \centering
    \includegraphics[width=0.48\textwidth]{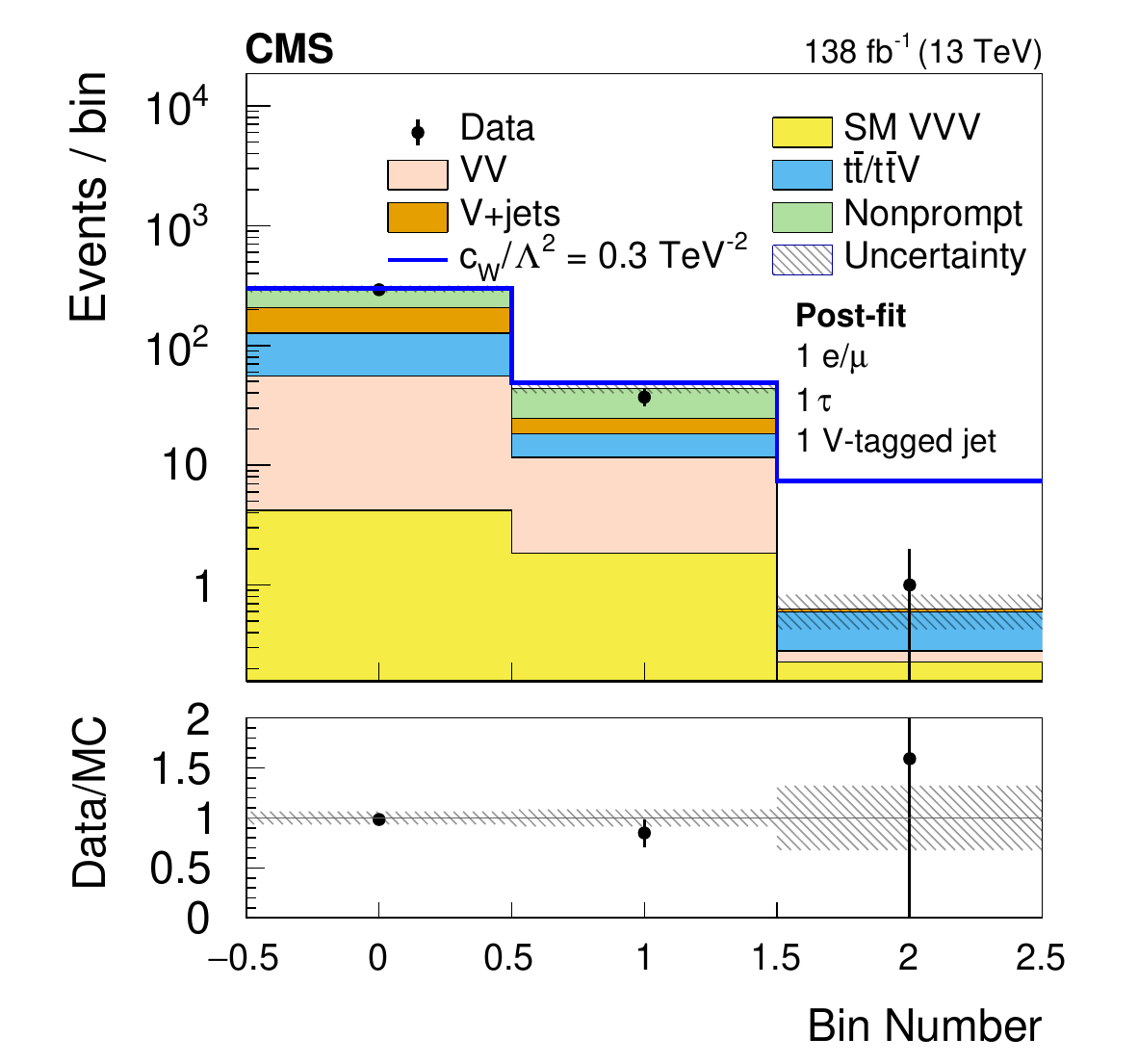}
    \includegraphics[width=0.48\textwidth]{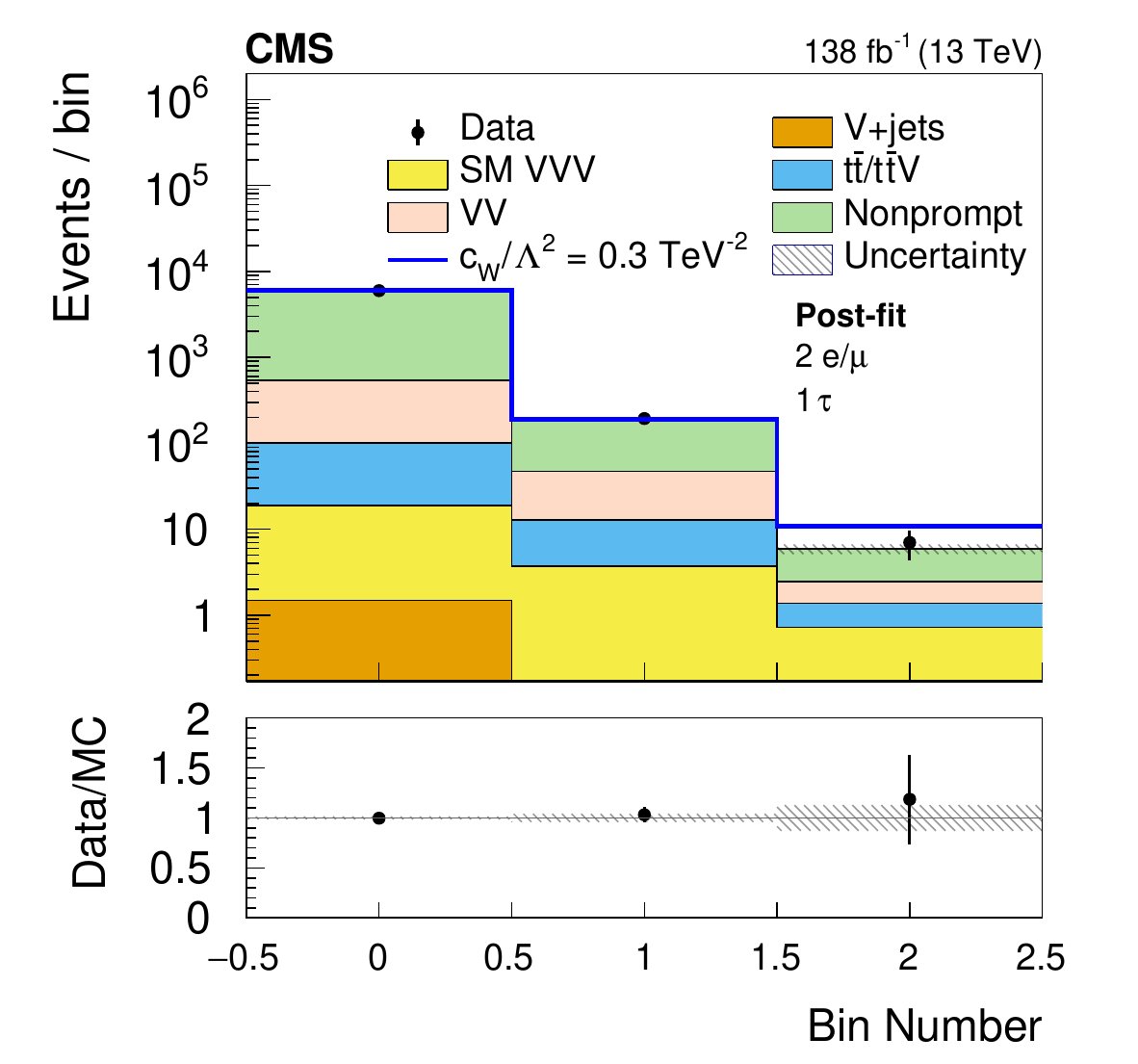}
    \caption{Comparison of the post-fit distributions binned in the BDT score and \ST
      for the \oneLEPoneTAUoneFJ (left) and \twoLEPoneTAUnoFJ (right) signal regions.
      The shaded bands in the ratio plots represent the MC total uncertainties.
      The black dots with error bars represent the data with statistical uncertainties.
    }
    \label{fig:tausr}
\end{figure*}

Three bins for each \oneLEPoneTAUoneFJ and \twoLEPoneTAUnoFJ are defined according to the corresponding BDT score and the \ST value. In the \oneLEPoneTAUoneFJ (\twoLEPoneTAUnoFJ) channel, the boundary separating low and high \ST is at 600 (300)\GeV.
The first bin includes events with low BDT scores and low \ST. While this bin is dominated by background and has very little signal sensitivity, its large sample size makes it useful for constraining background estimates. The second bin, with low BDT scores but high \ST, offers moderate sensitivity to the signal. The third bin, characterized by both high BDT scores and high \ST, is the most sensitive bin. Events with high BDT scores and low \ST suffer from small yields and poor signal-to-background ratio, and are therefore excluded from the analysis. A comparison of observed data and expected background yields in each region is shown in Fig.~\ref{fig:tausr}. No significant deviations from the background-only hypothesis are observed.

\section{Systematic uncertainties}
\label{sec:Systematics}

Systematic uncertainties are assigned following prescriptions common to all channels. Some uncertainties
relate to the way objects are reconstructed and calibrated, while others follow from theoretical considerations.
Unless otherwise noted, uncertainties on the result refer to event yields.

Scale factors for leptons are obtained using a tag-and-probe technique~\cite{CMS:2011aa} as a function of the probe lepton's \pt and $\eta$.
Systematic uncertainties are on the order of 1--2\% (3--5\%) for \Pe and \PGm ($\tauh$), and reflect the number of
probe leptons available, the impact of nonprompt leptons, and limitations of the method.
Uncertainties for a given lepton flavor are fully correlated, whereas uncertainties for different lepton flavors are uncorrelated.
In general, the impact of lepton scale factors uncertainties is small.

Events from top quark processes are suppressed by vetoing any event with a \PQb-tagged jet.
Yields depend on the efficiency for tagging jets that contain a \PQb hadron, as well as tagging jets that do not.
The scale factors are obtained as functions of \pt and $\eta$ for the efficiency to tag a jet as a \PQb jet
when the jet contains a \PQb hadron, contains a \PQc but no \PQb hadron, and contains only light quarks and gluons.
Uncertainties for these scale factors are typically 3--5\% and are taken to be fully correlated.

We derived scale factors for two categories of \fatjets: those that contain a hadronically decaying \PW boson
(which are characteristic of the signal) and those that contain an ISR gluon (typical of the
background, \ie, mistagged jets). The scale factors for \fatjets containing hadronically decaying \PZ bosons
or top quarks are assumed to be the same as those for \fatjets containing \PW bosons.
Special calibration samples of \ttbar events are defined.
First, boosted \PW jets are tagged by fully reconstructing a \ttbar event and tagging the two \PQb jets.
This sample is used to measure the scale factor for \PW jets.
Second, a sample rich in high-\pt ISR gluon jets is defined by a pair of OS SF leptons
that are not compatible with \PZ boson decays, two \PQb-tagged jets, and at least one high-\pt \fatjet candidate.
\fatjet scale factors are obtained in three bins of jet \pt for each data-taking era; these scale factors tend
to be smaller than unity for \PW bosons and larger than unity for \PQb quark and ISR gluon jets;
uncertainties are 5--35\% depending on \pt.

Scale factors for trigger efficiencies are calculated using events selected on the basis of other triggers or
through the tag-and-probe technique. The impact of trigger scale factor uncertainties is about 1\% for the
single-lepton channel, 1--3\% for the double-lepton and \tauh channels, and 5\% for the fully hadronic channel.
An additional correction is applied to account for mistiming of the trigger during the 2016--2017 data taking
period~\cite{CMS:2024ppo,CMS:2021yvr}. The associated systematic uncertainty is small, usually below 1\%.

Jet energy corrections (JECs)~\cite{CMS:2016lmd,CMS:2011shu} are applied to the energy and momentum of jets
reconstructed in the data so that they match on average those of particle-based jets in the MC.
Uncertainties associated with the JECs
are divided into 27 uncorrelated sources. They are evaluated by scaling the jet momenta in a correlated fashion.
JEC uncertainties are among the most significant uncertainties encountered in this analysis and can reach up to tens of percent
at high \ST.
Studies show that the jet energy resolution in data is slightly worse than in
the simulation~\cite{CMS:2016lmd}. Additional smearing is applied to the jets in simulation to correct this discrepancy.

The energy scale uncertainties for electrons and muons have a negligible impact on the results.
The energy scale uncertainty for \tauh leptons, however, must be taken into account.
Shifts in the \tauh energy scale lead to small changes in the numbers of selected events and in the BDT scores.
We assess the uncertainty on the tau energy scale on the basis of a calibration~\cite{TAU-16-003,TAU-20-001,CMS:2025kgf}
derived from $\PZ\to\PGt_{\PGm}\tauh$ events; the changes in the yields are typically a couple of percent.

Pileup has an impact on jet reconstruction and the isolation of prompt leptons. Weights are applied
to MC events so that the pileup distributions in the simulation match what is observed in the data.
The uncertainty assigned to the total inelastic cross section used to calculate these weights is 4.6\%~\cite{CMS:inelastic} and is
propagated to obtain yield uncertainties due to pileup, which are generally around~1\%.

The integrated luminosity is needed for predicting signal and background yields. It is well measured
using multiple luminometers calibrated in van der Meer scans~\cite{CMS:LUMI2021} with a precision
better than 2.5\%, varying slightly
from one data-taking period to another. When the 2016--2018 data are combined, the assigned uncertainty
is 1.6\%~\cite{CMS:LUMI2017,CMS:LUMI2018,CMS:LUMI2021}.

Predicted yields depend on theoretical cross sections for which the most precise available values are used.
The leading impact arises from missing QCD higher-order corrections, evaluated
by varying the factorization and renormalization scales up and down by a factor of two
(excluding opposite variations).
The envelope of these variations is taken as the theoretical cross section uncertainty.
These scale uncertainties can reach tens of percent and significantly impact the sensitivity of this analysis.

The PDFs and the value of the strong coupling
are key ingredients for calculating theoretical cross sections.
Their uncertainties are propagated to the yields using the method of replicas~\cite{NNPDF31}.
They have a substantial impact on the event yields, amounting to a few percent in the
leptonic channels and up to several tens of percent in the zero-lepton channels, \ie,
up to 50\% of the total uncertainty. The impact on the bounds obtained for Wilson coefficients is significant.
This sensitivity to PDF uncertainties comes from the tendency of the signal process at high \ST to be
initiated by incoming partons at high $x$, where PDF uncertainties are relatively large.

All sources of systematic uncertainty are evaluated for both signal and backgrounds.
For the signal, each source is incorporated in the likelihood fit as a nuisance parameter fully correlated
across channels and data-taking periods.
For some backgrounds, small numbers of MC events lead to extremely small or very large values for any given uncertainty.
Rather than propagating unrealistic values to the likelihood, reasonable overall background uncertainties
in the range 20--50\% are assigned; they are fully correlated across \ST bins.
In general, the impact of systematic uncertainties is moderate, weakening the expected bounds by 30\% or less.

\section{Results}
\label{sec:Results}

The main goal of this analysis is to constrain or measure certain dim-6 and dim-8 Wilson coefficients.
We derive 95\% \CL bounds on individual coefficients and on pairs of coefficients,
as detailed below. We also report best fit values and 68\% confidence intervals.
A so-called ``clipping'' procedure~\cite{CMS:clipping,Szleper:clipping} is implemented to reduce the impact of high-\ST events that cannot be accommodated by EFTs with $\Lambda = 1\TeV$.
A template fit is introduced that constrains the shape of the theoretical \mVVV distribution;
this treatment would shed light on the shape of the \mVVV distribution if an excess of events were observed.
It could also facilitate the combination of our results with others.
Finally, while this analysis is not designed to measure SM \VVV production, there is nonetheless some
statistical sensitivity to the SM process without any enhancements from new physics, as we demonstrate below.

The numbers of events selected in the 2016--2018 data are reported for all \nbinstotal~channels in Table~\ref{tab:observed}.
The yields agree with SM expectations and there is no indication for any excess or deficit.
A statistical test based on the Poisson distribution is performed, taking the SM expectation as the mean and
computing the probability to obtain the observed number of events or more.
The result of this test is consistent with the null hypothesis and the distribution
of $p$-values~\cite{Gross:2010qma} is uniform even when restricted to the most sensitive bins of all channels.
If a Wilson operator were active at a significant level, then deviations from expected yields would be seen in several channels, in contrast to a statistical fluctuation, which would appear only in one.
We conclude that there is no sign of new physics in our signal regions.

\begin{figure*}[hbpt!]
    \centering
    \includegraphics[width=0.98\textwidth]{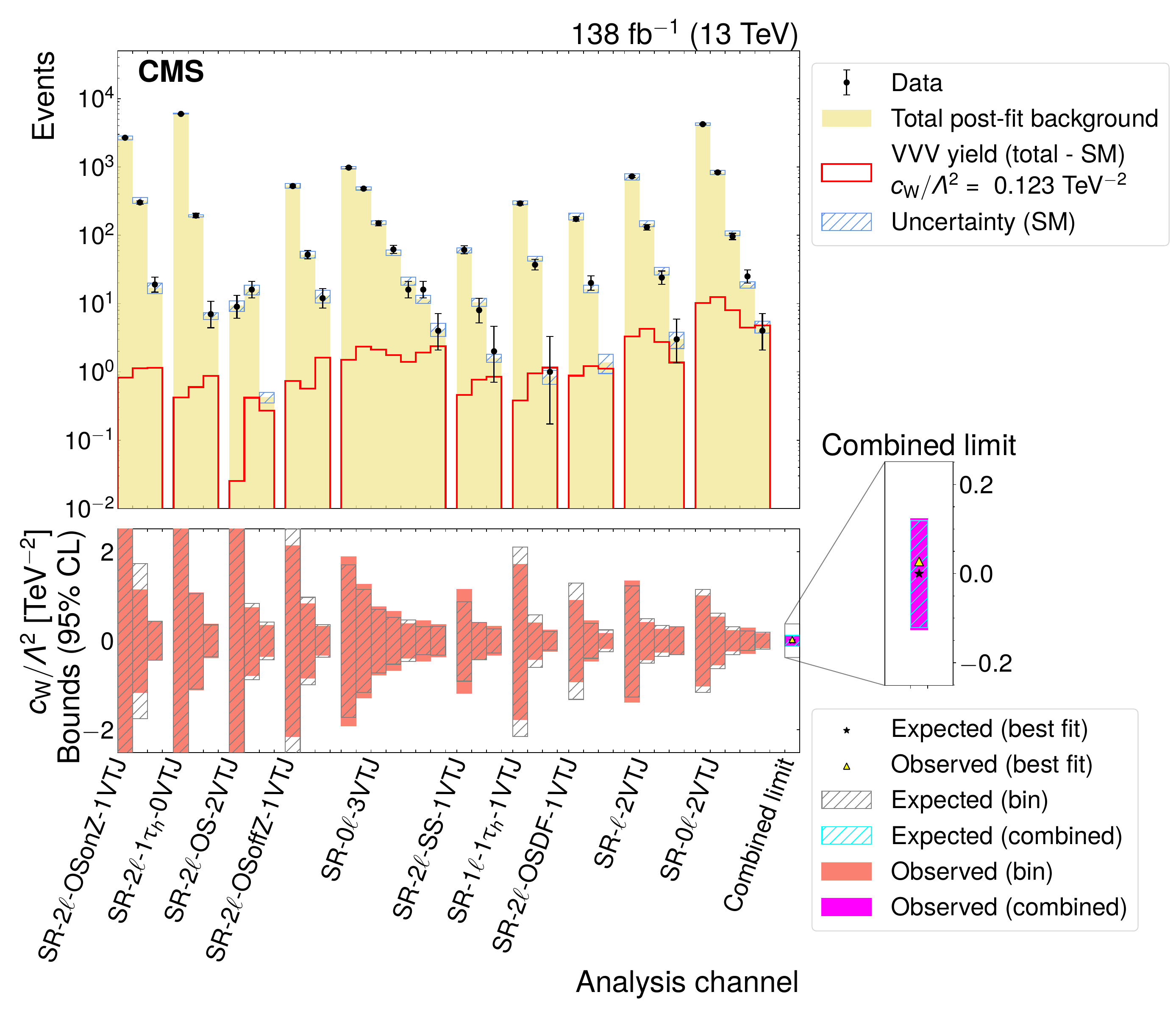}
    \caption{ \label{fig:observed}
      Summary of the bin-by-bin yields in all signal regions and associated limits on $\cW/\Lambda^2$.
      The channels are listed from left to right in order of increasing sensitivity to $\cW/\Lambda^2$.
      In the upper panel, the beige histogram shows the predicted SM yields including SM \VVV production
      while the red line represents the additional contribution expected when $\cW/\Lambda^2 = \cWvalue\TeV^{-2}$.
      The black dots with error bars represent the data with statistical uncertainties.
      The lower panel shows the expected and observed limits on $\cW/\Lambda^2$ for each bin in each channel.
      The 95\% \CL combined limit on $\cW/\Lambda^{-2}$ is obtained by performing the fit on all \nbinstotal bins;
      the observed (expected) result is represented by the magenta (cyan) bar.}
\end{figure*}

A graphical representation of event yields and the bounds on $\cW/\Lambda^2$ is given in Fig.~\ref{fig:observed}.
For this illustrative fit, all Wilson coefficients aside from \cW are fixed to zero.
The upper panel shows good agreement between the data and the SM predicted yields
which include SM \VVV production. A prediction for $\cW/\Lambda^2 = \cWvalue\TeV^{-2}$ is also shown, for comparison.
This value corresponds to the expected 95\% \CL.
The lower panel shows that sensitivity to $\cW/\Lambda^2$ is similar across channels
and that the highest \ST bin is generally the most sensitive one in each channel.
The combined bound on $\cW/\Lambda^2$ is much more stringent than any one channel.

\begin{table}[htp]
\centering
\topcaption{ \label{tab:observed}
Summary of the SM expected and observed numbers of events.
The post-fit uncertainties in the expected numbers of events include all
statistical and systematic uncertainties relating to the prediction.
The row ``Sum highest bins'' is computed by summing the expected and observed
numbers of events for the last bin in each channel;
the last bin in each channel is usually the most sensitive one.}
\begin{tabular}{l c c c}
\hline
Channel & Bin & Expected & Observed \\
\hline
\noLEPtwoFJ
& $\STnoMET<1.5\TeV$ & $4222 \pm 186$ & $4217$ \\
& $1.5<\STnoMET<2\TeV$ & $826 \pm 58$ & $830$ \\
& $2<\STnoMET<2.5\TeV$ & $106 \pm 9$ & $96$ \\
& $2.5<\STnoMET<3\TeV$ & $18.5 \pm 1.9$ & $25$ \\
& $3\TeV<\STnoMET$ & $4.5 \pm 0.9$ & $4$ \\[\cmsTabSkip]
\noLEPthreeFJ
& $\STnoMET<1.5\TeV$ & $973 \pm 39$ & $977$ \\
& $1.5<\STnoMET<1.75\TeV$ & $483 \pm 26$ & $481$ \\
& $1.75<\STnoMET<2\TeV$ & $152 \pm 10$ & $150$ \\
& $2<\STnoMET<2.25\TeV$ & $55 \pm 5$ & $62$ \\
& $2.25<\STnoMET<2.5\TeV$ & $21 \pm 3$ & $16$ \\
& $2.5<\STnoMET<3\TeV$ & $11.5 \pm 1.6$ & $16$ \\
& $3\TeV<\STnoMET$ & $4.0 \pm 0.9$ & $4$ \\[\cmsTabSkip]
\oneLEPtwoFJ
& $\MJJlnu<1.6\TeV$ & $705 \pm 78$ & $727$ \\
& $1.6<\MJJlnu<2.4\TeV$ & $145 \pm 15$ & $131$ \\
& $2.4<\MJJlnu<3.4\TeV$ & $29 \pm 4$ & $24$ \\
& $3.4\TeV<\MJJlnu$ & $2.9 \pm 0.8$ & $3$ \\[\cmsTabSkip]
\twoLEPOSONZoneFJ
& $\STnoMET<0.8\TeV$ & $2662 \pm 173$ & $2675$ \\
& $0.8<\STnoMET<1.3\TeV$ & $321 \pm 32$ & $302$ \\
& $1.3\TeV<\STnoMET$ & $17 \pm 3$ & $19$ \\[\cmsTabSkip]
\twoLEPOSOFFZoneFJ
& $\STnoMET<0.8\TeV$ & $527 \pm 43$ & $525$ \\
& $0.8<\STnoMET<1.05\TeV$ & $51 \pm 6$ & $52$ \\
& $1.05\TeV<\STnoMET$ & $12 \pm 3$ & $12$ \\[\cmsTabSkip]
\twoLEPOSOFoneFJ
& $\STnoMET<0.8\TeV$ & $181 \pm 21$ & $173$ \\
& $0.8<\STnoMET<1.35\TeV$ & $15 \pm 2$ & $20$ \\
& $1.35\TeV<\STnoMET$ & $1.2 \pm 0.4$ & $0$ \\[\cmsTabSkip]
\twoLEPOStwoFJ
& $\ST<0.8\TeV$ & $9.0 \pm 1.7$ & $9$ \\
& $0.8\leq\ST<1.9\TeV$ & $15 \pm 3$ & $16$ \\
& $1.9\TeV<\ST$ & $0.37 \pm 0.07$ & $0$ \\[\cmsTabSkip]
\twoLEPSSoneFJ
& $\ST<1\TeV$ & $54 \pm 5$ & $61$ \\
& $1<\ST<1.6\TeV$ & $8.5 \pm 1.4$ & $8$ \\
& $1.6\TeV<\ST$ & $1.2 \pm 0.2$ & $2$ \\[\cmsTabSkip]
\oneLEPoneTAUoneFJ
& low BDT score, low $\ST$& $297 \pm 19$ & $292$ \\
& low BDT score, high $\ST$& $44 \pm 4$ & $37$ \\
& high BDT score, high $\ST$& $0.6 \pm 0.2$ & $1$ \\[\cmsTabSkip]
\twoLEPoneTAUnoFJ
& low BDT score, low $\ST$& $5984 \pm 83$ & $5968$ \\
& low BDT score, high $\ST$& $188 \pm 8$ & $194$ \\
& high BDT score, high $\ST$& $5.9 \pm 0.8$ & $7$ \\[\cmsTabSkip]
\multicolumn{2}{l}{Sum highest bins} & $50 \pm 4$ & $52$ \\
\hline
\end{tabular}
\end{table}

Wilson coefficients are determined by simultaneously fitting the observed yields in all signal-region \ST bins.
Poisson probability functions are used for the observed yields; they depend on
signal and background expectations and include nuisance parameters for systematic uncertainties
modeled by log-normal functions.
The fit minimizes a negative log-likelihood function (NLL) to determine point values,
and parameter scans of $\twoDNLL = 2(\text{NLL} - \text{NLL}_{\text{min}})$ determine confidence intervals.
Signal contributions are parameterized in accord with Eq.~(\ref{eq:CrossSection}) for the Wilson coefficients
listed in Tables~\ref{tab:dim6} and~\ref{tab:dim8}.

For the first set of results, one Wilson coefficient at a time is allowed to vary while
all others are fixed to their SM value of zero.
Tables~\ref{tab:limit_summary_dim6} and~\ref{tab:limit_summary_dim8} list the expected and
observed 95\% \CL bounds for dim-6 and dim-8 Wilson coefficients.
Table~\ref{tab:measurement_summary_dim6} reports the measured dim-6 Wilson coefficients with their 68\% \CL intervals.

\begin{table}[hbtp!]
\renewcommand{\arraystretch}{1.1}
\centering
\topcaption{Summary of the 95\% \CL bounds on the dim-6 Wilson coefficients. We consider the case of a single varying Wilson coefficient (``Freeze other WCs'') as well as the case when the other Wilson coefficients are profiled (``Profile other WCs''). The Wilson coefficients are ordered by increasing confidence interval width. \label{tab:limit_summary_dim6}}
\begin{tabular}{r c c c c}
\hline
& \multicolumn{2}{p{7cm}}{\centering Freeze other WCs} & \multicolumn{2}{p{7cm}}{\centering Profile other WCs} \\
\hline
Wilson & \multicolumn{2}{p{7cm}}{\centering 95\% \CL Bounds $[\TeVns^{-2}]$} & \multicolumn{2}{p{7cm}}{\centering 95\% \CL Bounds $[\TeVns^{-2}]$} \\
coefficient & \multicolumn{1}{p{3.5cm}}{\centering Observed} & Expected & \multicolumn{1}{p{3.5cm}}{\centering Observed} & Expected \\
\hline
\rule{0pt}{2.6ex}
$\cW/\Lambda^{2}$ & $[-0.13, 0.12]$ & $[-0.12, 0.12]$ & $[-0.11, 0.12]$ & $[-0.13, 0.12]$\\
$\cHqthree/\Lambda^{2}$ & $[-0.24, 0.21]$ & $[-0.23, 0.20]$ & $[-0.44, 0.37]$ & $[-0.30, 0.25]$\\
$\cHqone/\Lambda^{2}$ & $[-0.34, 0.34]$ & $[-0.32, 0.32]$ & $[-0.39, 0.42]$ & $[-0.33, 0.33]$\\
$\cHu/\Lambda^{2}$ & $[-0.60, 0.59]$ & $[-0.61, 0.59]$ & $[-0.89, 0.83]$ & $[-0.74, 0.73]$\\
$\cHd/\Lambda^{2}$ & $[-0.79, 0.79]$ & $[-0.79, 0.79]$ & $[-0.98, 1.04]$ & $[-0.86, 0.88]$\\
$\cHW/\Lambda^{2}$ & $[-1.60, 1.55]$ & $[-1.63, 1.55]$ & $[-3.2, 3.6]$ & $[-2.1, 2.3]$\\
$\cHWB/\Lambda^{2}$ & $[-5.2, 5.0]$ & $[-5.5, 5.2]$ & $[-9.6, 9.7]$ & $[-7.6, 7.4]$\\
$\cHlthree/\Lambda^{2}$ & $[-3.7, 1.2] \cup [9, 17]$ & $[-3, 15]$ & $[-5, 22]$ & $[-4, 18]$\\
$\cHB/\Lambda^{2}$ & $[-11, 11]$ & $[-12, 12]$ & $[-11, 12]$ & $[-13, 13]$\\
$\cllone/\Lambda^{2}$ & $[-32, -13] \cup [-9, 10]$ & $[-30, 7]$ & $[-34, 10]$ & $[-32, 8]$\\
$\cHsquare/\Lambda^{2}$ & $[-76, 69]$ & $[-69, 61]$ & $[-71, 68]$ & $[-56, 54]$\\
$\cHDD/\Lambda^{2}$ & $[-114, 71]$ & $[-108, 68]$ & $[-164, 81]$ & $[-130, 72]$\\
\hline
\end{tabular}
\end{table}

\begin{table}[hbtp!]
\renewcommand{\arraystretch}{1.1}
\centering
\topcaption{Summary of the 95\% \CL bounds and measurements on the dim-8 Wilson coefficients, when considering a single varying Wilson coefficient at a time. The Wilson coefficients are ordered by increasing confidence interval width. \label{tab:limit_summary_dim8}}
\begin{tabular}{r c c c}
\hline
Wilson & \multicolumn{2}{p{7cm}}{\centering 95\% \CL bounds $[\TeVns^{-4}]$} & \multicolumn{1}{p{4cm}}{\centering Measurement $[\TeVns{-4}]$} \\
coefficient & \multicolumn{1}{p{3.5cm}}{\centering Observed} & Expected & Observed\\
\hline
\rule{0pt}{2.6ex}
$\fT{0}/\Lambda^{4}$ & $[-0.57, 0.63]$ & $[-0.48, 0.54]$ & $-0.05^{+0.34}_{- 0.28 }$ \\
$\fT{1}/\Lambda^{4}$ & $[-0.63, 0.70]$ & $[-0.54, 0.62]$ & $0.05^{+0.26}_{- 0.41 }$ \\
$\fT{3}/\Lambda^{4}$ & $[-1.22, 1.32]$ & $[-1.04, 1.18]$ & $0.10^{+0.48}_{- 0.82 }$ \\
$\fT{2}/\Lambda^{4}$ & $[-1.29, 1.39]$ & $[-1.10, 1.24]$ & $0.10^{+0.50}_{- 0.85 }$ \\
$\fM{0}/\Lambda^{4}$ & $[-3.8, 4.0]$ & $[-2.8, 3.2]$ & $-0.4^{+2.9}_{-1.8 }$ \\
$\fT{5}/\Lambda^{4}$ & $[-4.0, 3.9]$ & $[-3.1, 3.0]$ & $0.0^{+2.6}_{-2.7 }$ \\
$\fT{6}/\Lambda^{4}$ & $[-4.9, 4.8]$ & $[-3.8, 3.7]$ & $0.0^{+3.2}_{-3.3 }$ \\
$\fM{1}/\Lambda^{4}$ & $[-6.7, 6.4]$ & $[-5.0, 5.1]$ & $-1.0^{+4.7}_{-3.0 }$ \\
$\fT{4}/\Lambda^{4}$ & $[-9.2, 9.0]$ & $[-7.1, 7.0]$ & $0.0^{+6.0}_{-6.0 }$ \\
$\fT{7}/\Lambda^{4}$ & $[-10, 10]$ & $[-8, 8]$ & $0.1^{+6.8}_{- 7.1 }$ \\
$\fM{7}/\Lambda^{4}$ & $[-10, 11]$ & $[-8, 9]$ & $-1.0^{+6.7}_{- 4.4 }$ \\
$\fM{5}/\Lambda^{4}$ & $[-11, 11]$ & $[-9, 9]$ & $0.0^{+7.3}_{- 7.0 }$ \\
$\fT{8}/\Lambda^{4}$ & $[-11, 12]$ & $[-7, 7]$ & $5^{+3}_{- 13 }$ \\
$\fM{4}/\Lambda^{4}$ & $[-13, 13]$ & $[-11, 11]$ & $0.0^{+8.7}_{- 8.4 }$ \\
$\fM{2}/\Lambda^{4}$ & $[-14, 14]$ & $[-11, 11]$ & $0.0^{+9.7}_{- 9.7 }$ \\
$\fT{9}/\Lambda^{4}$ & $[-22, 23]$ & $[-14, 14]$ & $11^{+5}_{- 27 }$ \\
$\fM{3}/\Lambda^{4}$ & $[-23, 23]$ & $[-18, 18]$ & $0^{+16}_{- 15 }$ \\
$\fS{1}/\Lambda^{4}$ & $[-26, 26]$ & $[-24, 25]$ & $-2^{+16}_{- 13 }$ \\
$\fS{0}/\Lambda^{4}$ & $[-38, 38]$ & $[-35, 35]$ & $-2^{+22}_{- 20 }$ \\
$\fS{2}/\Lambda^{4}$ & $[-38, 39]$ & $[-36, 36]$ & $-2^{+22}_{- 20 }$ \\
\hline
\end{tabular}
\end{table}

\begin{table}[hbtp!]
\renewcommand{\arraystretch}{1.2}
\centering
\topcaption{Summary of the measurements of the dim-6 Wilson coefficients. We consider the case of a single varying Wilson coefficient (``Freeze other WCs'') as well as the case when the other Wilson coefficients are profiled (``Profile other WCs''). \label{tab:measurement_summary_dim6}}
\begin{tabular}{rcc}
\hline
& \multicolumn{1}{p{4cm}}{\centering Freeze other WCs} & \multicolumn{1}{p{4cm}}{\centering Profile other WCs} \\
\hline
Wilson & \multicolumn{1}{p{4cm}}{\centering Measurement $[\TeVns^{-2}]$} & \multicolumn{1}{p{4cm}}{\centering Measurement $[\TeVns^{-2}]$} \\
coefficient & Observed & Observed \\
\hline
$\cW/\Lambda^{2}$ & $0.03^{+0.053}_{- 0.11 }$ & $0.001^{+0.057}_{- 0.055 }$\\
$\cHqthree/\Lambda^{2}$ & $0.05^{+0.09}_{- 0.21 }$ & $-0.14^{+0.36}_{- 0.16 }$\\
$\cHqone/\Lambda^{2}$ & $0.10^{+0.12}_{- 0.33 }$ & $-0.17^{+0.13}_{- 0.11 }$ or $0.19^{+0.10}_{- 0.13 }$\\
$\cHu/\Lambda^{2}$ & $0.09^{+0.29}_{- 0.47 }$ & $0.04^{+0.40}_{- 0.60 }$\\
$\cHd/\Lambda^{2}$ & $0.10^{+0.39}_{- 0.59 }$ & $-0.03^{+0.65}_{- 0.46 }$\\
$\cHW/\Lambda^{2}$ & $0.17^{+0.84}_{- 1.22 }$ & $0.5^{+1.8}_{- 2.2 }$\\
$\cHWB/\Lambda^{2}$ & $0.5^{+2.7}_{- 3.8 }$ & $1.6^{+5.3}_{- 7.1 }$\\
$\cHlthree/\Lambda^{2}$ & $14.4^{+0.9}_{- 2.7 }$ & $16.1^{+3.7}_{- 5.9 }$\\
$\cHB/\Lambda^{2}$ & $0.7^{+6.4}_{- 7.8 }$ & $0.7^{+5.7}_{- 6.1 }$\\
$\cllone/\Lambda^{2}$ & $-25^{+4}_{-4}$ or $5.1^{+1.6}_{- 5.8 }$ & $-25^{+5}_{-5}$ or $1.6^{+4.7}_{- 3.1 }$\\
$\cHsquare/\Lambda^{2}$ & $29^{+23}_{- 85 }$ & $-32^{+82}_{- 21 }$\\
$\cHDD/\Lambda^{2}$ & $-58^{+104}_{- 32 }$ & $8^{+34}_{- 138 }$\\
\hline
\end{tabular}
\end{table}

The channels have comparable sensitivity though the relative sensitivity varies by Wilson coefficient;
for example, the \noLEPtwoFJ and \twoLEPSSoneFJ channels are the most sensitive for $\cW/\Lambda^2$ and $\cHqthree/\Lambda^2$.
The two \tauh channels, \oneLEPoneTAUoneFJ and \twoLEPoneTAUnoFJ, are individually less sensitive
but improve expected bounds by up to 3\%.
The kinematic features and SM backgrounds differ channel-by-channel so the combination
is robust against systematic effects.
Removing any single channel changes the combined expected bounds only marginally, confirming the stability of the results.

The single-parameter fits serve mainly to compare this analysis to others.
It seems unrealistic to assume that only one Wilson coefficient would be nonzero, however.
A more realistic approach allows all Wilson coefficients to vary simultaneously. In other words,
when considering one particular Wilson coefficient, we profile the likelihood over all the others.
(We do not consider both dim-6 and dim-8 coefficients at the same time.)
In general, the expected limits obtained by profiling are not much weaker than when fixing
Wilson coefficients to zero, as can be seen in Table~\ref{tab:limit_summary_dim6}.
For example, the expected limits on $\cW/\Lambda^2$ are $[-0.12,0.12]\TeV^{-2}$ when all coefficients except \cW are
fixed to zero, and $[-0.13,0.12]\TeV^{-2}$ when they are profiled.
For $\cHqthree/\Lambda^2$, the corresponding expected limits are $[-0.23,0.20]\TeV^{-2}$ and $[-0.30,0.25]\TeV^{-2}$.

\begin{figure*}[htbp!]
\centering
\includegraphics[width=0.48\textwidth]{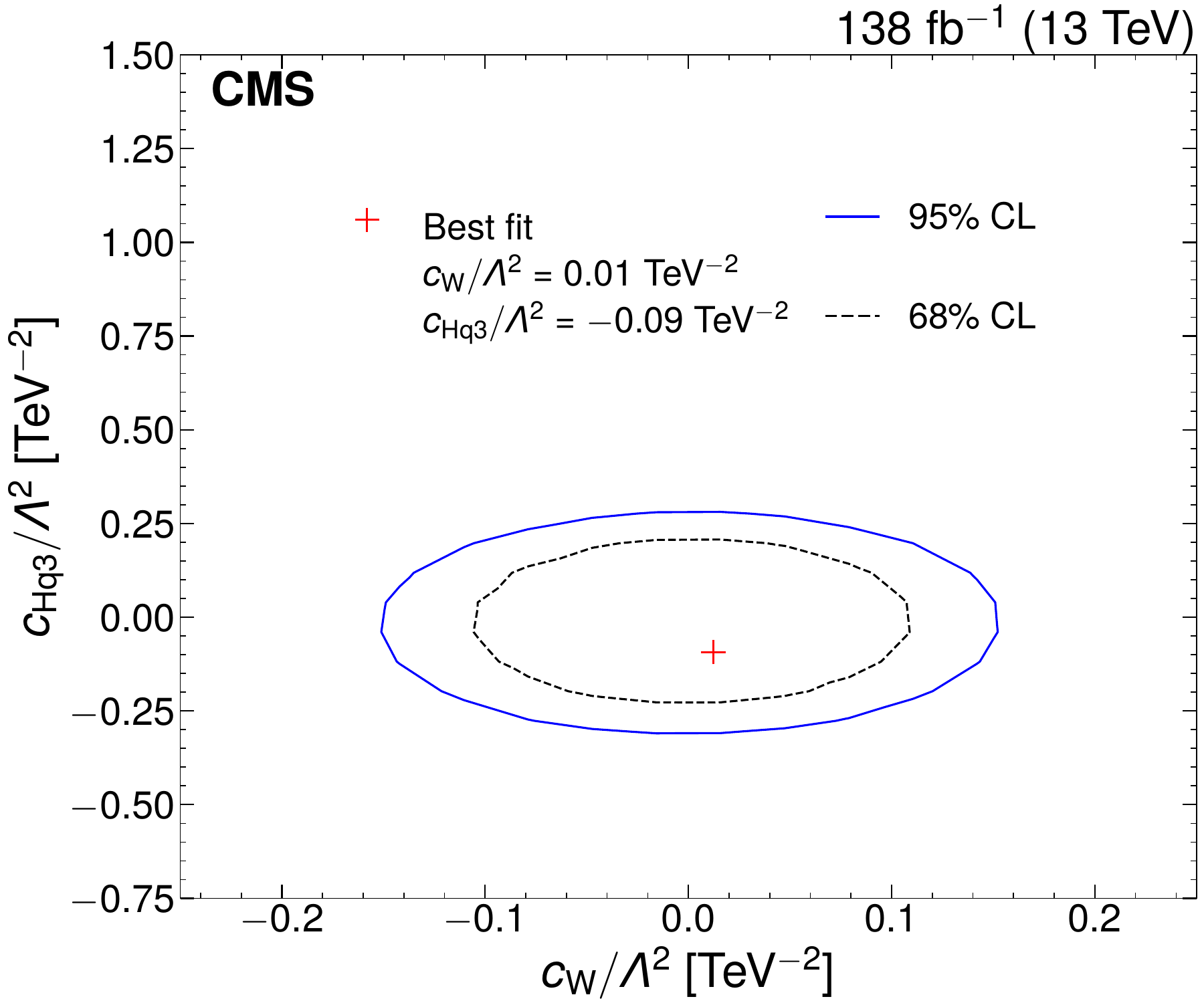}
\includegraphics[width=0.48\textwidth]{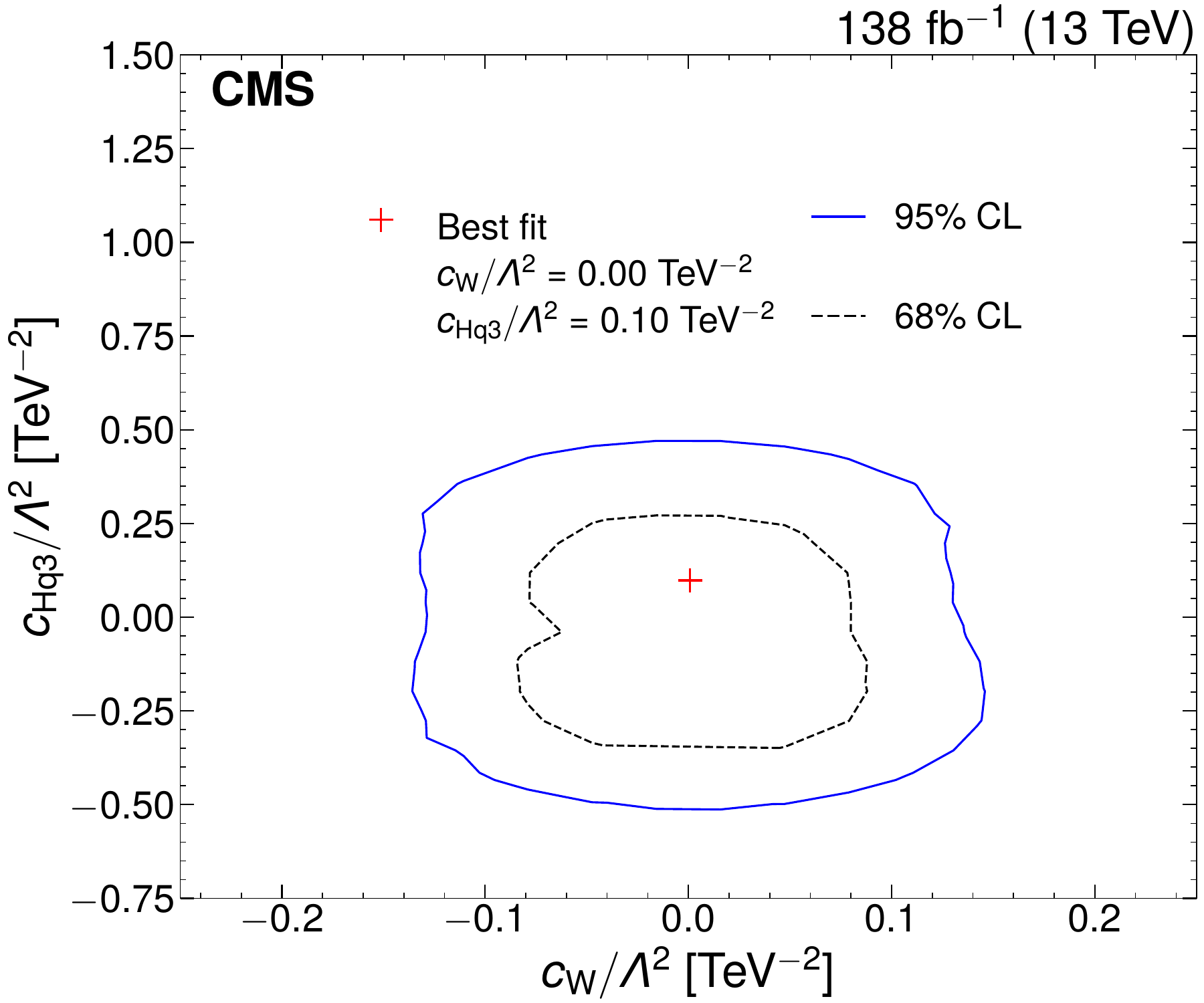}
\includegraphics[width=0.48\textwidth]{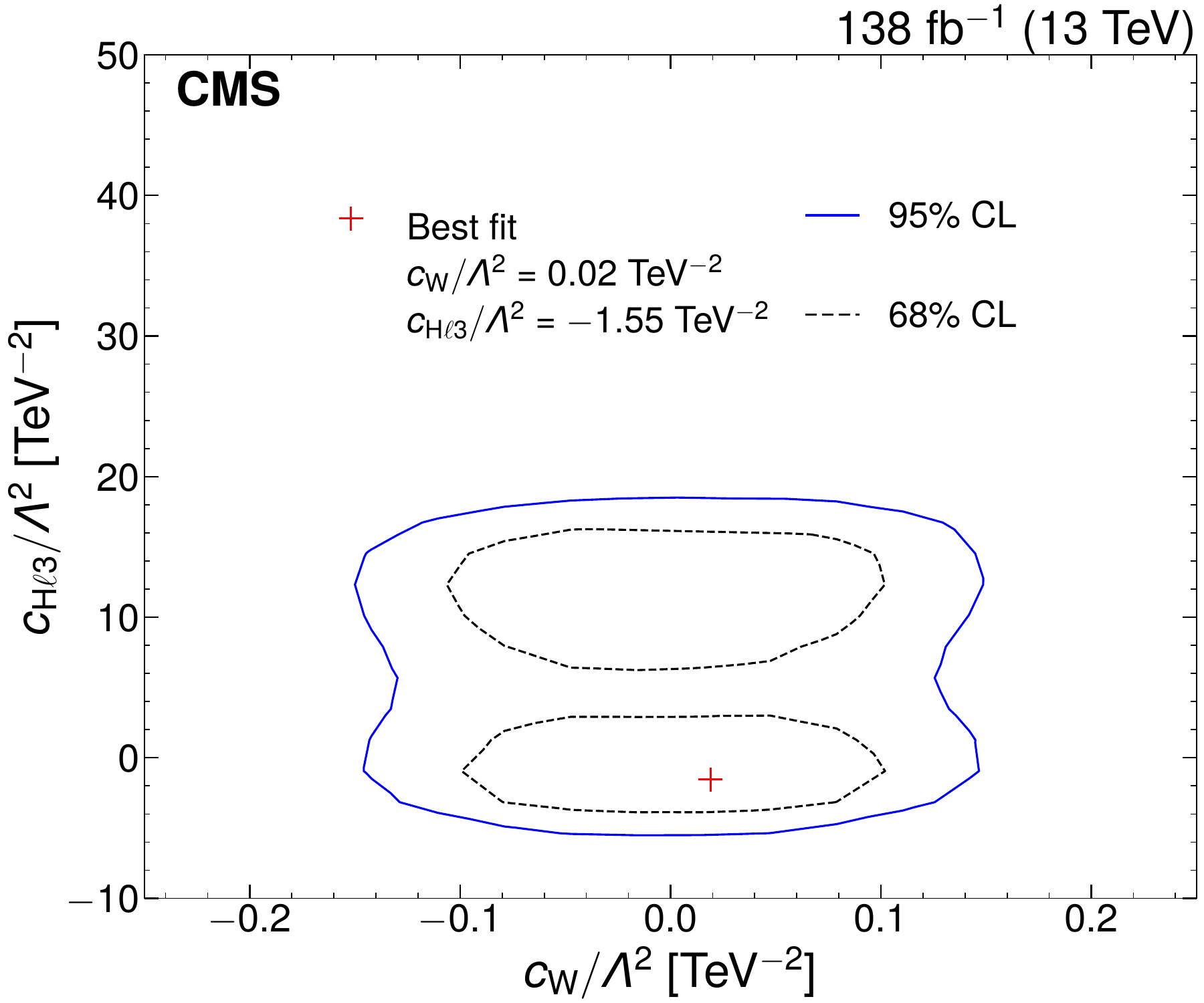}
\includegraphics[width=0.48\textwidth]{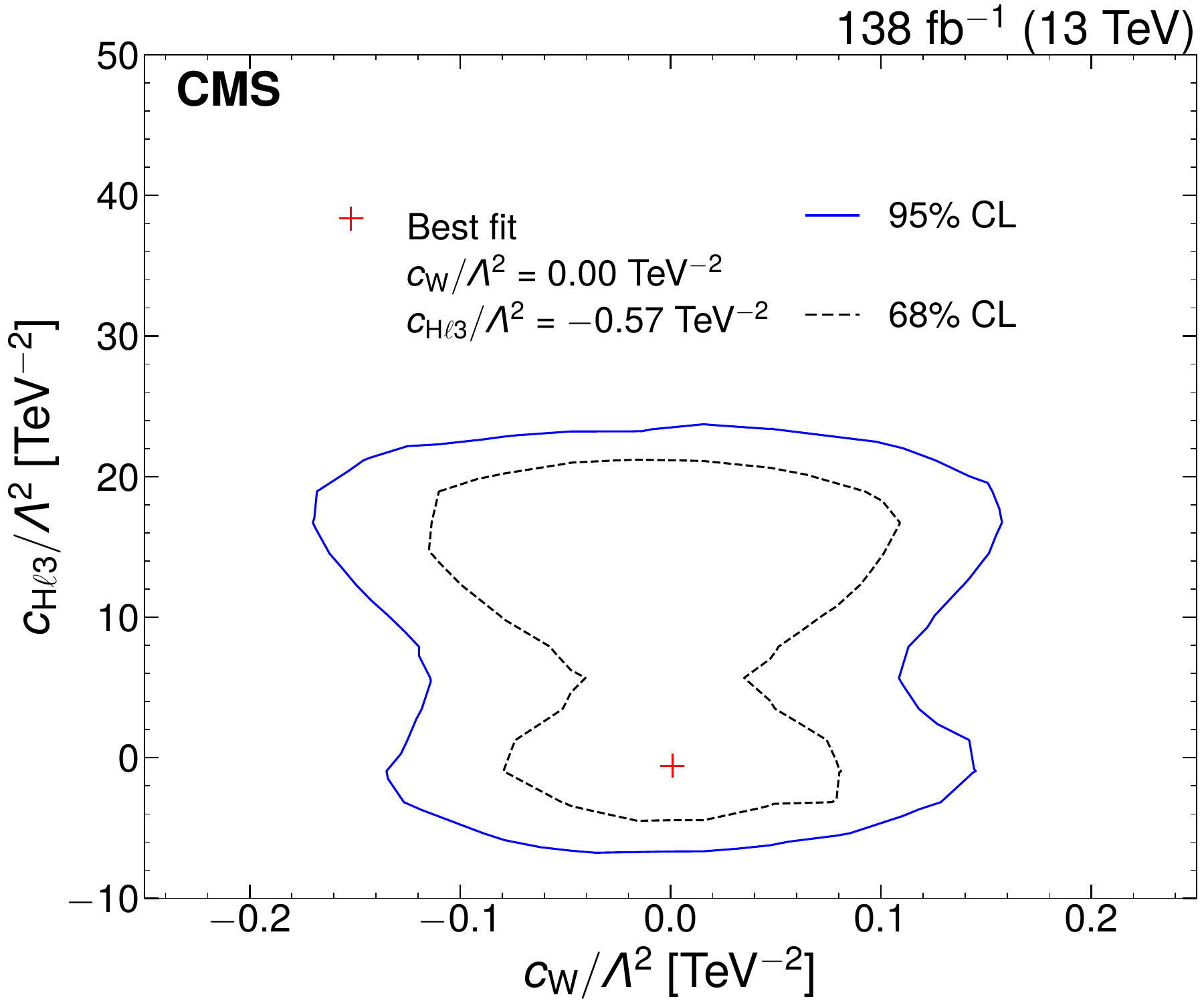}
\includegraphics[width=0.48\textwidth]{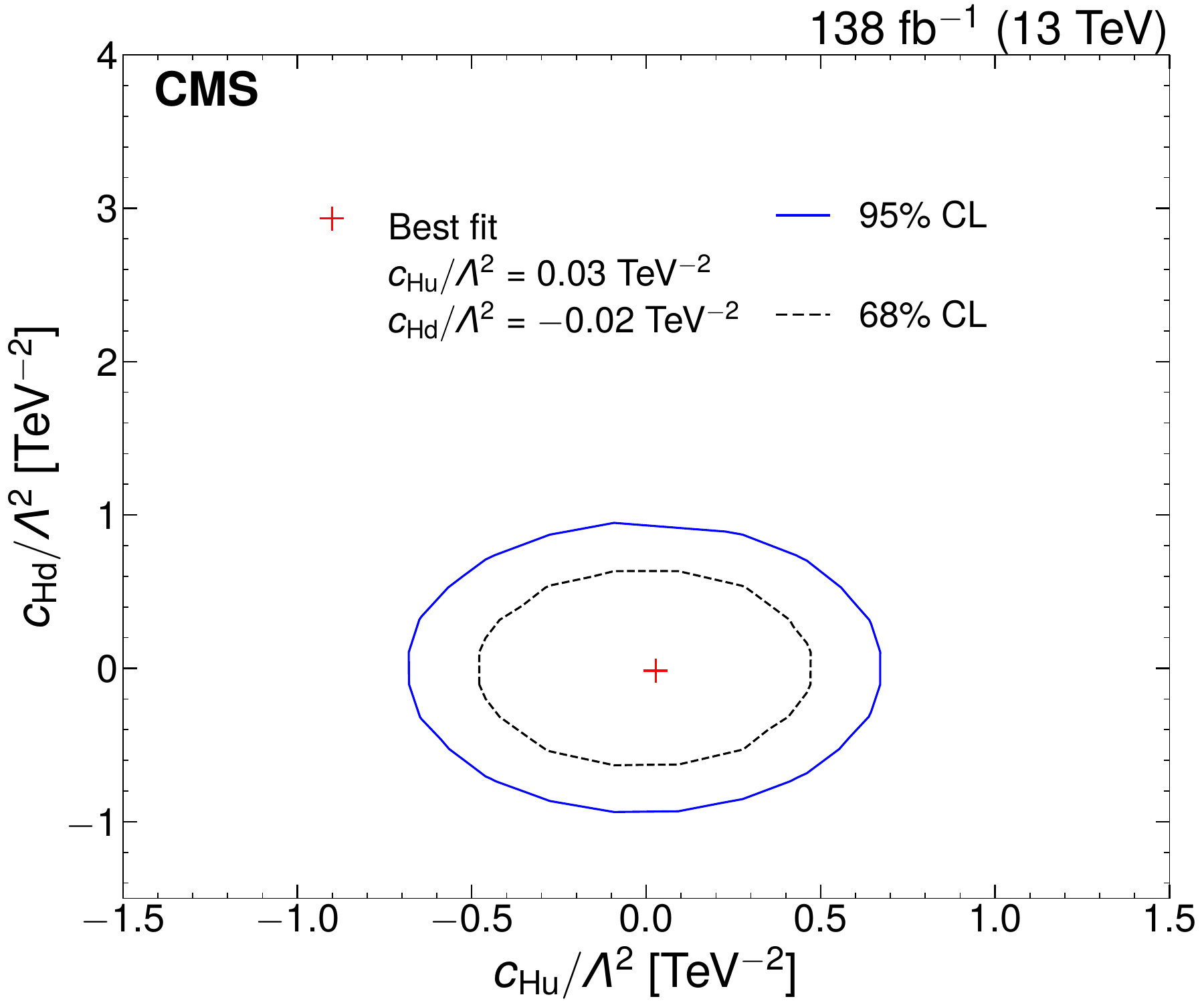}
\includegraphics[width=0.48\textwidth]{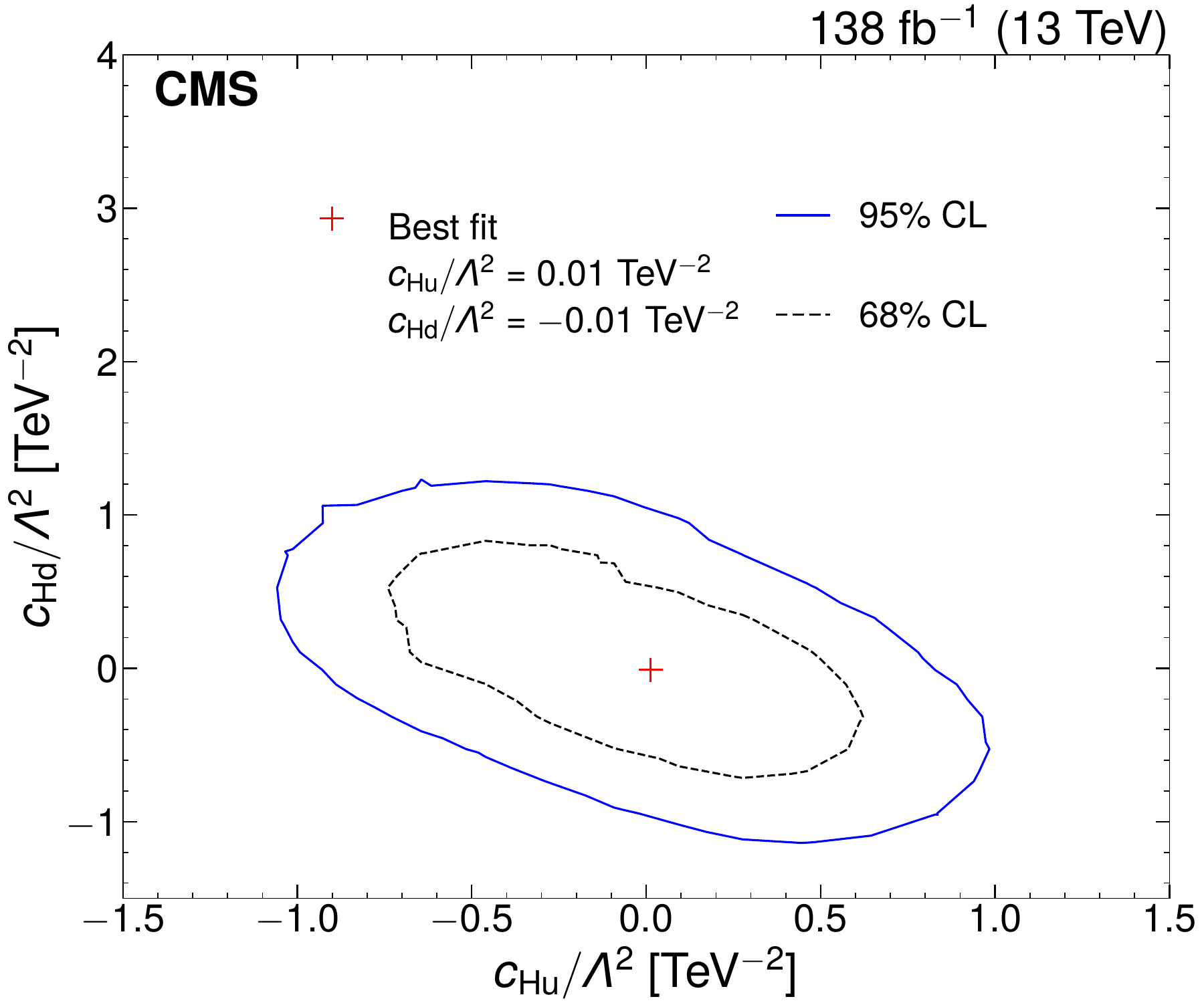}
\caption{ \label{fig:2D_bounds}
Bounds on pairs of Wilson coefficients.
The dashed black (solid blue) curves show the 68\% (95\%) \CL bounds determined by 2D likelihood quantiles.
The red plus sign indicates the minimum of \twoDNLL which can be compared to the
SM expectation (\ie, zero for both Wilson coefficients).
The three plots on the left are made freezing all Wilson coefficients to zero except for the two indicated on the plot.
The three plots on the right are made allowing all dim-6 Wilson coefficients to vary simultaneously. }
\end{figure*}

Two Wilson operators may have correlated constraints because their effects on kinematic distributions are similar.
In such a case, the impact of one is difficult to distinguish from the other -- one says that a ``flat direction''
arises when the Wilson coefficients are free to vary simultaneously. The bounds on each Wilson coefficient can be
much weaker in these double-parameter fits compared to the single-parameter fits described above.
We performed fits allowing two coefficients to vary simultaneously.
No significant flat dimensions appear.
Figure~\ref{fig:2D_bounds} shows illustrative examples.
As expected, the bounds obtained with profiling are somewhat weaker than those obtained when Wilson coefficients
are fixed to zero, but not drastically so.
The contours for $(\cW/\Lambda^2,\cHqthree/\Lambda^2)$ and $(\cHu/\Lambda^2,\cHd/\Lambda^2)$ are centered on the origin while the contours for $(\cW/\Lambda^2,\cHlthree/\Lambda^2)$ are asymmetric as a consequence of quantum mechanical interference between the SM and new physics matrix elements. For the case of the $(\cHu/\Lambda^2,\cHd/\Lambda^2)$ pair, allowing other Wilson coefficients to float leads to a change in the shape of the contours and the appearance of a mild anticorrelation between $\cHu/\Lambda^2$ and $\cHd/\Lambda^2$.

\begin{figure*}[htbp!]
  \centering
  \includegraphics[width=0.48\textwidth]{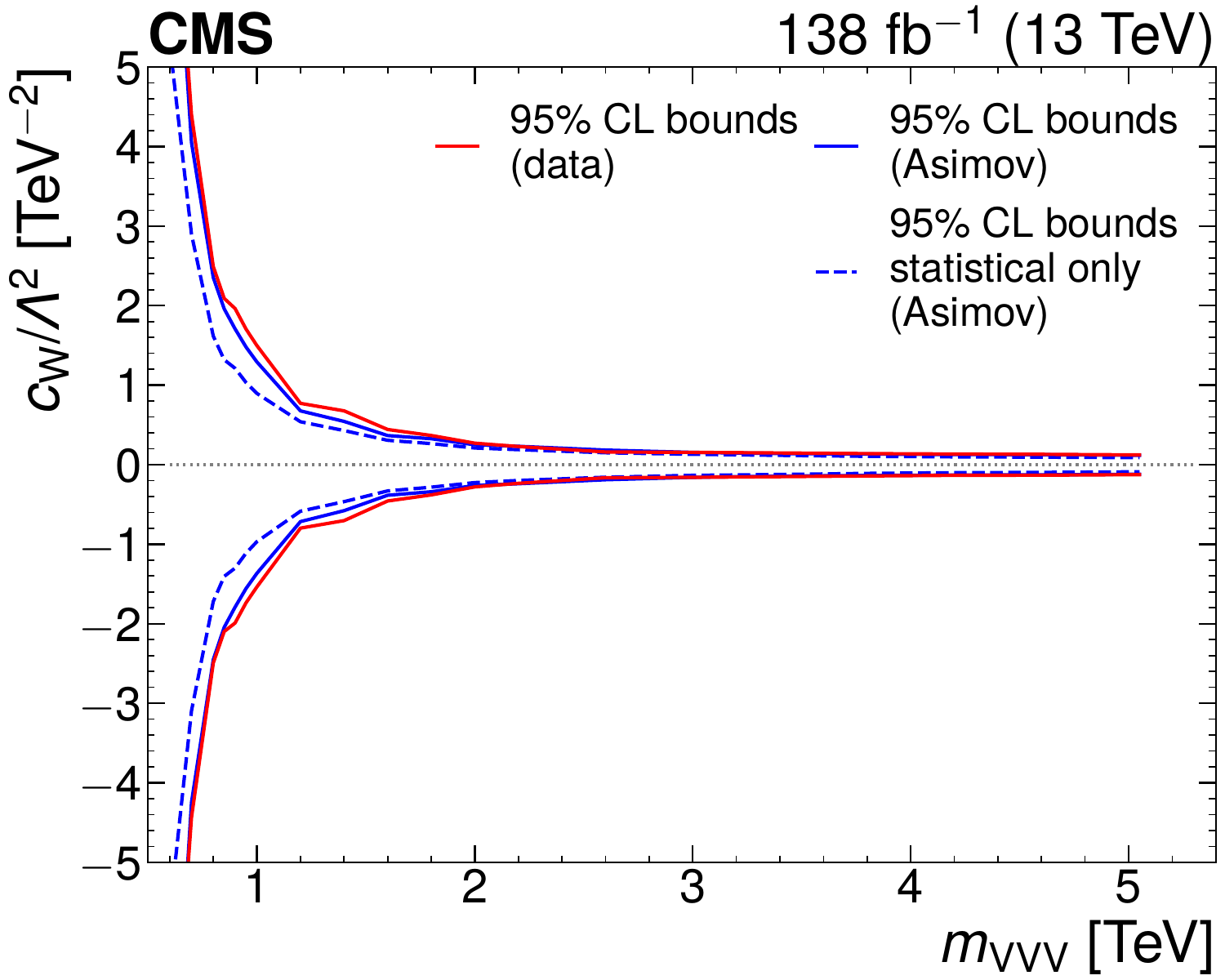}
  \includegraphics[width=0.48\textwidth]{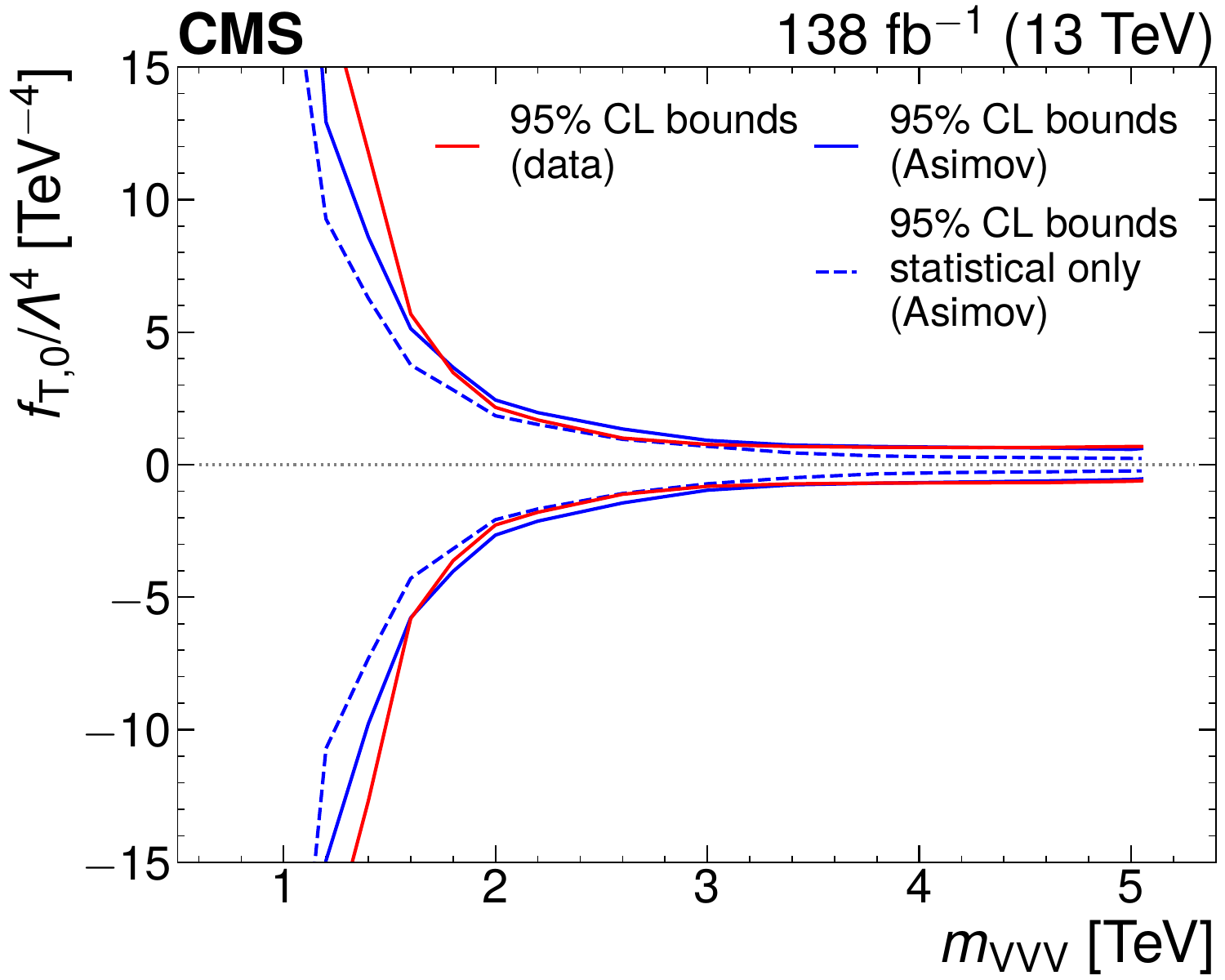}
  \caption{ \label{fig:clipping}
    Illustration of the impact of the clipping procedure.
    The horizontal axis indicates the threshold values placed in \mVVV (see text).
    When the threshold is high, there is little impact and the bounds on
    Wilson coefficients $\cW/\Lambda^2$ (left) and $\fT{0}/\Lambda^4$ (right)
    are essentially the same as reported in Tables~\ref{tab:limit_summary_dim6} and~\ref{tab:limit_summary_dim8}.
    As the upper bound on \mVVV is reduced, however, the bounds on Wilson coefficients weaken substantially.
    The solid (dashed) blue lines show the expected bounds computing using Asimov~\cite{Cowan:2010js}
    data sets including all (statistical) uncertainties.
  }
\end{figure*}

A key feature of the SM is that all cross sections for $2\to 2$ and $2\to 4$ scattering processes respect unitarity.
The same is not necessarily true for EFT-based extensions of the SM.
If $c / \Lambda^2$ or $f / \Lambda^4$ in Eq.~(\ref{eq:Lagrangian}) are too large in magnitude,
then unitarity may be violated, especially at high \mVVV.
Events at the highest \ST are the ones that lead to difficulties in an EFT-based interpretation.
One approach to circumventing this problem is to systematically remove events at high \ST through a procedure called ``clipping.''
In our implementation, we place a series of successive thresholds on \mVVV and reject MC events that fall above
the threshold. For each threshold value, we form templates in \ST and run the fit to obtain a bound on the given Wilson coefficient.
The key point is that the data are not changed -- only the signal expectation changes.
A series of \mVVV threshold values results in a series of upper and lower bounds on $c / \Lambda^2$ (or $f / \Lambda^4$)
which can be plotted in the $(\mVVV,c/\Lambda^2)$ (or $(\mVVV,f / \Lambda^4)$) plane.
The region between the curves is naturally narrowest at high \mVVV and significantly broader at lower \mVVV.
Figure~\ref{fig:clipping} shows the results of this clipping procedure for $\cW / \Lambda^2$ and
$\fT{0} / \Lambda^4$. Clearly, as the threshold for \mVVV decreases toward 1\TeV and below, the
bounds on the Wilson coefficients weaken considerably.

If one or more Wilson coefficients were nonzero and this analysis produced evidence for new physics
beyond the SM, an interpretation of the signal would begin with the \mVVV distribution.
We devised a fit that enables us to obtain the \mVVV distribution on the basis of the
observed \ST distribution.
The idea is to constrain bins of \mVVV using a template fit to \ST.
The templates in \ST are formed for bins in \mVVV using signal MC.
Since the correlation between \ST and \mVVV is good, the \ST templates for different \mVVV bins
are rather different and distinct:
the distribution in \ST reflects the boundaries of the given \mVVV bin.
All channels use the same bins in \mVVV but each channel has its own set of templates in \ST.
There is one multiplicative factor, similar to a signal strength, for each \mVVV bin.
Thus, all channels help determine a single set of multiplicative factors (in this case, there are five).
A maximum likelihood fit determines the values and uncertainties; all systematic uncertainties
described above are taken into account.

\begin{figure}[htbp!]
  \centering
  \includegraphics[width=0.48\textwidth]{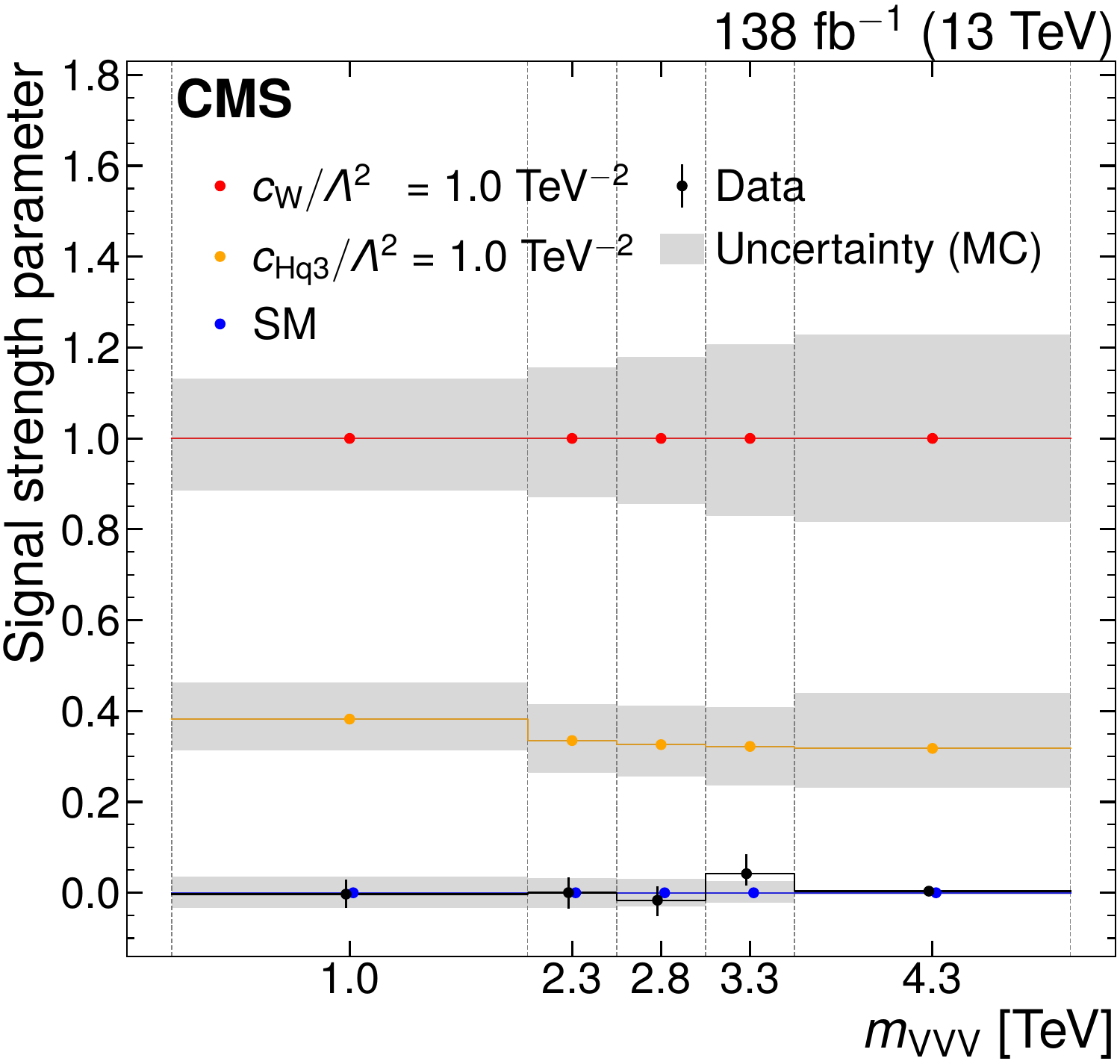}
  \caption{ \label{fig:templatefit}
    Visual summary of fitted multiplicative values obtained from the template fit.
    In essence, this plot shows the \mVVV distribution inferred from the data (black points)
    which agrees with the SM prediction (blue points).
    The red and orange points correspond to new physics scenarios (see text).
    The shaded regions centered on the red, orange, and blue points show the predicted total uncertainties
    based on fits to Asimov data sets. The error bars for the black dots show the total uncertainties
    when fitting the data. The black and blue points are slightly displaced to avoid overlap that would
    obscure them.
    The values are listed in Table~\ref{tab:templatefit}.
    }
\end{figure}

A reference model is needed to specify the \ST distribution when all multiplicative factors are set to unity.
The SM does not serve this purpose well because it contributes too few events in the highest \ST bins.
We defined our reference model by setting $\cW / \Lambda^2 = 1.0\TeV^{-2}$ and all other Wilson coefficients fixed to zero.
This choice corresponds to a well-defined scenario in which we would demonstrate the presence of new physics
if the scenario were true.
If nature corresponds to a different model, then the multiplicative factors would be different from zero and would
generally differ from one another.
We use $\cHqthree/\Lambda^2 = 1.0\TeV^{-2}$ to illustrate this point: the fitted values of the signal strength parameters
range from $0.32$ to $0.38\TeV^{-2}$ and vary gently across the \mVVV range.
For the SM, the values would be close to zero.
The results of the template fit to the data are reported in Table~\ref{tab:templatefit}
and summarized in Fig.~\ref{fig:templatefit}.

\begin{table}[hbtp!]
  \centering
  \topcaption{\label{tab:templatefit}
    Summary of the fitted multiplicative values in the template fit.
    The case $\cW/\Lambda^2 = 1\TeV^{-2}$ defines the reference model and
    $\cHqthree/\Lambda^2 = 1\TeV^{-2}$ is an alternative scenario resulting in
    multiplicative values less than unity.
    The SM expectation is zero and the measured values are consistent with the SM.
  }
  \renewcommand{\arraystretch}{1.3}
  \begin{tabular}{rcccr}
    \hline
    \mVVV [\TeVns{}] & \multicolumn{3}{c}{Expected} & Measured \\
    & $\cW / \Lambda^2 = 1\TeV^{-2}$ & $\cHqthree / \Lambda^2 = 1\TeV^{-2}$ & SM & \\
    \hline
    0 -- 2.0 & $1.00^{+0.13}_{-0.12}$ & $0.38^{+0.08}_{-0.07}$ & $0.00^{+0.04}_{-0.03}$ & $0.00 \pm 0.03$ \\
    2.0 -- 2.5 & $1.00^{+0.16}_{-0.13}$ & $0.34^{+0.08}_{-0.07}$ & $0.00 \pm 0.03$ & $0.00^{+0.03}_{-0.04}$ \\
    2.5 -- 3.0 & $1.00^{+0.18}_{-0.15}$ & $0.33^{+0.09}_{-0.07}$ & $0.00 \pm 0.03$ & $-0.02^{+0.03}_{-0.04}$ \\
    3.0 -- 3.5 & $1.00^{+0.21}_{-0.17}$ & $0.32 \pm 0.09$ & $0.00^{+0.03}_{-0.02}$ & $0.04^{+0.04}_{-0.03}$ \\
    3.5 -- ~$\infty$ & $1.00^{+0.23}_{-0.18}$ & $0.32^{+0.12}_{-0.09}$ & $0.000^{+0.007}_{-0.004}$ & $0.00 \pm 0.01$ \\
    \hline
  \end{tabular}
\end{table}

\begin{figure*}[htbp!]
    \centering
    \includegraphics[width=0.8\textwidth]{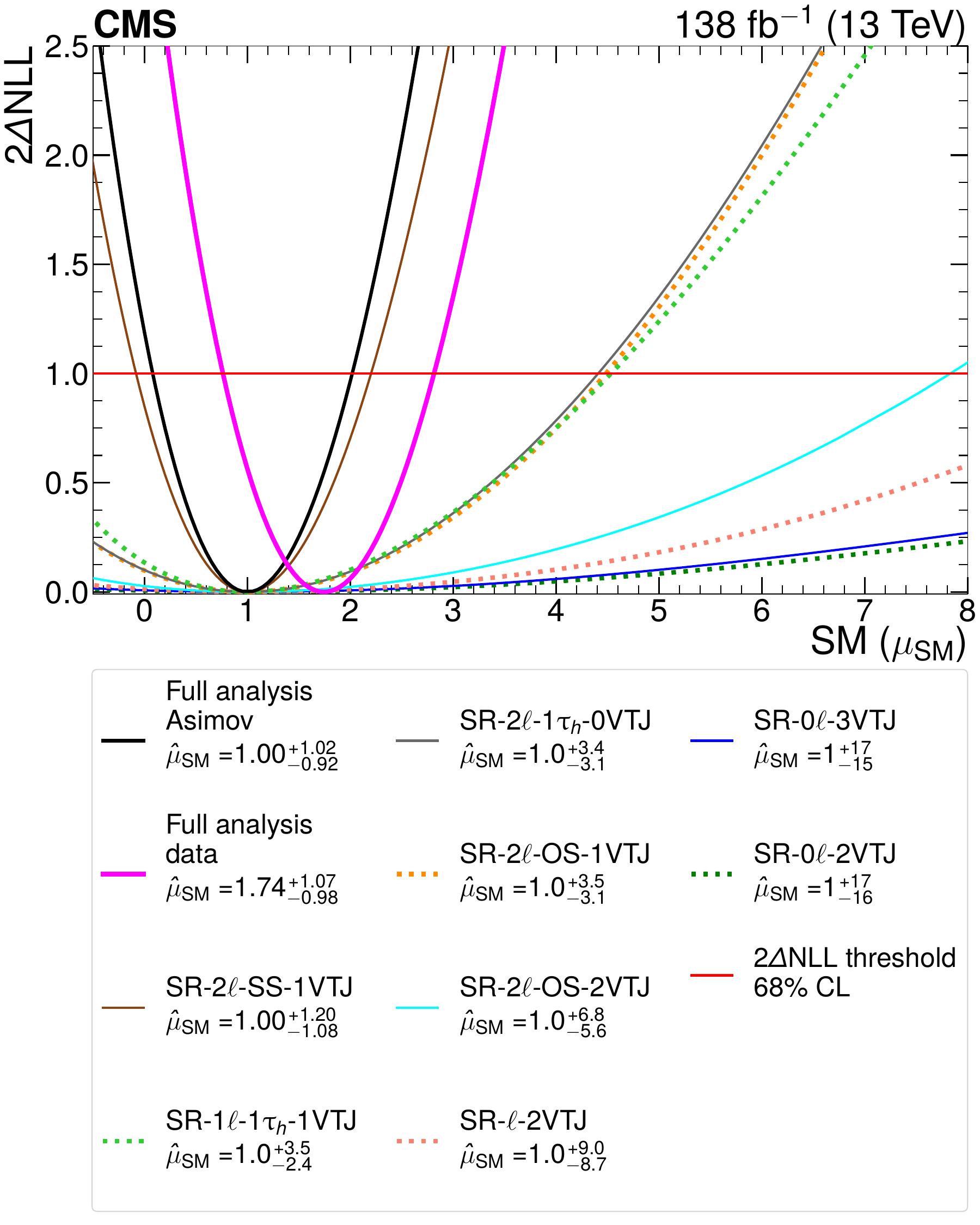}
    \caption{ \label{fig:SM}
      Sensitivity to the SM \VVV production process. The curves show the variation of \twoDNLL
      with the SM signal strength, $\muSM$.
      The Asimov curves for all the channels are shown, and the solid black curve shows the combined Asimov result.
      The solid magenta curve shows the combined result based on CMS data.
      Numerical values for 68\% \CL Asimov intervals and point values are listed in the box below the plot.
    }
\end{figure*}

Our selection retains some sensitivity to the SM \VVV production process.
We performed a fit in which there are no Wilson coefficients but there is a signal strength
parameter $\muSM$ for the SM cross section. If the SM prediction is correct then $\muSM$ should be consistent with unity.
Figure~\ref{fig:SM} shows \twoDNLL as $\muSM$ is varied, for the Asimov~\cite{Cowan:2010js} data sets and for the data,
combining all channels. Most of the sensitivity comes from the \twoLEPSSoneFJ channel.
The result of the fit to data is $\muSM = 1.74^{+1.07}_{-0.98}$.

\section{Summary}
\label{sec:Conclusion}

A search for new physics in the production of three massive gauge bosons in proton-proton collisions
($\Pp\Pp\to\VVV$, with $\PV = \PW$ or $\PZ$)
has been reported. The analysis targets the boosted regime in which the bosons have transverse momentum $\pt > 200\GeV$.
When they decay hadronically, large-radius jets with substructure
are formed; we identify such \fatjets using the \ParticleNet algorithm.
Signal \fatjets have a soft-drop mass consistent with the \PW or \PZ boson mass.
Several analysis channels are defined according to the multiplicities of leptons and \fatjets in an event;
two channels feature hadronically decaying \PGt leptons.
Signal regions are defined by a suitable kinematic variable that correlates well with the triboson invariant mass, \mVVV.
The observed signal yields are interpreted in a standard model effective field theory framework with twelve dimension-6 and twenty dimension-8 Wilson operators.
Agreement with the SM predictions is good, and bounds are placed on Wilson coefficients at 95\% \CL
in two scenarios.
In the first, all Wilson coefficients are fixed to zero except the one under consideration, and in the second,
all coefficients are allowed to float. Examples of bounds on pairs of Wilson coefficients are given, as well.
Potential difficulties with unitarity are handled using a clipping procedure: the bounds on individual Wilson coefficients
weaken as the threshold on \mVVV is lowered.
A template fit is introduced to infer the \mVVV distribution; the result is consistent with the SM.
Finally, the result of a fit for the signal strength for SM \VVV production is, again, consistent with the SM.

\begin{acknowledgments}
We congratulate our colleagues in the CERN accelerator departments for the excellent performance of the LHC and thank the technical and administrative staffs at CERN and at other CMS institutes for their contributions to the success of the CMS effort. In addition, we gratefully acknowledge the computing centers and personnel of the Worldwide LHC Computing Grid and other centers for delivering so effectively the computing infrastructure essential to our analyses. Finally, we acknowledge the enduring support for the construction and operation of the LHC, the CMS detector, and the supporting computing infrastructure provided by the following funding agencies: SC (Armenia), BMBWF and FWF (Austria); FNRS and FWO (Belgium); CNPq, CAPES, FAPERJ, FAPERGS, and FAPESP (Brazil); MES and BNSF (Bulgaria); CERN; CAS, MoST, and NSFC (China); MINCIENCIAS (Colombia); MSES and CSF (Croatia); RIF (Cyprus); SENESCYT (Ecuador); ERC PRG and PSG, TARISTU24-TK10 and MoER TK202 (Estonia); Academy of Finland, MEC, and HIP (Finland); CEA and CNRS/IN2P3 (France); SRNSF (Georgia); BMFTR, DFG, and HGF (Germany); GSRI (Greece); MATE and NKFIH (Hungary); DAE and DST (India); IPM (Iran); SFI (Ireland); INFN (Italy); MSIT and NRF (Republic of Korea); MES (Latvia); LMTLT (Lithuania); MOE and UM (Malaysia); BUAP, CINVESTAV, CONACYT, LNS, SEP, and UASLP-FAI (Mexico); MOS (Montenegro); MBIE (New Zealand); PAEC (Pakistan); MSHE, NSC, and NAWA (Poland); FCT (Portugal); MESTD (Serbia); MICIU/AEI and PCTI (Spain); MOSTR (Sri Lanka); Swiss Funding Agencies (Switzerland); MST (Taipei); MHESI (Thailand); TUBITAK and TENMAK (T\"{u}rkiye); NASU (Ukraine); STFC (United Kingdom); DOE and NSF (USA).

\hyphenation{Rachada-pisek} Individuals have received support from the Marie-Curie program and the European Research Council and Horizon 2020 Grant, contract Nos.\ 675440, 724704, 752730, 758316, 765710, 824093, 101115353, 101002207, 101001205, and COST Action CA16108 (European Union); the Leventis Foundation; the Alfred P.\ Sloan Foundation; the Alexander von Humboldt Foundation; the Science Committee, project no. 22rl-037 (Armenia); the Fonds pour la Formation \`a la Recherche dans l'Industrie et dans l'Agriculture (FRIA) and Fonds voor Wetenschappelijk Onderzoek contract No. 1228724N (Belgium); the Beijing Municipal Science \& Technology Commission, No. Z191100007219010, the Fundamental Research Funds for the Central Universities, the Ministry of Science and Technology of China under Grant No. 2023YFA1605804, the Natural Science Foundation of China under Grant No. 12535004, and USTC Research Funds of the Double First-Class Initiative No.\ YD2030002017 (China); the Ministry of Education, Youth and Sports (MEYS) of the Czech Republic; the Shota Rustaveli National Science Foundation (Georgia); the Deutsche Forschungsgemeinschaft (DFG), among others, under Germany's Excellence Strategy -- EXC 2121 ``Quantum Universe" -- 390833306, and under project number 400140256 - GRK2497; the Hellenic Foundation for Research and Innovation (HFRI), Project Number 2288 (Greece); the Hungarian Academy of Sciences, the New National Excellence Program - \'UNKP, the NKFIH research grants K 131991, K 138136, K 143460, K 143477, K 147557, K 146913, K 146914, K 147048, TKP2021-NKTA-64, and 2025-1.1.5-NEMZ\_KI-2025-00004, and MATE KKP and KKPCs Research Excellence and Flagship Research Groups grants (Hungary); the Council of Science and Industrial Research, India; ICSC -- National Research Center for High Performance Computing, Big Data and Quantum Computing, FAIR -- Future Artificial Intelligence Research, and CUP I53D23001070006 (Mission 4 Component 1), funded by the NextGenerationEU program, the Italian Ministry of University and Research (MUR) under Bando PRIN 2022 -- CUP I53C24002390006, PRIN PRIMULA 2022RBYK7T (Italy); the Latvian Council of Science; the Ministry of Science and Higher Education, project no. 2022/WK/14, and the National Science Center, contracts Opus 2021/41/B/ST2/01369, 2021/43/B/ST2/01552, 2023/49/B/ST2/03273, and the NAWA contract BPN/PPO/2021/1/00011 (Poland); the Funda\c{c}\~ao para a Ci\^encia e a Tecnologia (Portugal); the National Priorities Research Program by Qatar National Research Fund; MICIU/AEI/10.13039/501100011033, ERDF/EU, ``European Union NextGenerationEU/PRTR", projects PID2022-142604OB-C21, PID2022-139519OB-C21, PID2023-147706NB-I00, PID2023-148896NB-I00, PID2023-146983NB-I00, PID2023-147115NB-I00, PID2023-148418NB-C41, PID2023-148418NB-C42, PID2023-148418NB-C43, PID2023-148418NB-C44, PID2024-158190NB-C22, RYC2021-033305-I, RYC2024-048719-I, CNS2023-144781, CNS2024-154769 and Plan de Ciencia, Tecnolog{\'i}a e Innovaci{\'o}n de Asturias, Spain; the Chulalongkorn Academic into Its 2nd Century Project Advancement Project, the National Science, Research and Innovation Fund program IND\_FF\_68\_369\_2300\_097, and the Program Management Unit for Human Resources \& Institutional Development, Research and Innovation, grant B39G680009 (Thailand); the Eric \& Wendy Schmidt Fund for Strategic Innovation through the CERN Next Generation Triggers project under grant agreement number SIF-2023-004; the Kavli Foundation; the Nvidia Corporation; the SuperMicro Corporation; the Welch Foundation, contract C-1845; and the Weston Havens Foundation (USA).
\end{acknowledgments}\section*{Data availability} Release and preservation of data used by the CMS Collaboration as the basis for publications is guided by the  \href{https://doi.org/10.7483/OPENDATA.CMS.1BNU.8V1W}{CMS data preservation, re-use and open access policy}.

\bibliography{auto_generated}
\cleardoublepage \appendix\section{The CMS Collaboration \label{app:collab}}\begin{sloppypar}\hyphenpenalty=5000\widowpenalty=500\clubpenalty=5000\cmsinstitute{Yerevan Physics Institute, Yerevan, Armenia}
{\tolerance=6000
A.~Hayrapetyan, V.~Makarenko\cmsorcid{0000-0002-8406-8605}, A.~Tumasyan\cmsAuthorMark{1}\cmsorcid{0009-0000-0684-6742}
\par}
\cmsinstitute{Institut f\"{u}r Hochenergiephysik, Vienna, Austria}
{\tolerance=6000
W.~Adam\cmsorcid{0000-0001-9099-4341}, L.~Benato\cmsorcid{0000-0001-5135-7489}, T.~Bergauer\cmsorcid{0000-0002-5786-0293}, M.~Dragicevic\cmsorcid{0000-0003-1967-6783}, P.S.~Hussain\cmsorcid{0000-0002-4825-5278}, M.~Jeitler\cmsAuthorMark{2}\cmsorcid{0000-0002-5141-9560}, N.~Krammer\cmsorcid{0000-0002-0548-0985}, A.~Li\cmsorcid{0000-0002-4547-116X}, D.~Liko\cmsorcid{0000-0002-3380-473X}, M.~Matthewman, J.~Schieck\cmsAuthorMark{2}\cmsorcid{0000-0002-1058-8093}, R.~Sch\"{o}fbeck\cmsAuthorMark{2}\cmsorcid{0000-0002-2332-8784}, M.~Shooshtari\cmsorcid{0009-0004-8882-4887}, M.~Sonawane\cmsorcid{0000-0003-0510-7010}, W.~Waltenberger\cmsorcid{0000-0002-6215-7228}, C.-E.~Wulz\cmsAuthorMark{2}\cmsorcid{0000-0001-9226-5812}
% ...
\par}
\vskip\cmsinstskip
\dag:~Deceased\\
$^{1}$Also at Yerevan State University, Yerevan, Armenia\\
$^{2}$Also at TU Wien, Vienna, Austria\\
% ...
\end{sloppypar}
\end{document}